\documentclass[11pt,preprint]{aastex}
\usepackage{psfig,lscape,rotate,natbib}
\received{}
\begin{document}
\newcommand{\caii}{Ca{\,\sc ii}}
\newcommand{\msig}{$\mbh$--$\sigma$}
\newcommand{\kms}{{\rm km\;s^{-1}}}
\newcommand{\msol}{{\rm M_{\odot}}}
\newcommand{\mbh}{M_{BH}}
\newcommand{\ledd}{L_{Edd}}
\newcommand{\mdot}{\dot{m}}
\newcommand{\mdotedd}{\dot{m}_{Edd}}
\newcommand{\rg}{r_{g}}
\newcommand{\ang}{\rm \AA}
\newcommand{\oxi}{[O{\,\sc i}]}
\newcommand{\oxii}{[O{\,\sc ii}]}
\newcommand{\oxiii}{[O{\,\sc iii}]}
\newcommand{\feii}{[Fe{\,\sc ii}]}
\newcommand{\fek}{Fe~K$\alpha$}
\newcommand{\sii}{[S{\,\sc ii}]}
\newcommand{\nii}{[N{\,\sc ii}]}
\newcommand{\ergs}{{\rm erg\;s^{-1}}}
\newcommand{\ergscm}{{\rm erg\;s^{-1}\;cm^{-2}}}
\newcommand{\fbha}{{\rm F_{broad\;H\alpha}}}

\def\ls{\lower 2pt \hbox{$\;\scriptscriptstyle \buildrel<\over\sim\;$}} 
\def\gs{\lower 2pt \hbox{$\;\scriptscriptstyle \buildrel>\over\sim\;$}} 

\title{Long-Term Profile Variability in Active Galactic Nuclei with
  Double-Peaked Balmer Emission Lines}

\shorttitle{Long-term Profile Variability in Double-Peaked Emitters}

\author{Karen T. Lewis} 

\affil{Department of Physics and Astronomy, Dickinson College,
Carlisle, PA 17013, e-mail: {\tt lewiska@dickinson.edu}}

\author{Michael Eracleous\altaffilmark{1}}

\affil{Department of Astronomy and Astrophysics, The Pennsylvania
State University, 525 Davey Laboratory, University Park, PA 16802}

\altaffiltext{1}{Center for Gravitational Wave Physics, The
Pennsylvania State University, 104 Davey Laboratory, University Park,
PA 16802}

\author{Thaisa Storchi-Bergmann} 

\affil{Instituto de Fisica, Universidade Federal do Rio Grande do Sul,
Campus do Vale, Porto Alegre, RS, Brazil}

\begin{abstract}

An increasing number of Active Galactic Nuclei (AGNs) exhibit broad,
double-peaked Balmer emission lines,which represent some of the best
evidence for the existence of relatively large-scale accretion disks
in AGNs.  A set of 20 double-peaked emitters have been monitored for
nearly a decade in order to observe long-term variations in the
profiles of the double-peaked Balmer lines. Variations generally occur
on timescales of years, and are attributed to physical changes in the
accretion disk. Here we characterize the variability of a subset of
seven double-peaked emitters in a model independent way.  We find that
variability is caused primarily by the presence of one or more
discrete ``lumps'' of excess emission; over a timescale of a year (and
sometimes less) these lumps change in amplitude and shape, but the
projected velocity of these lumps changes over much longer timescales
(several years). We also find that all of the objects exhibit red
peaks that are stronger than the blue peak at some epochs and/or
blueshifts in the overall profile, contrary to
the expectations for a simple, circular accretion disk model, thus
emphasizing the need for asymmetries in the accretion
disk.  Comparisons with two simple models, an elliptical accretion disk
and a circular disk with a spiral arm, are unable to reproduce all
aspects of the observed variability, although both account for some of
the observed behaviors. Three of the seven objects have robust
estimates of the black hole masses. For these objects the observed
variability timescale is consistent with the expected precession
timescale for a spiral arm, but incompatible with that of an
elliptical accretion disk. We suggest that with the simple
modification of allowing the spiral arm to be fragmented, many of the
observed variability patterns could be reproduced.
\end{abstract}

\section{INTRODUCTION}\label{variability_intro}

It is now generally accepted that at the heart of an Active Galactic
Nucleus (AGN) lies a supermassive black hole with a mass in excess of
$10^{6}\;\msol$ that is accreting matter from its host galaxy
\citep[see, e.g.,][]{salpeter64,rees84,petersonagn}.  As matter
spirals inward, it forms an equatorial accretion disk around the
black hole (such as that described by Shakura \& Sunyaev 1973), whose
inner portions are heated to UV-emitting temperatures by ``viscous''
stresses which tap into the potential energy of the accreting gas.

The most direct kinematic evidence for {\it large-scale} accretion
disks in AGNs comes from the broad, double-peaked Balmer emission
lines detected in some AGNs.  Double-peaked emission lines were first
observed in the broad-line radio galaxies (BLRGs) Arp~102B
\citep*{ssk83,ch89,chf89}, 3C~390.3 \citep{o87,p88}, and 3C~332
\citep{h90}. Given the similarity between these lines and those
observed in cataclysmic variables (CVs), these authors suggested that
the lines originated in an accretion disk surrounding the black hole
at distances of hundreds to thousands of gravitational radii
($\rg$~=~$G\;m_{\rm BH}/c^{2}$, where {\rm $m_{\rm BH}$} is the mass of the black
hole). As discussed in great detail in \citet{eh03} and
\citet{gezari07}, alternative scenarios for the origin of the
double-peaked lines are not consistent with the observed variability
and multi-wavelength properties of double-peaked emitters.  Thus we
adopt this interpretation hereafter.  A later survey of (mostly
broad-lined) radio-loud AGN showed that $\sim$~20\% of the observed
objects exhibited double-peaked Balmer emission lines
\citep{eh94,eh03}. \citet{s03} found 116 double-peaked emitters in the
Sloan Digital Sky Survey \citep[SDSS;][]{y00}, representing 3\% of the
$z<0.332$ AGN population. In addition, several Low Ionization Nuclear
Emission-line Regions \citep[LINERs,][]{h80} were found to have
double-peaked Balmer lines, including NGC~1097 \citep*{sb93}, M81
\citep{b96}, NGC~4203 \citep{s00}, NGC~4450 \citep{h00}, and NGC~4579
\citep{b01}. Most recently, \citet{mbek07} discovered that the
obscured AGN NGC~2110 shows a double-peaked H$\alpha$ emission line in
polarized light.

The prototypical double-peaked emitters (Arp~102B, 3C~390.3, and
3C~332) exhibited variations in their line profiles that took place on
timescales of years \citep*[for examples,
see][]{vz91,zvg91,g99,shap01,sps00,serg02,gezari07}, with the most
striking variability being occasional reversals in the relative
strength of the red and blue peaks.  This long-term variability is
intriguing because it is un-related to more rapid changes in
luminosity but might be related to physical changes in the accretion
disk. This profile variability can be exploited to test various models
for physical phenomena in the outer accretion disk.

The observed {\it long-term} variability of the double-peaked line
profiles (on time scales of several months to a few years) and the
fact that in many objects the red peak is occasionally stronger than
the blue peak (in contrast to the expectations from relativistic
Doppler boosting) mandates that the accretion disk is
non-axisymmetric. A few of the best-monitored objects (Arp~102B,
3C~390.3, 3C~332, and NGC~1097) have been studied in detail using
models that are simple extensions to a circular accretion disk. These
include circular disks with orbiting bright spots or ``patches''
\citep*[Arp~102B and 3C~390.3,][]{ne97,sps00,zvg91}, a disk which is
composed of randomly distributed clouds that rotate in the same plane
\citep[Arp~102B,][]{sps00}, spiral emissivity perturbations
\citep[3C~390.3, 3C~332, and NGC~1097,][]{g99,sb03} and precessing
elliptical disks \citep[NGC~1097,][]{e95,sb95,sb97,sb03}. All of these
non-axisymmetric models naturally lead to modulations in the ratio of
the red to blue peak flux; in the case of the first two models,
modulations occur on the dynamical timescale whereas for the latter
two, the variations occur on the longer precession timescale of the
spiral density wave or the elliptical disk. However, in all cases,
these simple models do not reproduce the fine details of the profile
variability.

\begin{figure}
\includegraphics[angle=-90,scale=0.6]{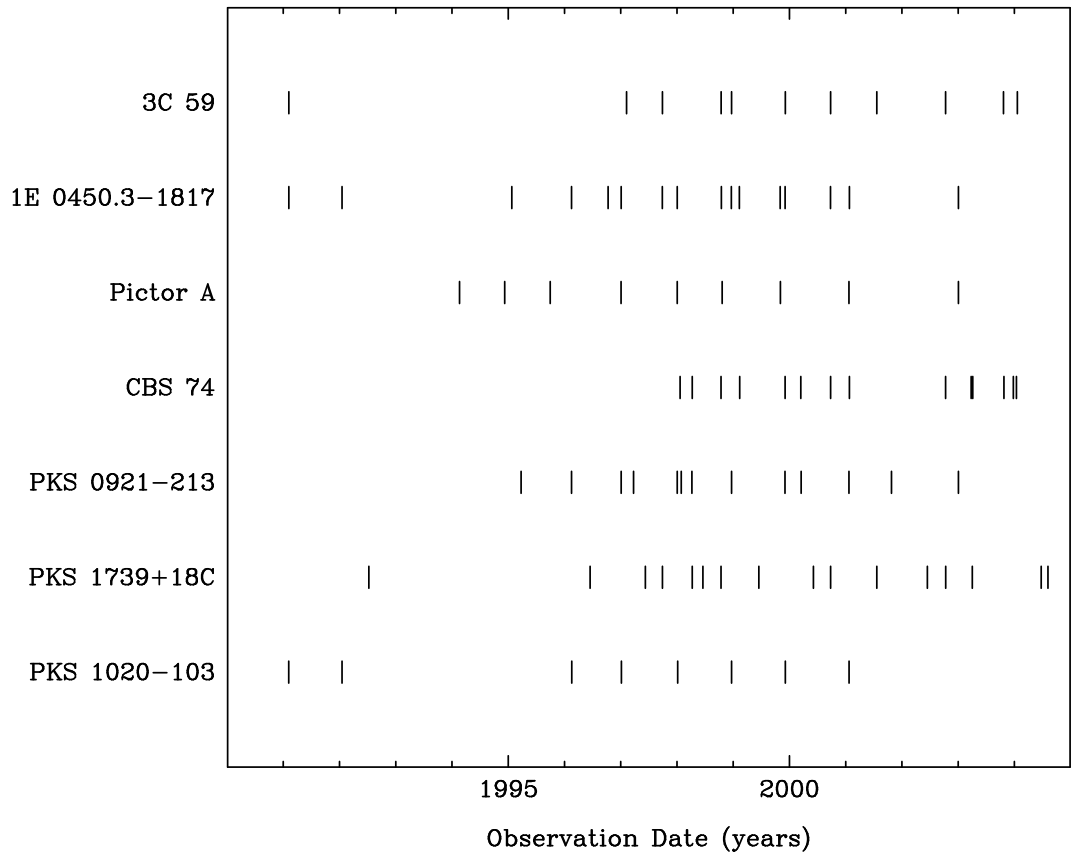} 
\caption{\label{obslog_fig} Plot of temporal coverage for the seven
objects in this study.  Each vertical bar indicates an observation
described in Table~\ref{all_obs}.}
\end{figure}

It is extremely important to determine whether the profile variability
observed in the small number of well-monitored objects is common to
all double-peaked emitters. With a larger sample, one can determine if
there are {\it universal} variability patterns that point to a global
phenomenon that occurs in AGN accretion disks.  Thus, a campaign was
undertaken to observe a larger sample of $\sim$20 radio-loud
double-peaked emitters from the \citet{eh94} sample, in addition to a
few more recently discovered objects. This campaign spanned nearly a
decade and many objects were observed 2--3 times per year, although
some objects, particularly those in the southern hemisphere, were only
observed once per year. In this paper we present results for seven
objects whose long-term variability has not been studied in any
detail: 3C~59, 1E~0450.3--1817, Pictor A, CBS~74, PKS~0921--213,
PKS~1020--103, and PKS~1739+18C. The properties of these objects are
given in Table~\ref{variability_objects} and a plot of the temporal
coverage of the observing campaign is provided in
Figure~\ref{obslog_fig}. \citet{gezari07} present results for
Arp~102B, 3C~390.3, 3C~332, PKS~0235+023, Mkn~668, 3C~227, and
3C~382.  The most recent profile variability results for NGC~1097 are
presented by \citet{sb03}, while the remaining objects from this
monitoring program will be presented in Flohic et al. (2010, in
preparation).

The primary goal of this paper is to characterize the observed
variability patterns in these seven objects in a {\it
model-independent} way. Currently, the models which have been used to
fit the line profiles represent the most simple extensions to a
circular disk. As mentioned above, these models fail to reproduce the
fine details of the profile variability.  A detailed characterization
of the variability will help guide the refinement of these models and
more importantly inspire ideas for new families of models. In essence,
these results will serve as benchmark that models can be easily tested
against.

This paper is organized as follows. In \S\ref{variability_reductions},
the observations and data reductions are
described. \S\ref{variability_analysis} is devoted to describing the
methods for characterizing the profile variability in a
model-independent way and noting the common trends in the data.  In
\S\ref{variability_models} we detail the two simple models we will
compare with the observations, including their physical motivation,
the model profile calculation, and the common variability trends. In
\S\ref{variability_discussion} we assess the viability of these two
models through a comparison of the observed and predicted variability
trends and timescales. We also suggest refinements to the models
which might afford a better description of the observed variability
trends. Finally, in \S\ref{variability_conclusions} we summarize the
primary findings.

\section{OBSERVATIONS AND DATA REDUCTIONS}\label{variability_reductions}

The majority of the spectra were obtained over the period from
1991--2004 using the 2.1m and 4m telescopes at Kitt Peak National
Observatory (KPNO), the 1.5m and 4m telescopes at Cerro-Tololo
Inter-American Observatory (CTIO), the 3m Shane Telescope at Lick
Observatory, the 1.3m and 2.4m telescopes at the MDM Observatory, and
the 9.2m Hobby-Eberly Telescope (HET) at the McDonald Observatory. A
complete list of the spectrographs and gratings used is provided in
Table \ref{variability_instruments} and a journal of the observations
is given in Table~\ref{all_obs}. The spectra were taken through a
narrow slit (1.$\!\!^{\prime\prime}$25 -- 2.$\!\!^{\prime\prime}$0)
and the seeing varied from 1$\arcsec$--3$\arcsec$.  The spectra were
extracted from a 4$\arcsec$-8$\arcsec$ wide region along the slit and
the resulting spectral resolution ranged from 3$\arcsec$--8$\arcsec$;
the resolution for each instrumental configuration is listed in Table
\ref{variability_instruments}. Most of the data reductions were
carried out by us, although some spectra were provided in reduced form
by A.~V. Filippenko (see also the notes to Table \ref{all_obs}.)  The
primary data reductions---the bias level correction, flat fielding,
sky subtraction, removal of bad columns and cosmic rays, extraction of
the spectra, and wavelength calibration---were performed with the Image
Reduction and Analysis Facility (IRAF)\footnotemark\footnotetext{IRAF
is distributed by the National Optical Astronomy Observatory, which
is operated by the Association of Universities for Research in
Astronomy, Inc., under cooperative agreement with the National Science
Foundation.}. Flux calibration was performed using spectrophotometric
standard stars drawn from \citet{phot1}, \citet{phot2} or
\citet{phot3} that were observed at the beginning and end of each
night. Although the relative flux scale is accurate to better than
5\%, the absolute flux scale is uncertain by up a factor of 2 due to
the small slit used and the occasional non-photometric sky conditions.
There are several atmospheric features that contaminate the spectra,
namely the ${\rm O_{2}}$ bands near 6890\ang~ and 7600\ang~ (the B-
and A- bands, respectively) and the water-vapor bands near 7200\ang~
and 8200\ang. These were corrected using templates constructed from
the (featureless) spectra of the standard stars.

\section{PROFILE ANALYSIS}\label{variability_analysis}

\begin{figure}
\epsscale{0.5}
\plotone{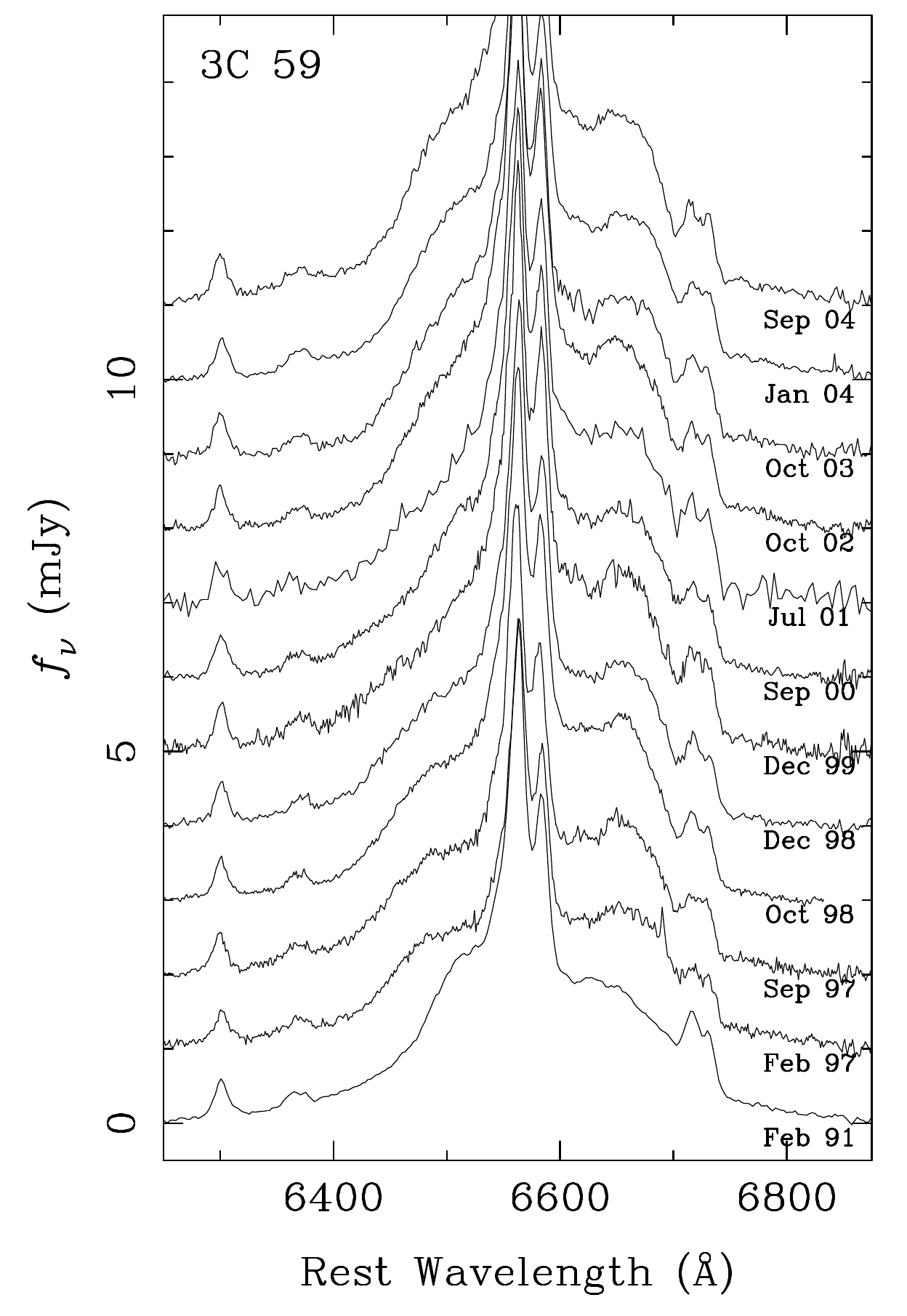}
\caption{\label{3c59_prof} Observed double-peaked Balmer line
profiles of 3C~59, after continuum subtraction. The spectra have
been scaled such that the narrow line flux is constant, so as to show
the relative changes in the total broad-line flux. Arbitrary vertical
offsets have been applied for clarity.}
\end{figure}

\begin{figure}
\epsscale{0.75}
\plotone{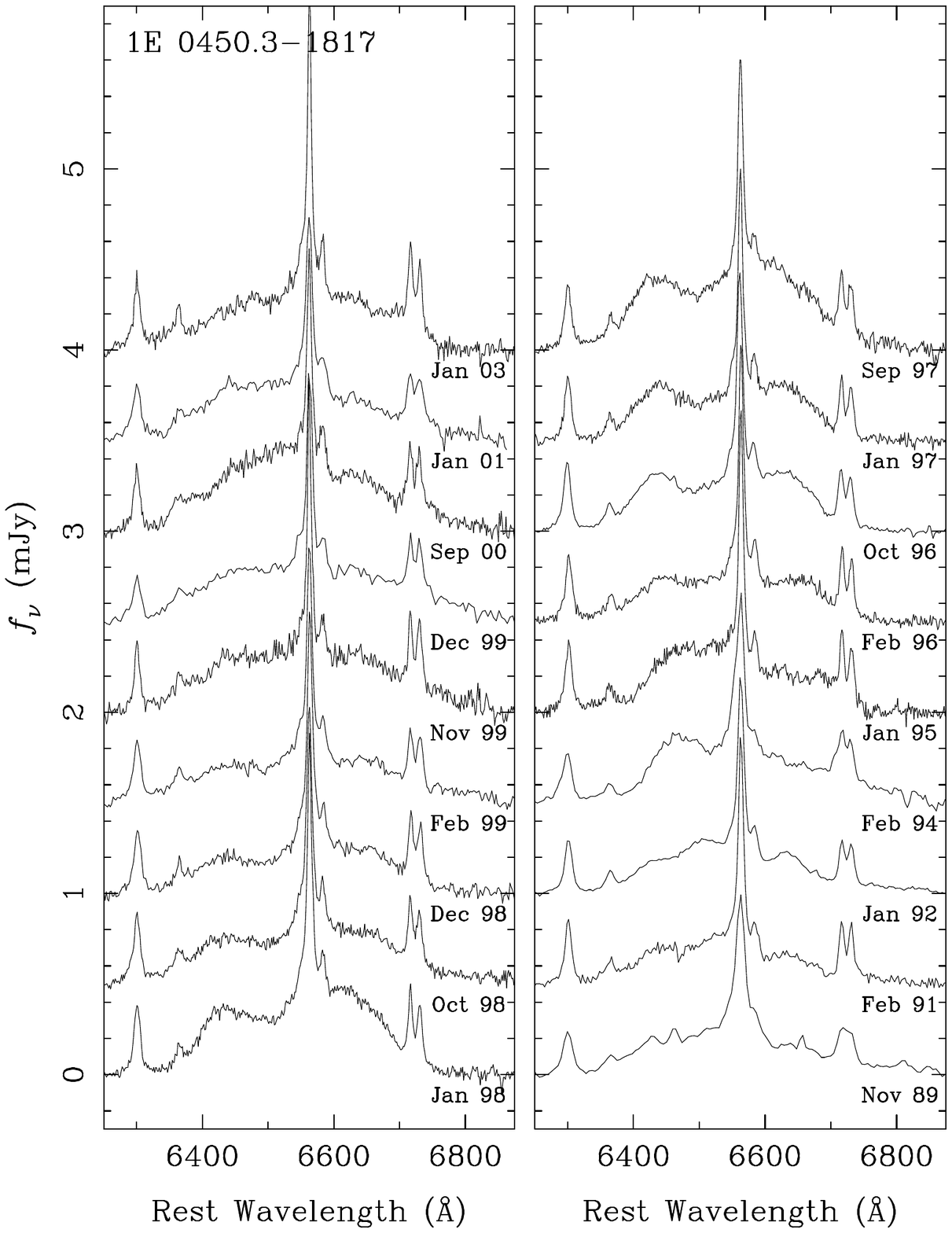}
\caption{\label{1e0450_prof} Same as Figure \ref{3c59_prof}, but
for 1E~0450.3--1817.}
\end{figure}

\begin{figure}
\epsscale{0.5}
\plotone{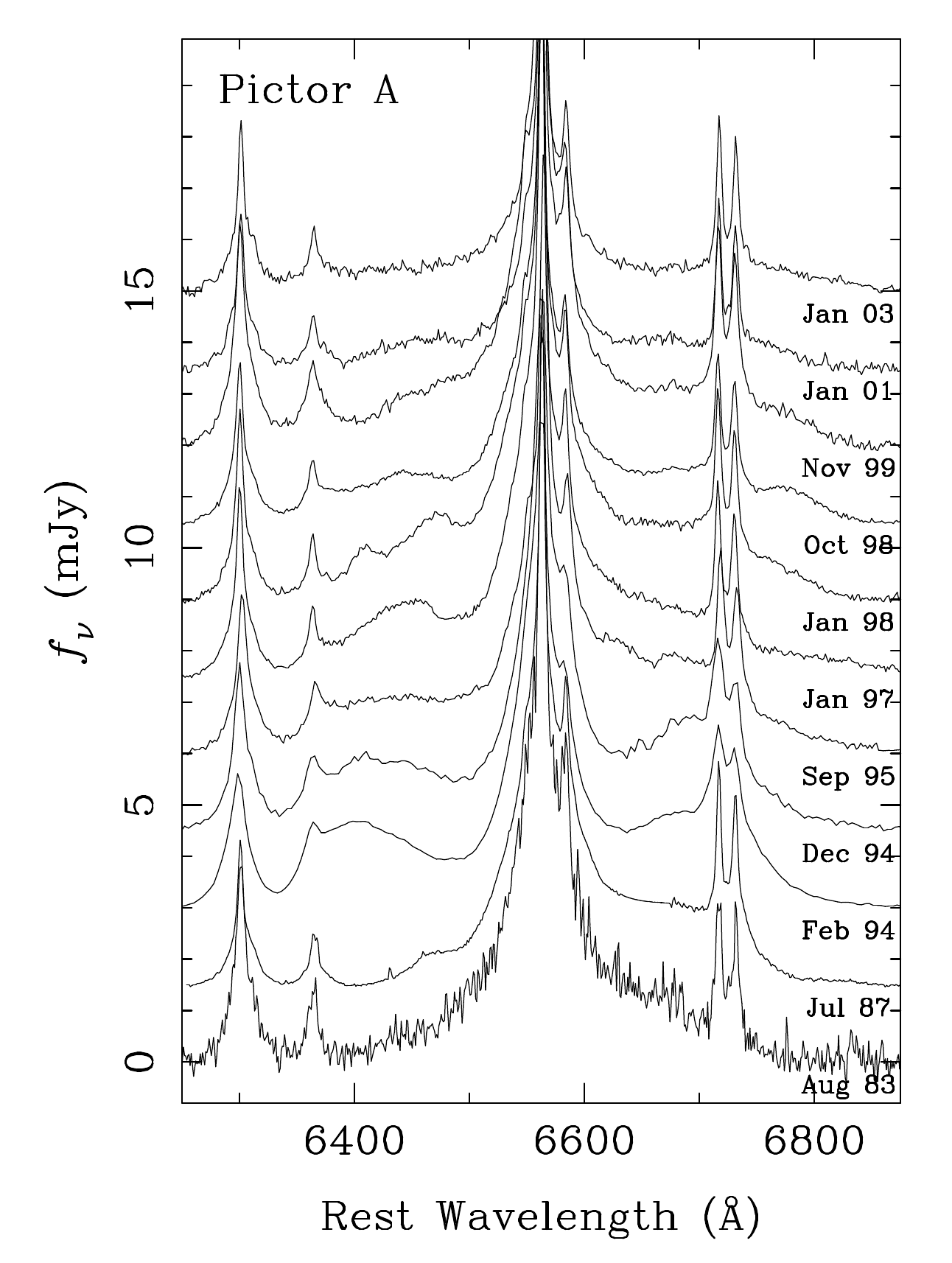}
\caption{\label{pictora_prof} Same as Figure \ref{3c59_prof}, but for Pictor~A.}
\end{figure}

\begin{figure}
\epsscale{0.5}
\plotone{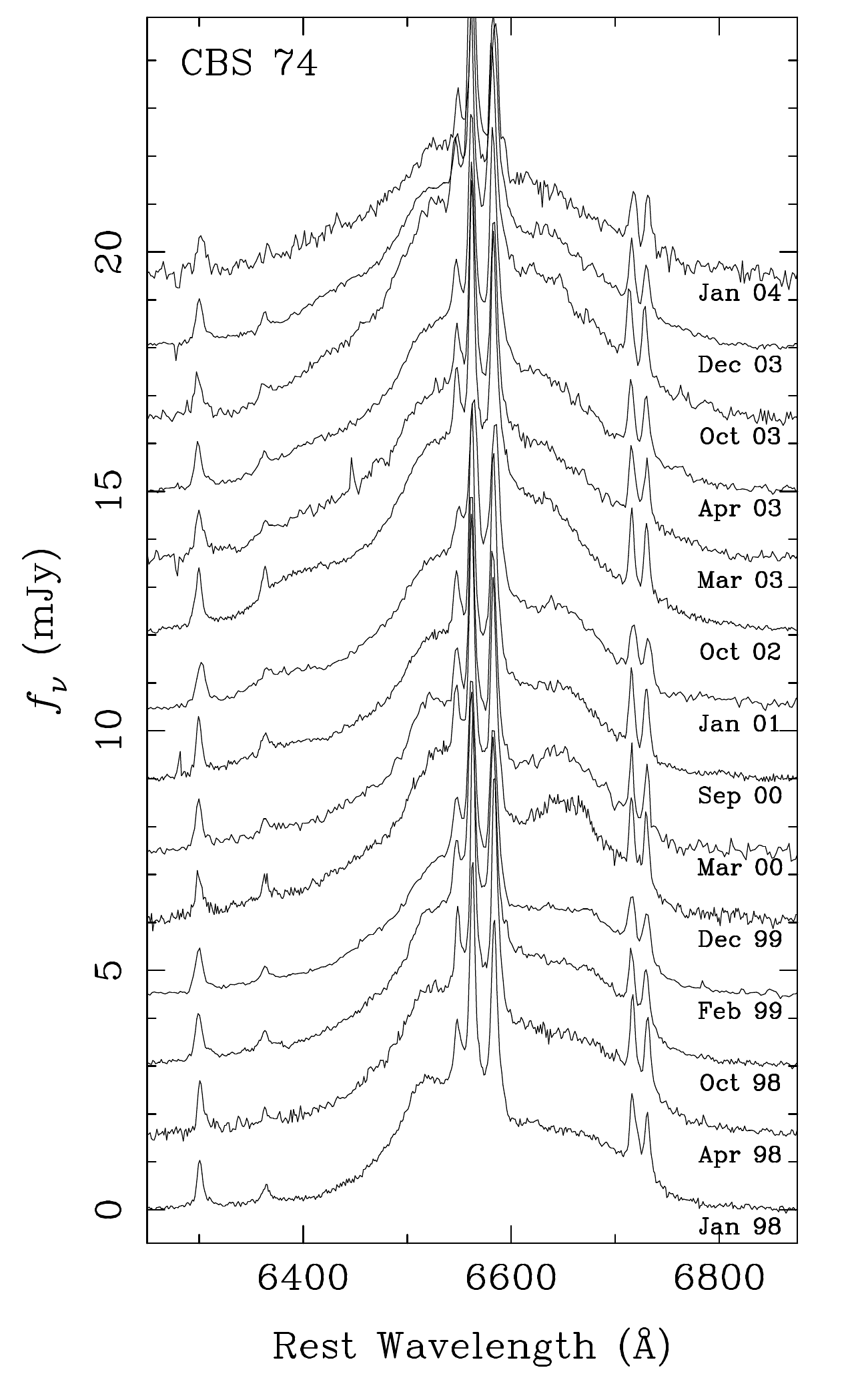}
\caption{\label{cbs74_prof} Same as Figure \ref{3c59_prof}, but for
CBS~74.}
\end{figure}

\begin{figure}
\epsscale{0.5}
\plotone{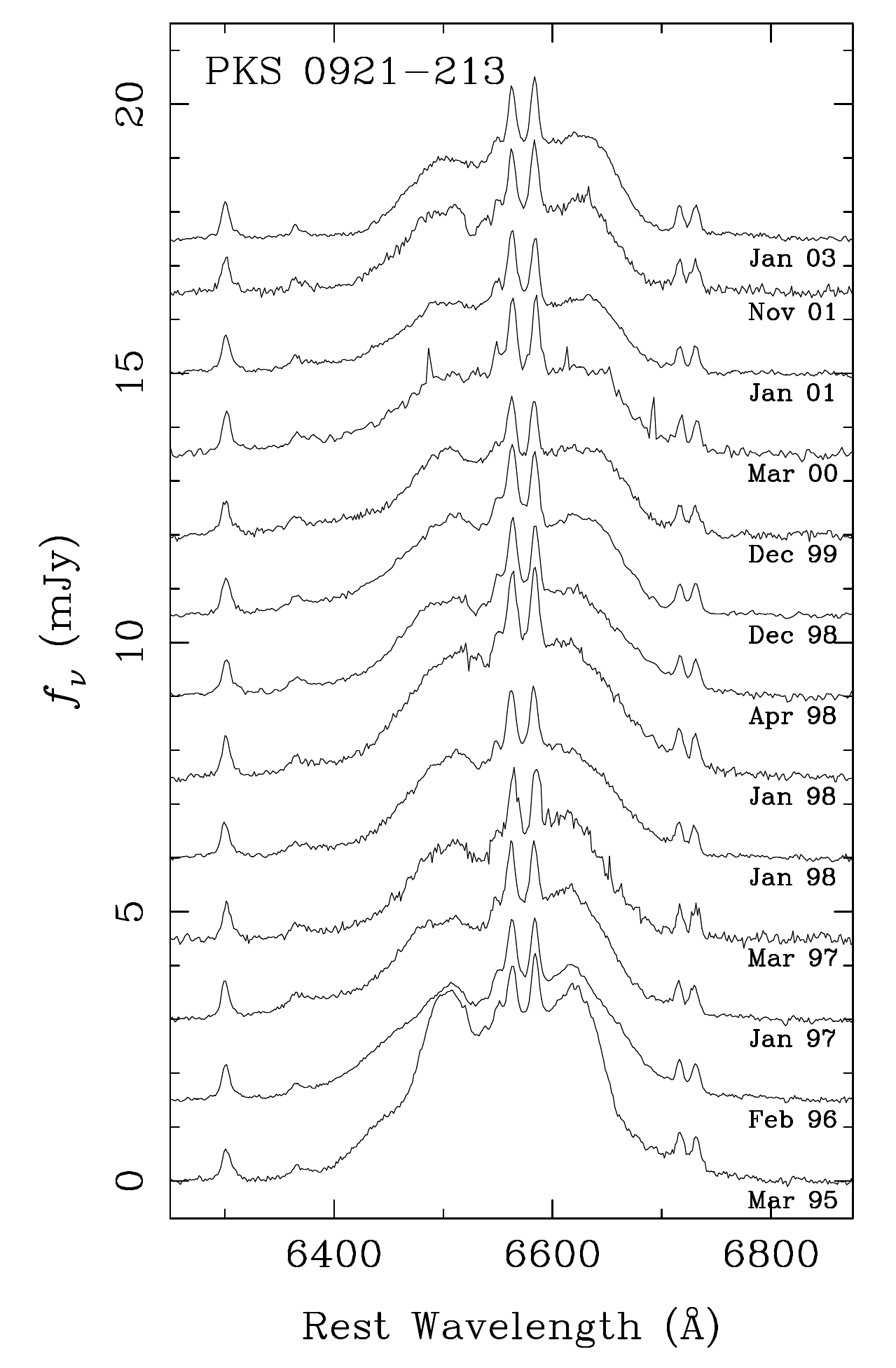}
\caption{\label{pks0921_prof} Same as Figure \ref{3c59_prof}, but
for PKS~0921--213. }
\end{figure}

\begin{figure}
\epsscale{0.5}
\plotone{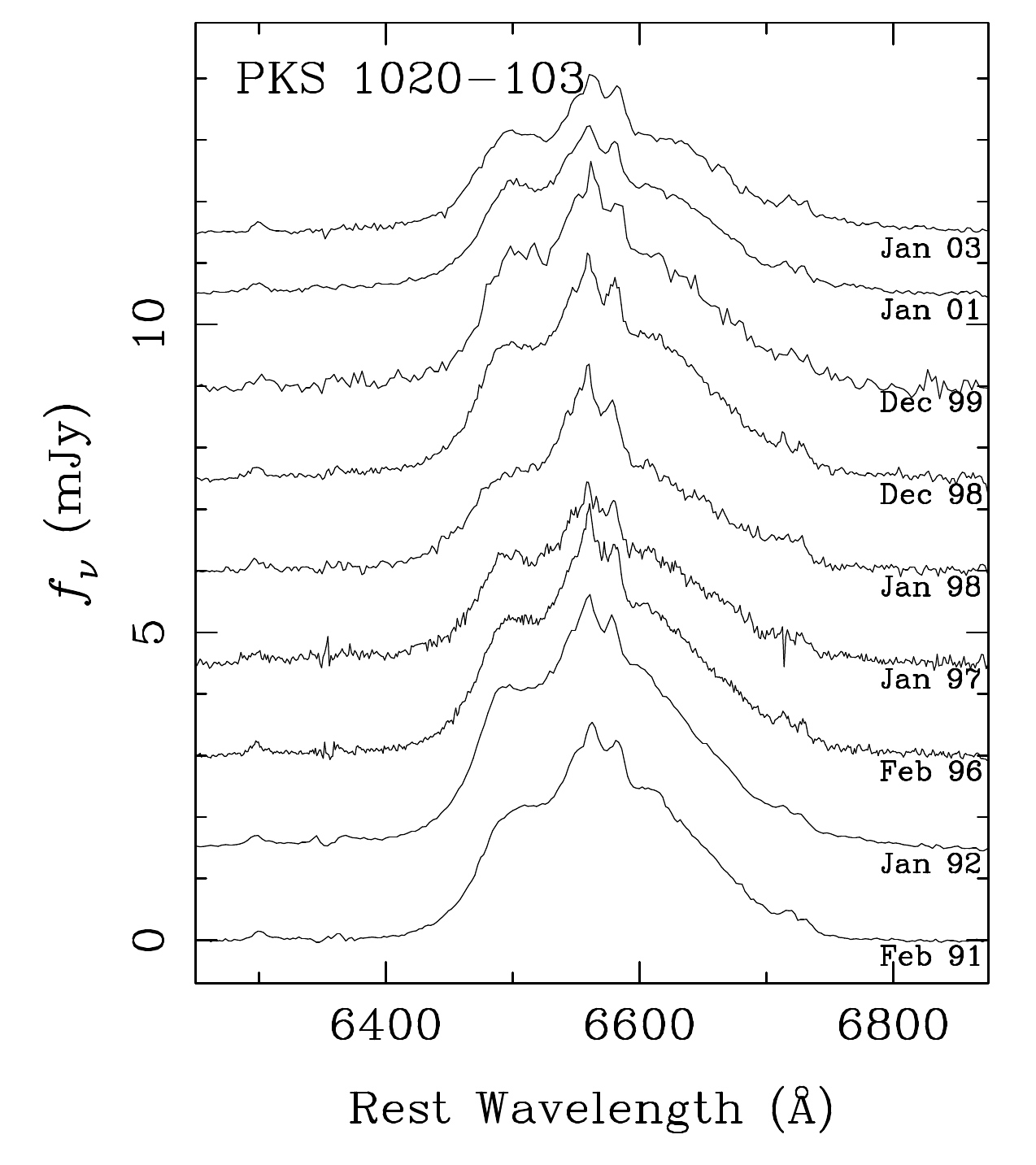}
\caption{\label{pks1020_prof} Same as Figure \ref{3c59_prof}, but
for PKS~1020-103.}
\end{figure}

\begin{figure}
\epsscale{0.5}
\plotone{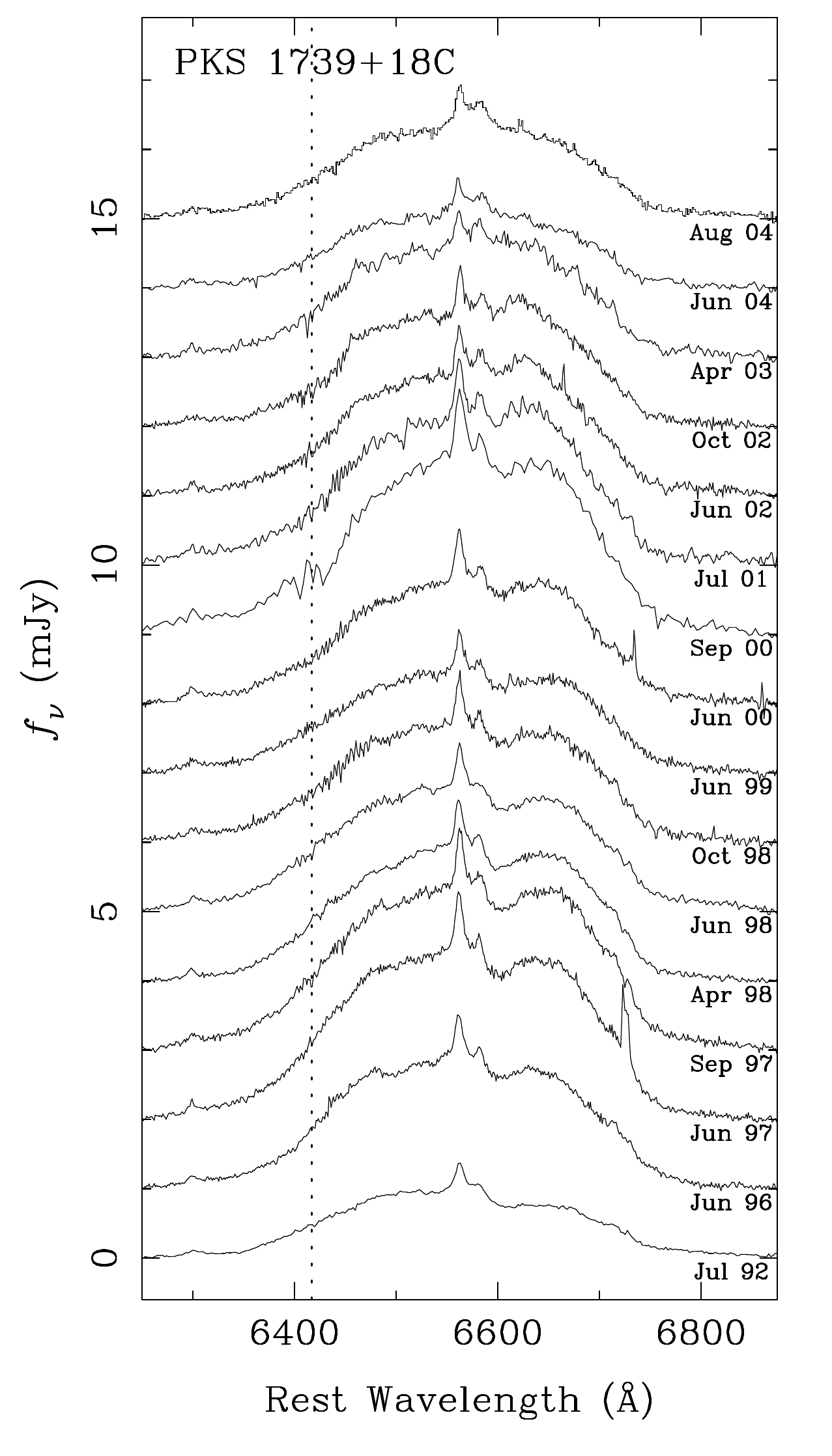}
\caption{\label{pks1739_prof} Same as Figure \ref{3c59_prof}, but
for PKS~1739+18C. The vertical dotted line indicates the position of
the telluric B-band whose correction is sometimes imperfect.}
\end{figure}

Before analyzing the spectra, they were shifted into the rest frame of
the source and corrected for Galactic reddening, using the redshifts
and color excesses given in Table \ref{variability_objects}.  Then the
underlying continuum (from both stellar and non-stellar sources) was
modeled, using the combination of a spectrum of an inactive elliptical
galaxy and a powerlaw continuum, and subtracted. There was often
residual curvature in the continuum around the broad H$\alpha$ profile
that was removed using a low (1$^{\rm st}$ or 2$^{\rm nd}$) order
polynomial. These processes of continuum subtraction, particularly
that of the local continuum, introduce some uncertainty which will be
discussed and quantified below as necessary.  In addition, small
wavelength shifts were applied based on the measured wavelength of the
narrow H$\alpha$ emission line and the spectra were re-sampled to have
a uniform wavelength scale. The profiles after continuum subtraction
are displayed in Figures \ref{3c59_prof}--\ref{pks1739_prof}. The
spectra have been scaled such that the narrow emission line fluxes are
identical in each observation, as described in
\S\ref{variability_fluxscale}. These profiles, and their variations
with time, will be characterized in the following sections.

\subsection{Measurement of Narrow and Broad-Line Fluxes}\label{variability_fluxscale}

We use the narrow emission lines to provide a {\it relative} flux
calibration between spectra, since these lines are thought to
originate in a very extended region around the AGN and they are not
expected to respond to rapid variations in continuum flux \citep[see,
  e.g.,][]{p93}. The narrow emission lines covered by these spectra
include \oxi$\lambda\lambda$6300,6363, \nii$\lambda\lambda$6548,6583,
\sii$\lambda\lambda$6716,6730, and the narrow component of H$\alpha$.

The narrow emission lines of interest lie on top of the broad
double-peaked H$\alpha$ line, which must be removed prior to measuring
the flux of the narrow lines. The \oxi$\lambda$6300,\oxi$\lambda$6363
and \sii$\lambda\lambda$6716,6730 lines lie on the wings of the
double-peaked ${\rm H\alpha}$ emission line. Repeated subtraction of the
underlying broad wings, fitted with a 3$^{\rm rd}$ order polynomial, indicates 
that the flux of the \oxi$\lambda$6300 and the \sii$\lambda\lambda$6716,6730 
can be measured robustly (with $\lesssim$ 10\% variation). The weak
\oxi$\lambda$6363 has a larger uncertainty (20\%).

The \nii\ and narrow H$\alpha$ emission lines are often blended, at
least at the base. The blend is decomposed by first fitting the
\nii$\lambda$6583 and \nii$\lambda$6548 emission lines in a 3:1 flux
ratio and then fitting the residual narrow H$\alpha$ line. For any
given fit to the underlying broad-line, the line fluxes of the \nii\
and narrow H$\alpha$ line can be measured with an accuracy similar to
the other narrow emission lines ($\sim$~10\%--15\%). However, the
process of subtracting the underlying broad-line is more ambiguous
because these lines lie on the trough of the double-peaked profile. We
found that the process of removing the underlying broad component can
introduce a {\it systematic} uncertainty of up to 30\%.

It is possible to guard against the systematic uncertainty by
considering the ratios of the narrow line fluxes (i.e. \oxi/\sii,
\sii/H$\alpha$, \oxi/H$\alpha$, and \nii/H$\alpha$) which should not
change significantly with time.  It is still possible that the
H$\alpha$+\nii\ flux are systematically uncertain, but no significant
{\it variations} in the narrow line fluxes will be introduced, because
all spectra are affected in the same manner.

To determine the broad-line flux ($\fbha$) the total
narrow line flux determined above was simply subtracted from flux of
the entire profile (narrow + broad).  The estimated errors on the
total flux and narrow line flux were added in quadrature to obtain an
approximate error in the broad-line flux.  The narrow lines in
PKS~1739+18C, with the exception of \oxi$\lambda6300$, could not be
measured because the profile is dominated by the broad double-peaked
component. Because the narrow lines probably contribute only 5--10\%
of the total flux in this object, we simply used the total measured
flux in lieu of the broad-line flux. We also consider {\it absolute}
changes in the broad-line flux by normalizing each broad-line flux by
the \oxi$\lambda6300$ line flux. The reported error on the
normalized broad-line flux include those from both the broad-line and
\oxi$\lambda6300$ fluxes.

In the next three sub-sections, we present three different methods for
characterizing the variations in double-peaked line profiles in a
model-independent way. After the results for each method are
presented, we first describe general trends and then highlight results
from specific objects.

\subsection{RMS and Correlation Plots}\label{rms_corr}

\begin{figure}
\epsscale{0.6}
\plotone{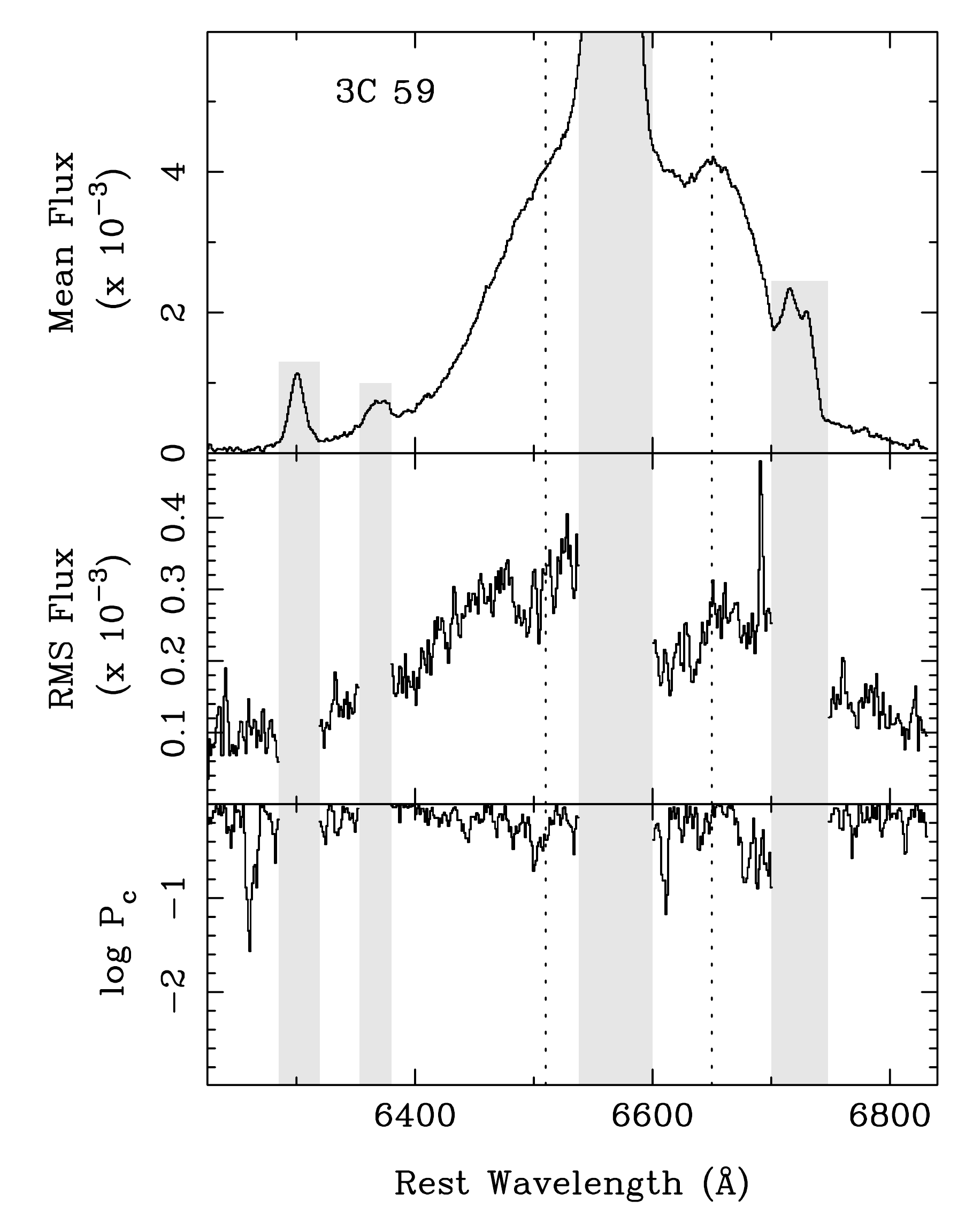}
\caption{\label{3c59_rms}{\it Top: } the mean profile of 3C~59,
constructed from profiles normalized by their total broad H$\alpha$
flux. {\it Middle: } the rms variability, as a function of
wavelength. {\it Bottom: } the log of the chance probability that
there is a spurious correlation between the total broad-line flux and
the broad-line normalized profile flux, determined with a Spearman's
rank-order correlation test. A small chance probability indicates that
there may be a correlation. The vertical dashed lines indicate the
position of the peaks in the mean profile and the vertical gray
stripes indicate the positions of the narrow emission lines, which
were not removed. See \S\ref{rms_corr} for more details. }
\end{figure}

\begin{figure}
\epsscale{0.6}
\plotone{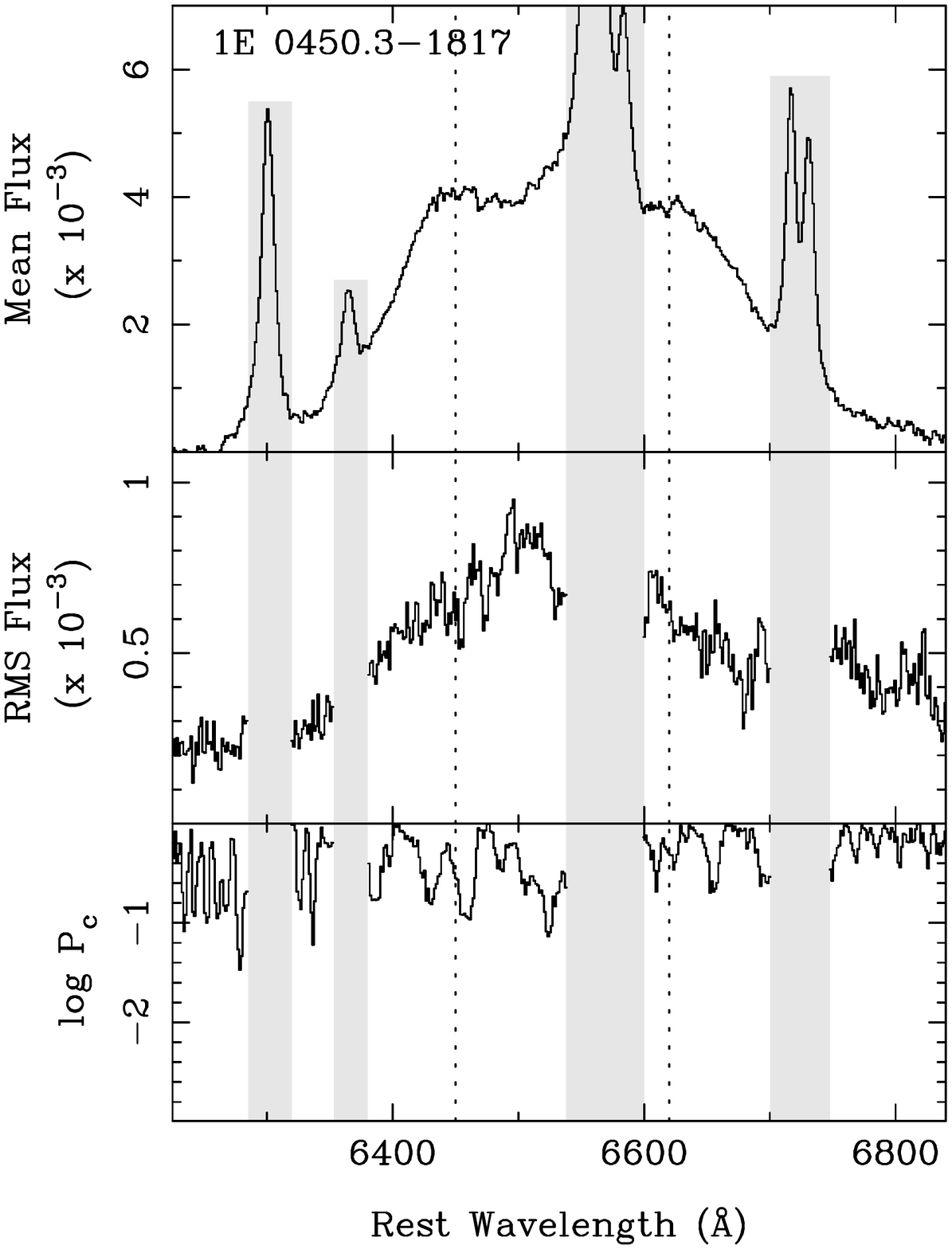}
\caption{\label{1e0450_rms} Same as Figure \ref{3c59_rms}, but for
1E~0450.3--1817.}
\end{figure}

\begin{figure}
\epsscale{0.6}
\plotone{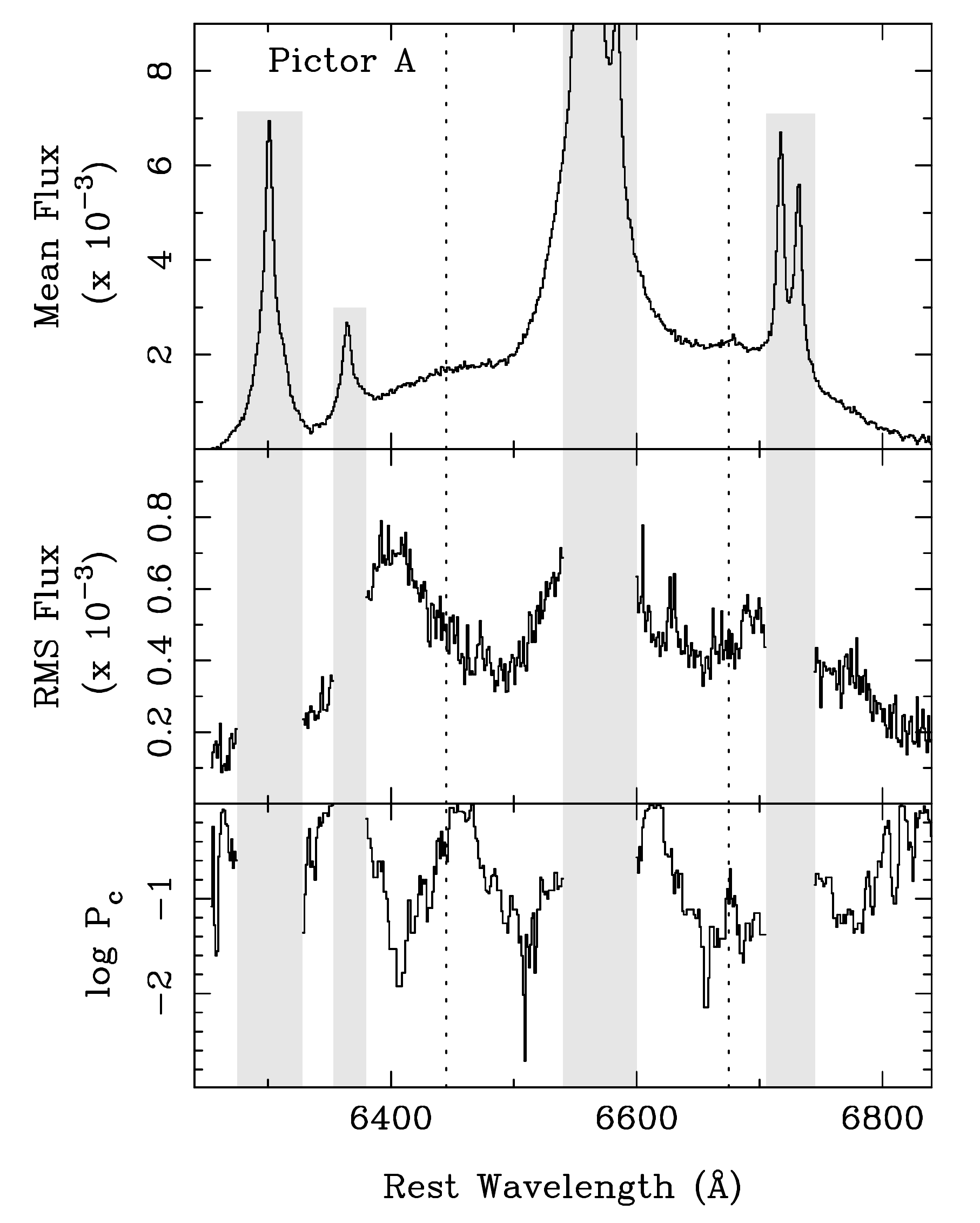}
\caption{\label{pictora_rms} Same as Figure \ref{3c59_rms}, but for Pictor~A. }
\end{figure}

\begin{figure}
\epsscale{0.6}
\plotone{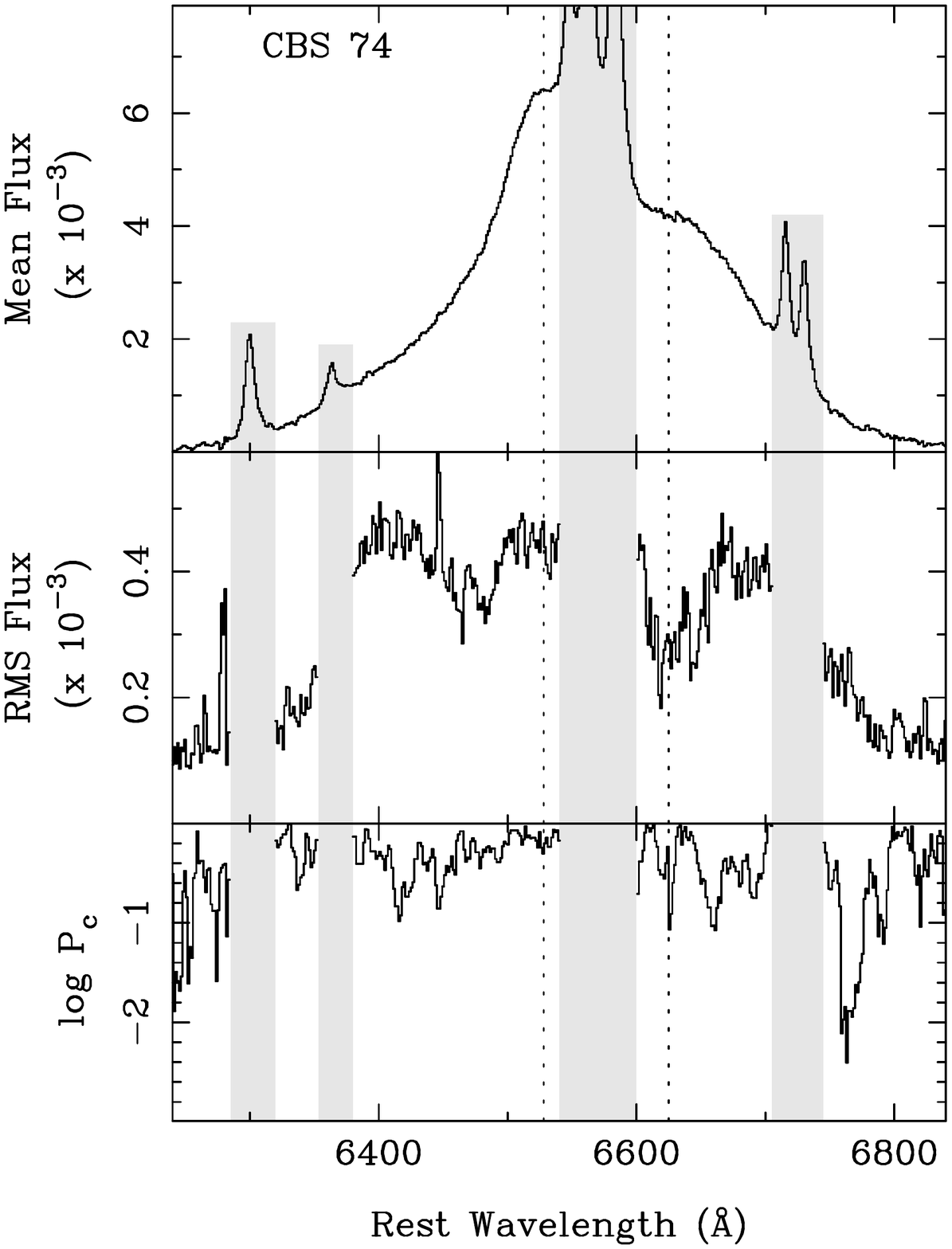}
\caption{\label{cbs74_rms} Same as Figure \ref{3c59_rms}, but for CBS~74.}
\end{figure}

\begin{figure}
\epsscale{0.6}
\plotone{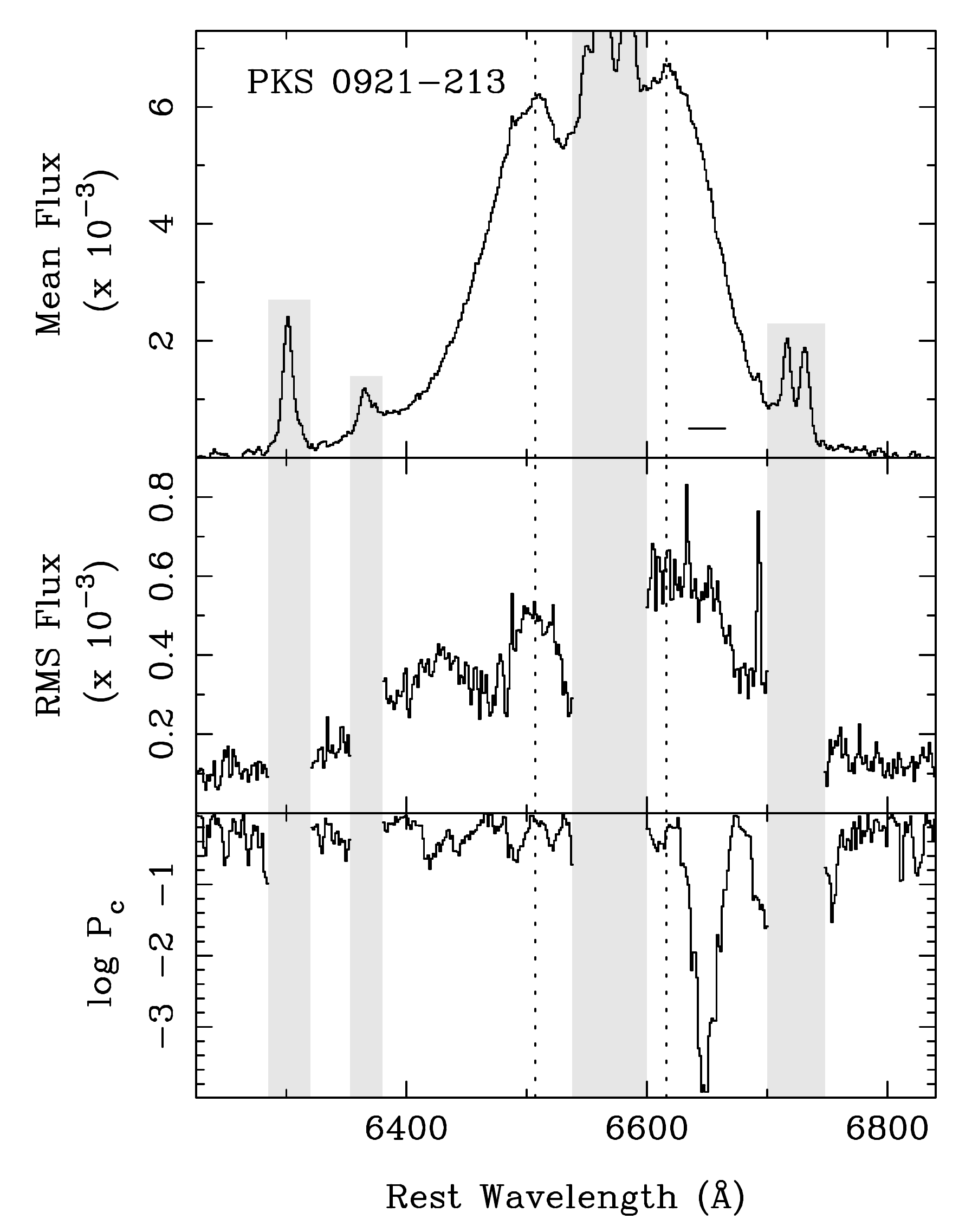}
\caption{\label{pks0921_rms} Same as Figure \ref{3c59_rms}, but for
PKS~0921--213.  The plot of the chance probability indicates that
there may be a correlation between the broad-line normalized flux 
integrated over the 6635--6665\AA\ wavelength interval (indicated by the 
horizontal bar in the top panel) and $\fbha/{\rm F_{[O\;I]}}$
is explored further in Figure \ref{pks0921_corr}.}
\end{figure}

\begin{figure}
\epsscale{0.6}
\plotone{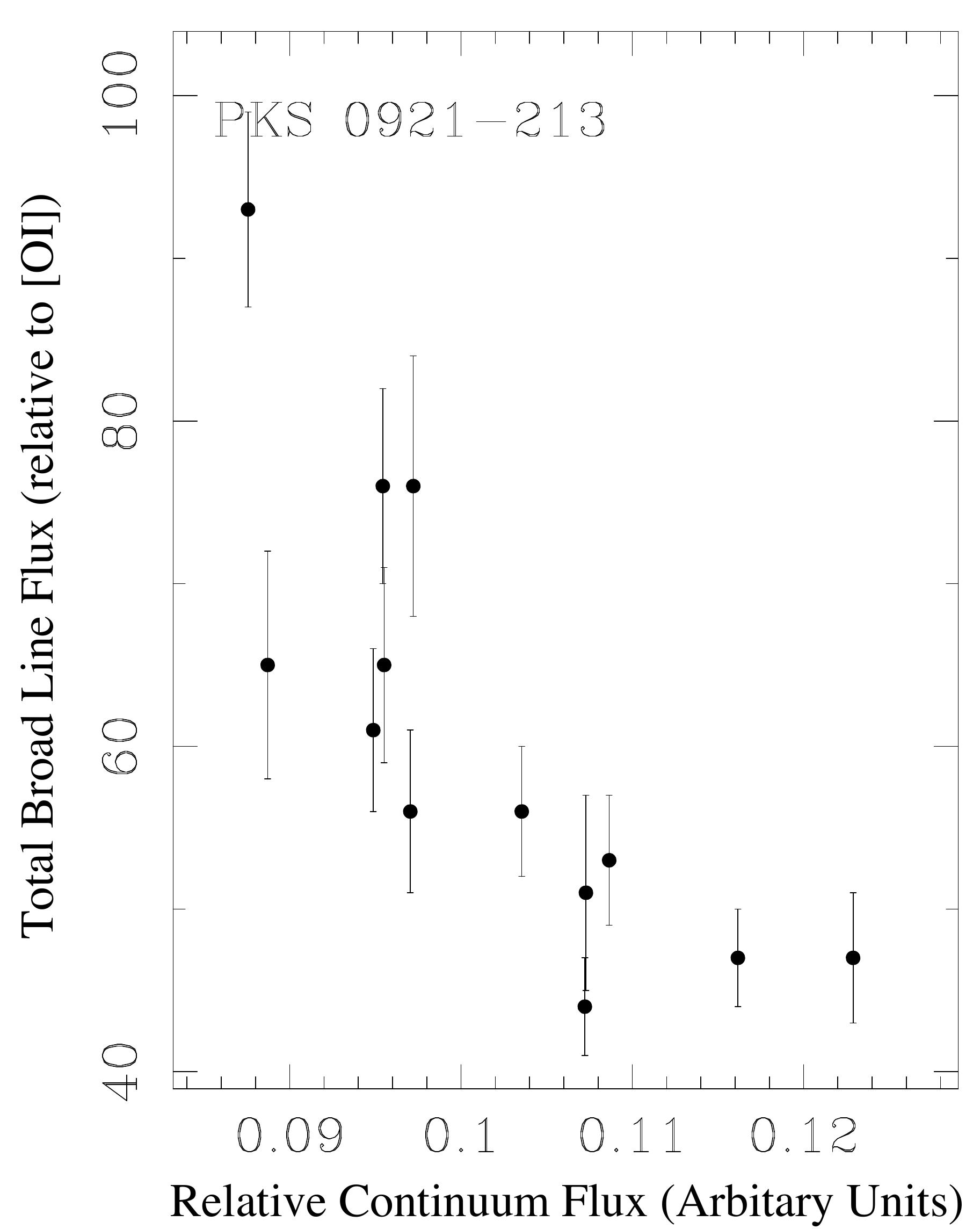}
\caption{\label{pks0921_corr} Plot of $\fbha/{\rm F_{[O\;I]}}$ versus
  the normalized continuum flux integrated over the wavelength
  interval of 6635--6655\AA\ (i.e. $\int
  f_{\lambda}\;d\lambda/\fbha$).}
\end{figure}

\begin{figure}
\epsscale{0.75}
\plotone{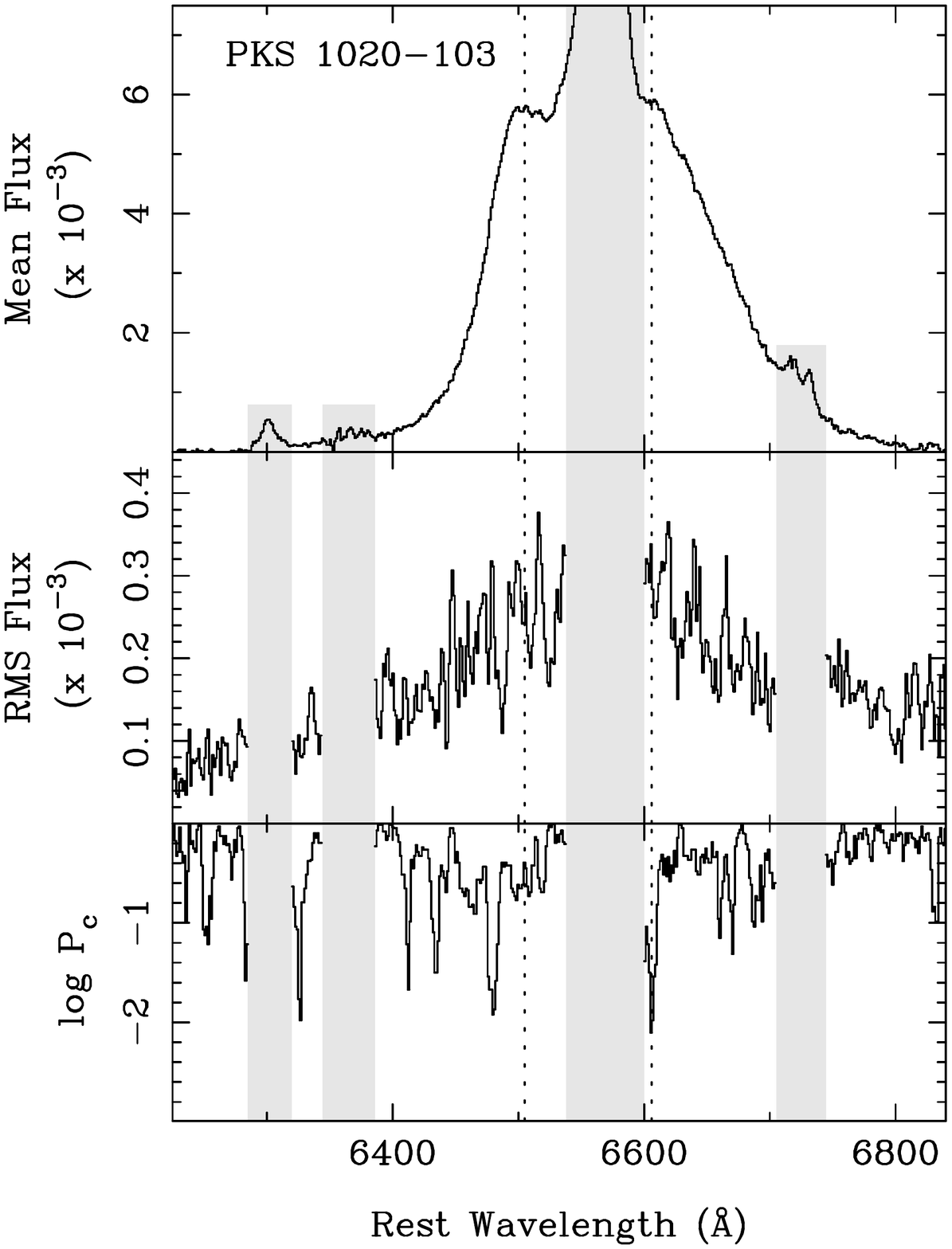}
\caption{\label{pks1020_rms} Same as Figure \ref{3c59_rms}, but for
PKS~1020--103.}
\end{figure}

\begin{figure}
\epsscale{0.6}
\plotone{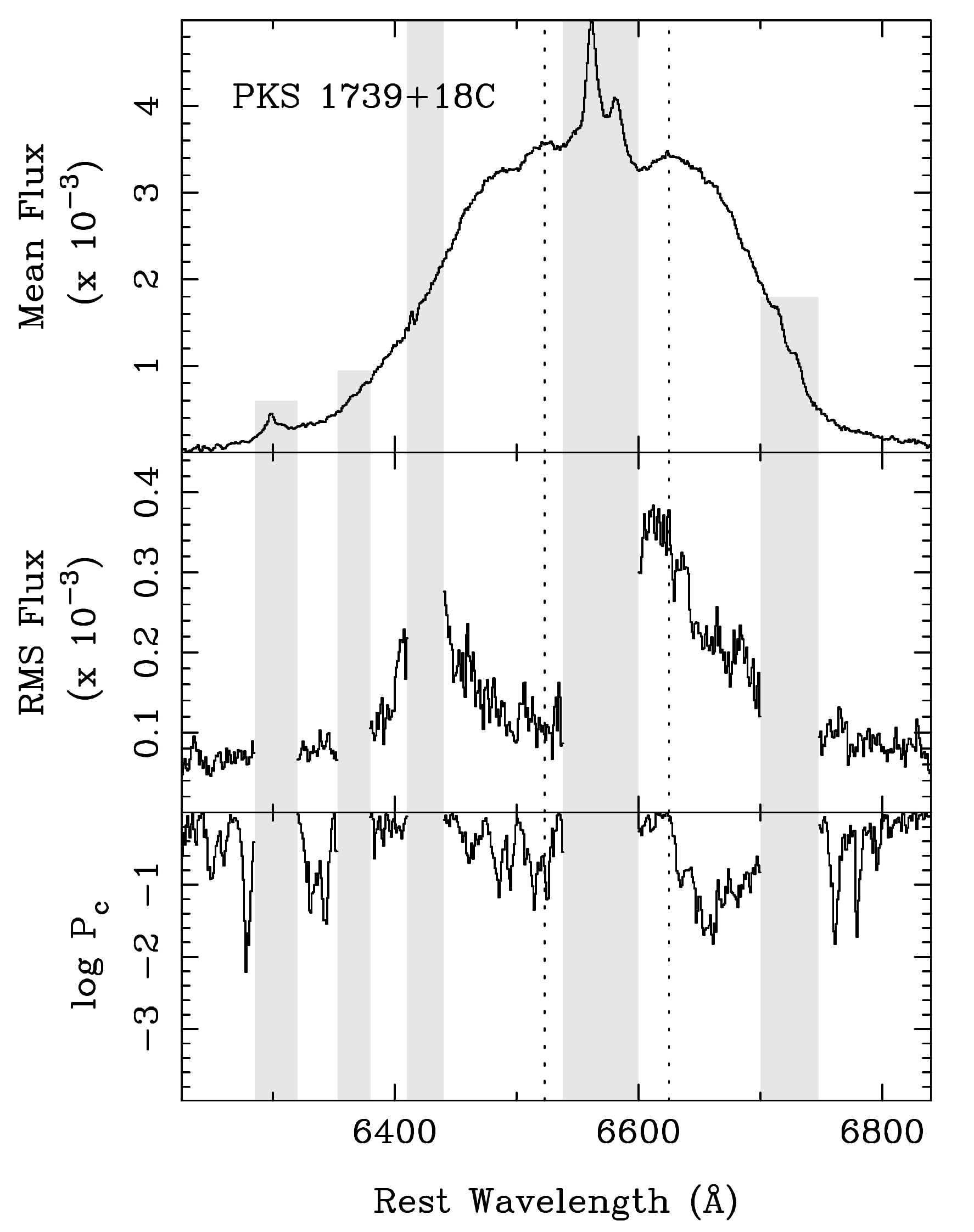}
\caption{\label{pks1739_rms} Same as Figure \ref{3c59_rms}, but for
  PKS~1739+18C. An additional gray stripe indicates the region of the
  spectrum affected by telluric A-band ($\sim$7400--7440$\AA$ at the
  redshift of this object.) Imperfect removal of the telluric features
  may result in spurious rms variability. Imperfect removal of the
  telluric features may result in spurious features in the difference
  spectra.}
\end{figure}

As shown in Figures \ref{3c59_prof}--\ref{pks1739_prof}, the broad
H$\alpha$ profiles clearly change with time. We first investigate
whether some portions of the profile are more variable than others by
plotting the root mean square (rms) flux variations as a function of
wavelength. Short-term fluctuations in the illumination will cause the
total broad-line flux to change (i.e. through light reverberation), as
observed in 3C 390.3 \citep{d98,shap01}. Because we are primarily
concerned with changes in the {\it shape} of the profile rather than
changes in the overall flux, the spectra were normalized by the broad
line flux (i.e.we convert $f_\lambda$ to
$f_{\lambda}\;\Delta\lambda/\fbha$, where $\Delta\lambda$ is the
resolution element) before constructing the mean and rms profiles. In
Figures \ref{3c59_rms}--\ref{pks1739_rms}, we show the mean profile in
the top panel, and the rms profile in the middle panel.

The rms profile for most objects is quite different from the mean
profile, indicating that the profile is not uniformly variable. Many
objects do show considerable variability at wavelengths that roughly
coincide with a peak in the mean profile (e.g. both peaks of 3C~59,
PKS~0921--213, and the red peaks of Pictor~A and PKS~1739+18C, and the
blue peak of CBS~74). In some objects however, there are minima in the
rms profiles at locations of the peaks (the red peak of CBS~74 and the
blue peak of PKS~1739+18C) indicating that the flux of these peak
regions are unusually stable compared with the rest of the
profile. 

For some objects, there are distinct features in the rms
profile that are at significantly larger projected velocity than the
peaks in the mean profile (e.g. 3C~59, Pictor A, CBS~74,
PKS~0921--213, PKS~1739+18C). Increased variability at large projected
velocity is consistent with the idea that the inner portions of the
line-emitting disk (which give rise to the wings of the profile) could
vary on a shorter timescale than the outer regions of the disk. For
1E~0450.3--1817 and PKS~1020--103, the rms profiles do not show many
distinct features. We do note that 1E 1E~0450.3--1817 is slightly more
variable at negative projected velocities, with a maximum at $\sim
6500$~\AA\ (see Figure \ref{1e0450_rms}).

One could also expect to find a narrower, single-peaked variable
component in the middle of the profile between the two peaks. However,
because the narrow H$\alpha$+\nii\ lines were not subtracted, it is
difficult to detect a single-peaked component unless it has
${\rm FWHM}\gtrsim 2000\;\kms$ or the line is significantly blue- or
redshifted with respect to the narrow H$\alpha$ line. In Pictor~A
there is a central peak in the rms variability profile that is
slightly blue-shifted with respect to narrow H$\alpha$; this is most
likely due to the single-peaked, broad H$\alpha$ component (${\rm FWHM}\sim
3000\;\kms$) which is very obvious in this object. Three other objects
(3C~59, CBS~74, and 1E~0450.3--1817) show features in the central
regions of the rms profile ($\lambda \sim 6500-6530 {\rm \AA}$).In all
three objects, there is a blue peak or shoulder at a similar
wavelength. If related to a single-peaked H$\alpha$ line, the line
would be blue-shifted by $\sim 1000-3000 {\rm km\;s^{-1}}$ relative to
the narrow H$\alpha$ line; although such blue-shifts are observed in
high-ionization lines (such as \ion{C}{4}), they are not common in the
low-ionization Balmer lines \citep{sulentic2000}. Thus these features
are more likely to be associated with the blue peak or shoulder seen
in the mean profile, rather than a single-peaked broad H$\alpha$
component.

We note that \citet{gezari07} also find that structure in the rms
profile often occurs at larger projected velocity than the peaks in
the mean profile. These authors were able to perform narrow-line
subtraction and, as a result, were able to detect central peaks in the
rms profiles of Arp~102B and 3C~332 with velocity widths of $\sim
1000\;\kms$. This suggests that a ``standard'' single-peaked component
may be present in those objects.

It is possible that the double-peaked line profile might vary in
response to long-term, secular variations in the illuminating flux
which could alter the physical properties of the disk (such as the
ionization state). Such long-term effects have been noticed in
NGC~1097 \citep{sb03} and 3C~332 \citep{gezari07}.  To investigate
whether the profile varies in response to long-term changes in the
broad-line flux, we looked for correlations between $\fbha/{\rm F_{
    [O\;I]}}$ and the flux in the normalized line profile
($f_{\lambda}\;\Delta\lambda/\fbha$) integrated over five-pixel bins
centered on $\lambda$.  We used the non-parametric Spearman's
rank-order correlation test \citep[see, e.g.][]{numercf}, since we do
not know what functional form any potential correlation will take. In
the bottom panel of Figures \ref{3c59_rms}--\ref{pks1739_rms}, the log
of the chance probability ($P_{c}$) is plotted as a function of
wavelength, where $P_{c}$ is the chance probability that the
correlation is spurious.

We plotted the broad H$\alpha$ flux (relative to \oxi) versus the
integrated normalized profile flux for any region in which
$P_{c}<0.1$, but in most instances there was either little obvious
correlation or the correlation was driven by a few data points. A
wavelength interval redward of the red peak ($\lambda$=6635--6665${\rm
  \AA}$ in PKS~0921-0213 does show a possible negative correlation
(see Figure \ref{pks0921_corr}). It must be noted that the broad-line
flux in PKS~0921--213 decreases nearly monotonically; thus it is
impossible to ascertain whether the flux in this wavelength interval
is decreasing in response to a decrease in the total broad H$\alpha$
flux (and presumably the illuminating flux) or some other, perhaps
dynamical, effect which is also steadily varying with time. We come
back the issue of long-term correlations between the broad-line
profile and the illuminating flux in
\S\ref{variability_parameter_plots}.

\subsection{Difference Spectra}\label{variability_diff_spec}

In order to better visualize the profile variability, we present here
difference spectra that result from subtracting either the mean or the
``minimum'' profile. The latter is constructed by selecting, at each
wavelength, the minimum flux density from among all the spectra of the
same object. This minimum spectrum represents a base profile which is
common to all the profiles. As an idealized example, if the emission
arises from a circular accretion disk with an emissivity enhancement
(such as a hot spot or spiral arm), then the minimum profile would be
that of the underlying circular accretion disk. The difference spectra
from the minimum would clearly reveal the (positive) emissivity
enhancement as it travels through the disk and affects different
portions of the emission line profile. However, even in this idealized
scenario, one must use caution in interpreting the minimum profile.
In the simple additive models mentioned above, the variability is
expected to be cyclical. The minimum profile will not be equivalent to
the profile of the underlying disk if only a portion of the cycle has
been observed (see \S\ref{compare_observed}).

\begin{figure}
\epsscale{1}
\plotone{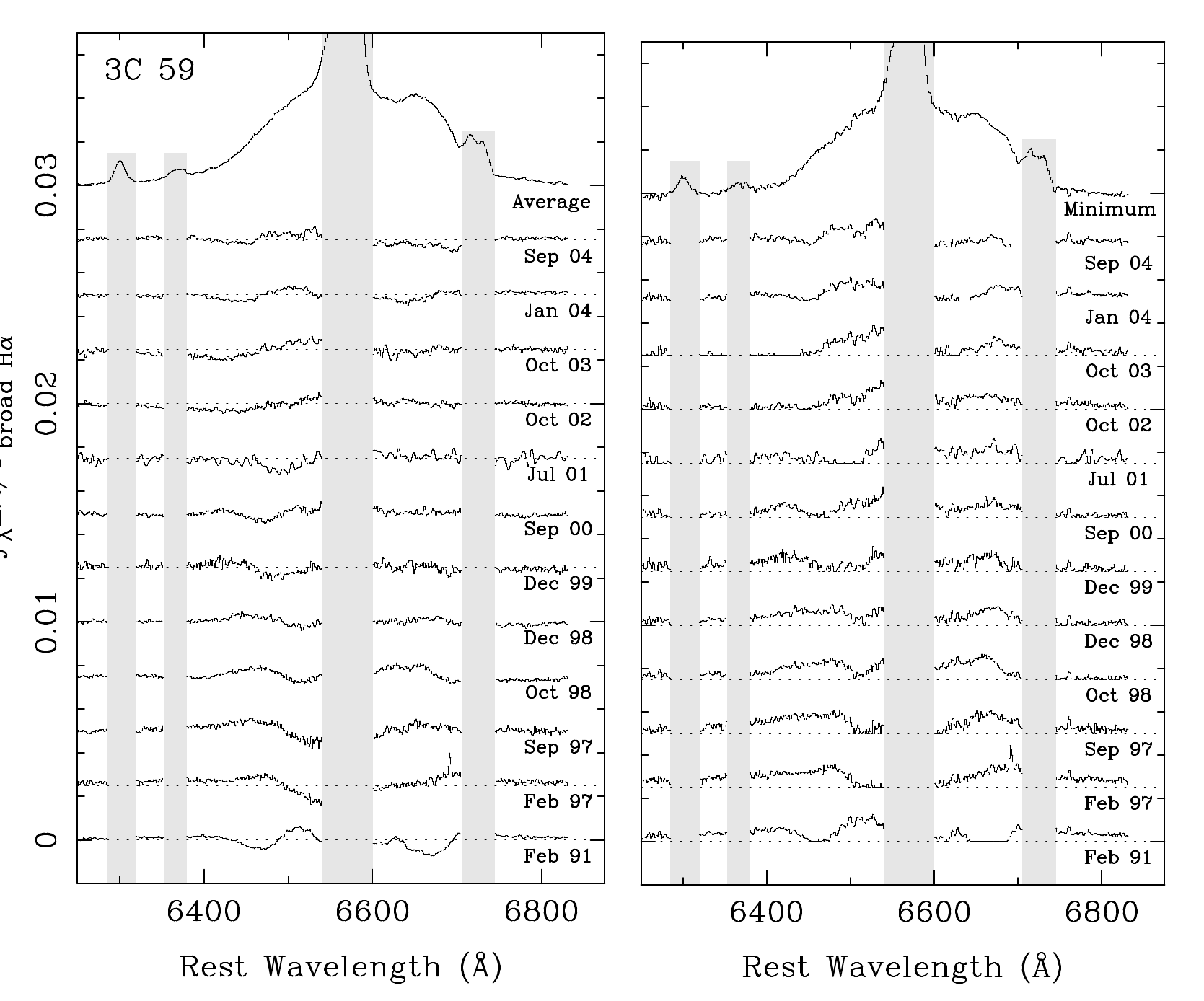}
\caption{\label{3c59_diff} Difference spectra for 3C~59. In the
left panel, the average spectrum and the difference between each
spectrum and the average are shown, whereas the minimum and the
differences from it are shown in the right panel. Arbitrary vertical
offsets were applied for clarity. The vertical gray stripes indicate
the positions of the narrow emission lines. See
\S\ref{variability_diff_spec} for a description of how these spectra
were constructed.}
\end{figure}

\begin{figure}
\epsscale{0.85}
\plotone{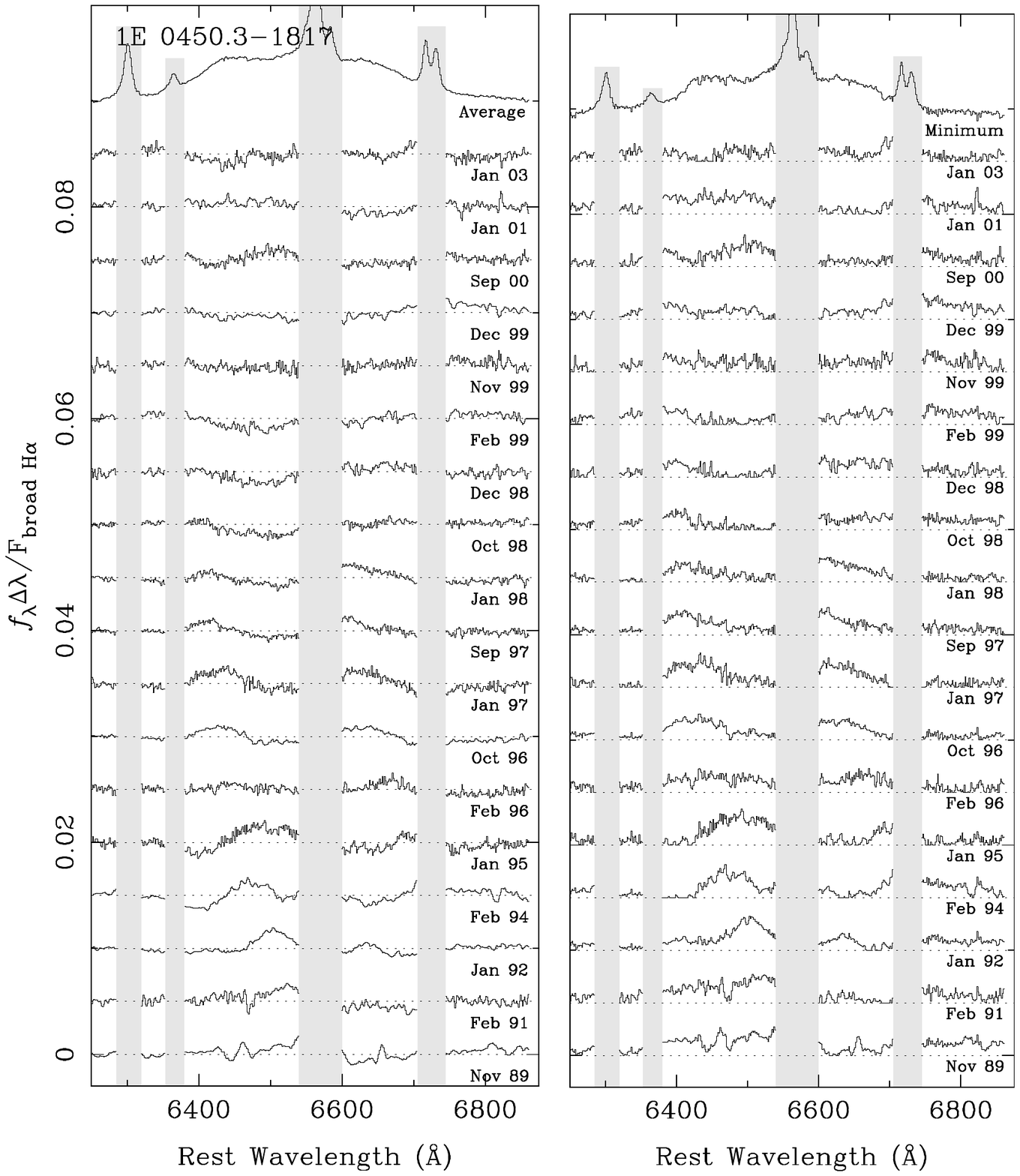}
\caption{\label{1e0450_diff} Same as Figure \ref{3c59_diff}, but
for 1E~0450.3--1817.}
\end{figure}

\begin{figure}
\epsscale{1}
\plotone{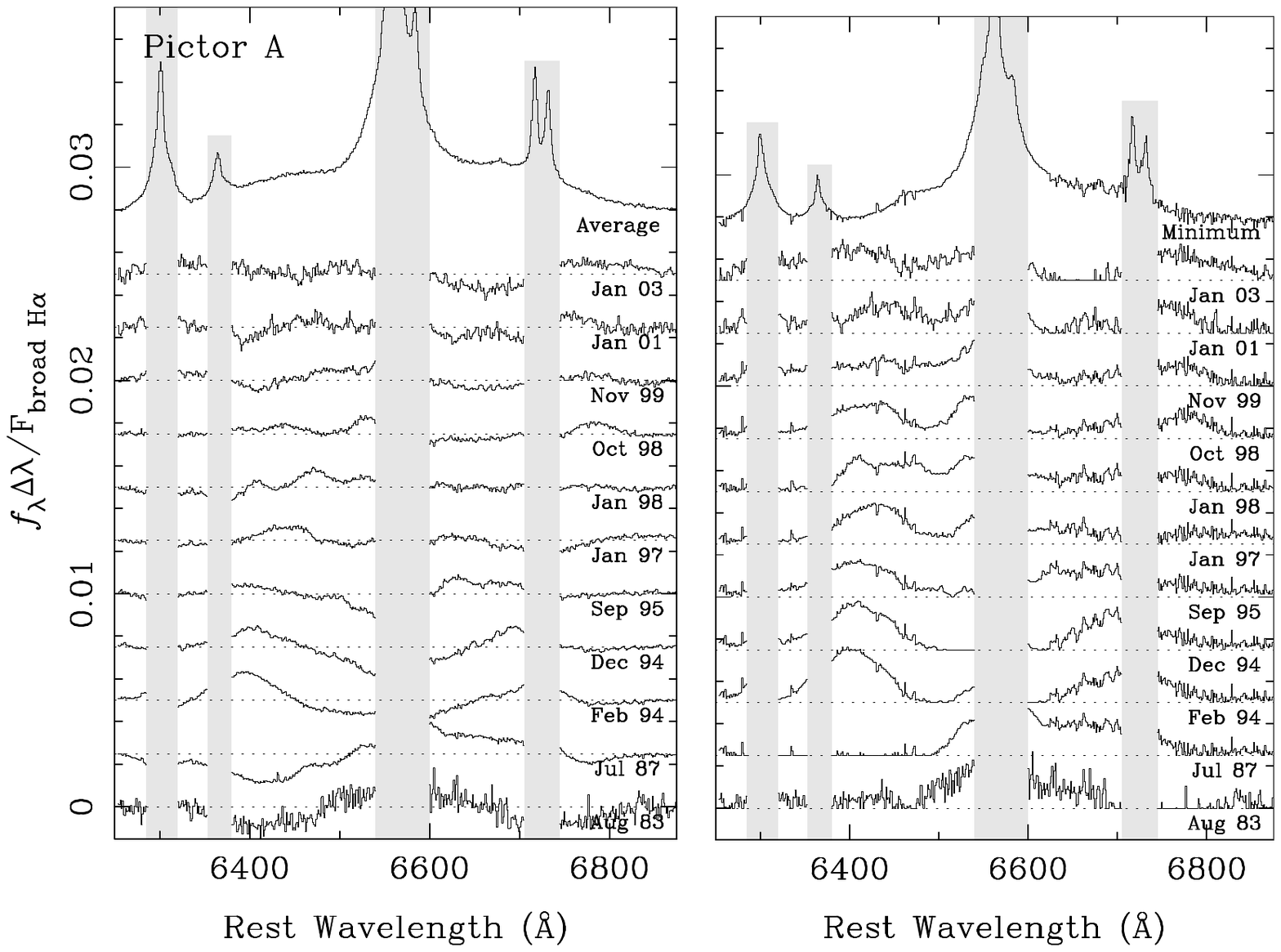}
\caption{\label{pictora_diff} Same as Figure \ref{3c59_diff}, but
for Pictor~A.}
\end{figure}

\begin{figure}
\epsscale{1}
\plotone{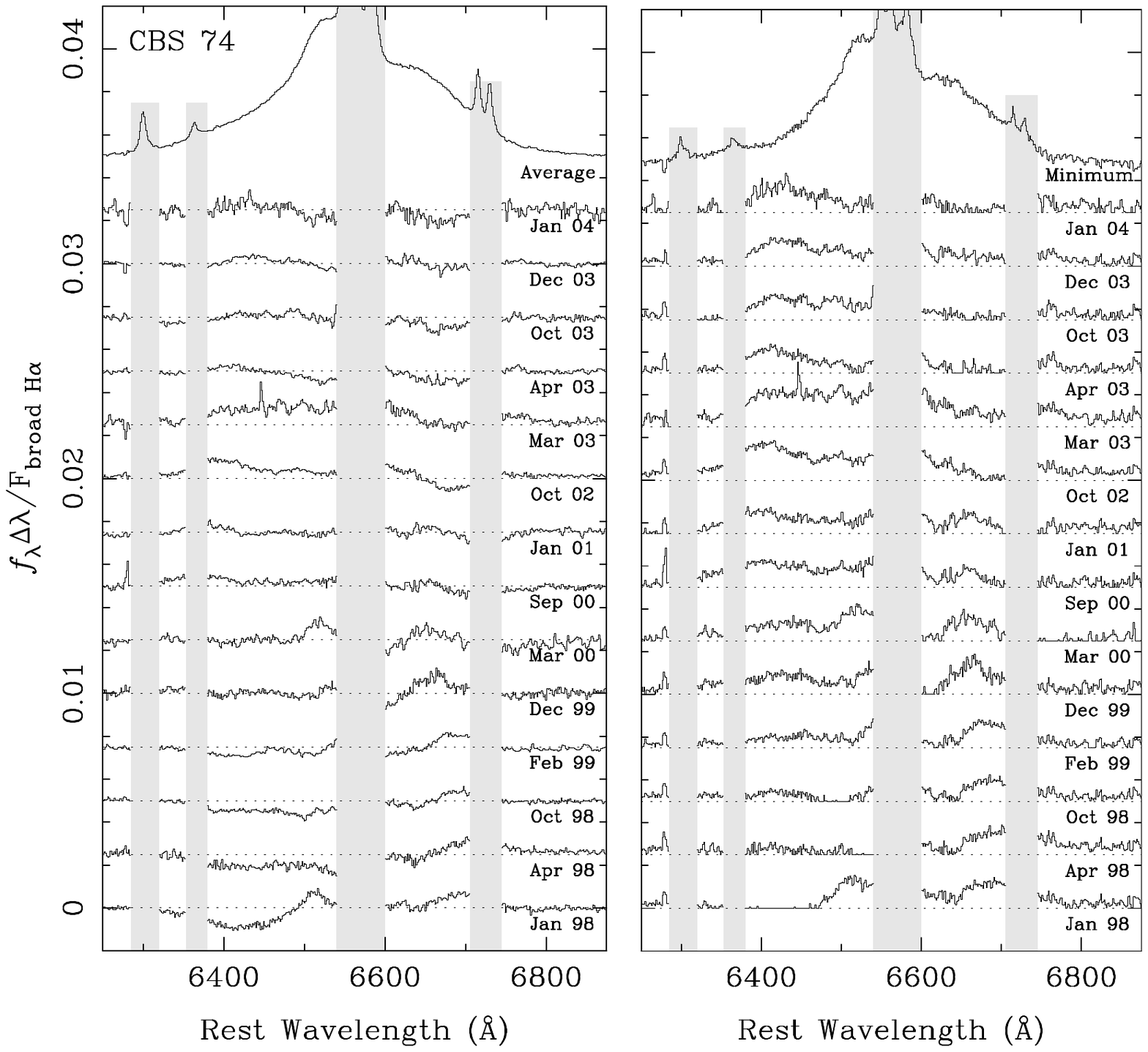}
\caption{\label{cbs74_diff} Same as Figure \ref{3c59_diff}, but for
CBS~74.}
\end{figure}

\begin{figure}
\epsscale{1}
\plotone{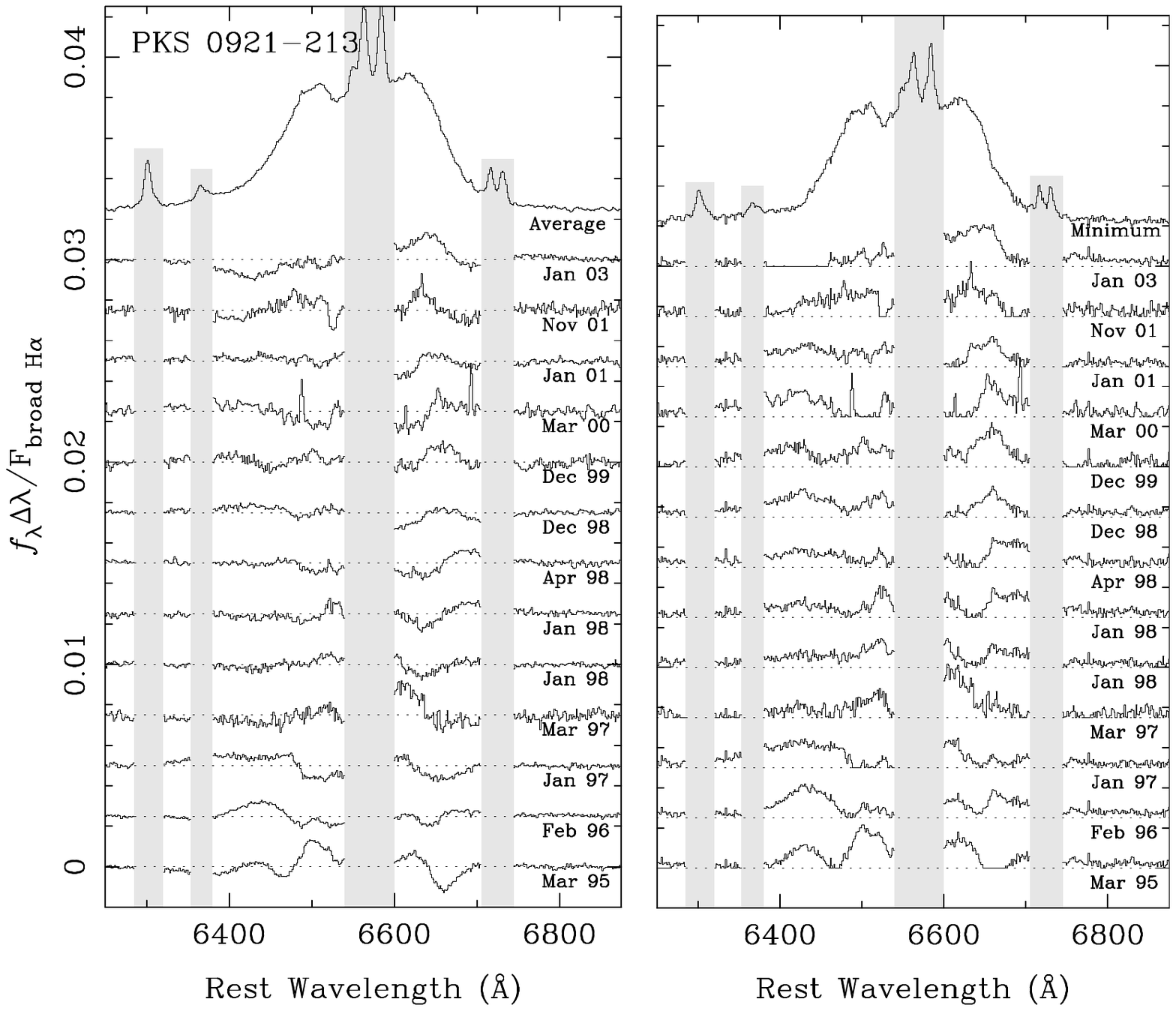}
\caption{\label{pks0921_diff} Same as Figure \ref{3c59_diff}, but
for PKS~0921--213.}
\end{figure}

\begin{figure}
\epsscale{1}
\plotone{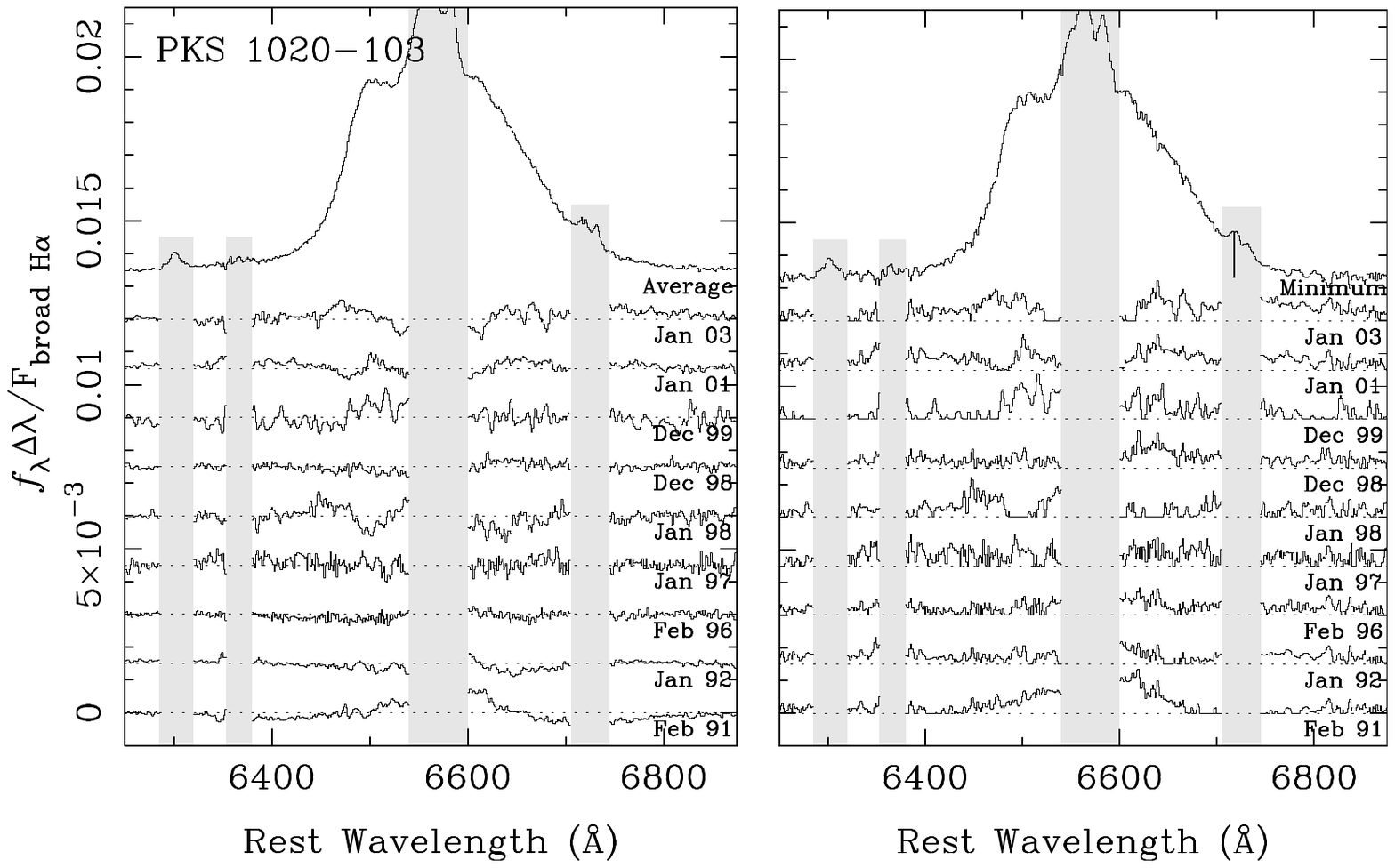}
\caption{\label{pks1020_diff} Same as Figure \ref{3c59_diff}, but
for PKS~1020--103.}
\end{figure}

\begin{figure}
\epsscale{0.85}
\plotone{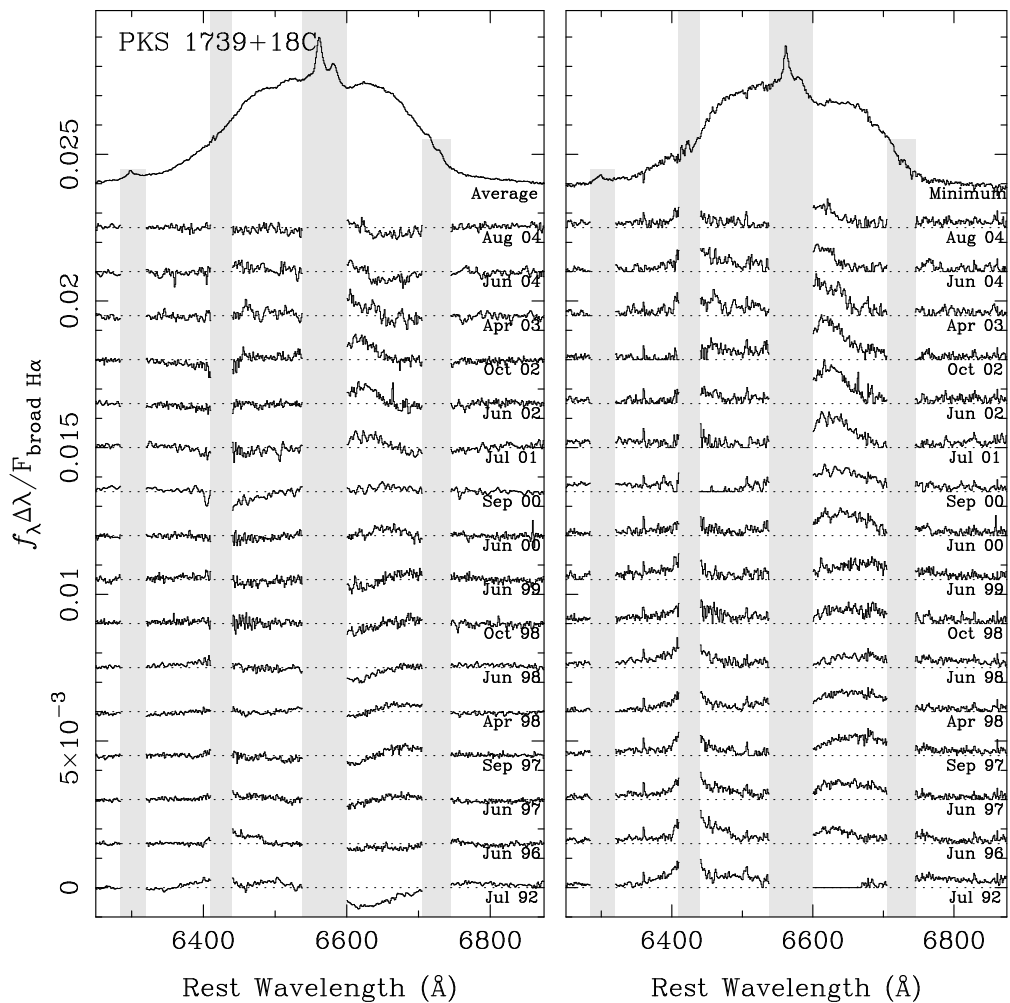}
\caption{\label{pks1739_diff} Same as Figure \ref{3c59_diff}, but
for PKS~1739+18C.  An additional gray stripe indicates the region
of the spectrum affected by telluric A-band ($\sim$7400--7440$\AA$
at the redshift of this object.)}
\end{figure}

The mean and minimum comparison spectra were subtracted from each
individual spectrum and the resulting difference spectra are shown in
Figures \ref{3c59_diff}--\ref{pks1739_diff}.  As with the rms
profiles, we are not concerned with the variations in the broad-line
flux, thus the profiles were normalized as described above. Because
the spectra were renormalized by their broad-line flux, the narrow
emission lines will vary from one observation to the next.  Rather
than remove the narrow lines, which would require assumptions about
the shape of the underlying profile, these regions were simply not
plotted in the difference spectra and the omitted regions are
indicated by vertical gray stripes. As can be seen in Figures
\ref{3c59_diff}--\ref{pks1739_diff}, the prominent features in the
difference spectra are generally broader than the omitted spectral
regions, thus these gaps do not greatly impair the interpretation of
the difference spectra.

There are several trends commonly seen in the difference spectra of
most objects. In general, trends are best illustrated by the differences
from the minimum spectrum.

\begin{enumerate}

\item{The difference spectra suggest the presence of distinct ``lumps''
of emission, and typically there are multiple lumps, sometimes at both
negative and positive projected velocities, at any given time. The
amplitude of these lumps, relative to the flux in the minimum spectrum
at the position of the lump, is typically 20--30\% although when the
lump is located at the wings of the profile, the relative flux is much
larger ($\sim$ 50\%). The lumps typically represent 5--10\% of the
{\it total} broad H$\alpha$ flux. An exception is Pictor~A, in which
the spectrum has undergone significant changes in profile shape; the
flux in the ``lumps'' of emission is several times larger than in
either the minimum or the average spectrum, and the lumps are as much
as 20\% of the total broad-line flux. The changes in these distinct
``lumps'' of excess emission, either in shape, amplitude, or projected
velocity, are primarily responsible for the variability in the line
profiles.}

\item{Neither the average nor, more importantly, the minimum spectra
  are consistent with the expected profile from a circular accretion
  disk, because the red peak is often as strong as, or stronger than,
  the blue peak. This may be due to the fact that if only a partial
  variability cycle has been observed, the minimum profile will not be
  that of the underlying disk; we explore this further in
  \S\ref{compare_observed}.  However, another possibility is that if
  the profile variability can indeed be explained as an underlying
  disk with extra, transient lumps of emission, the underlying disk
  itself may not be axisymmetric and thus the ``base'' (i.e. minimum)
  profile would vary on long timescales (e.g. if the underlying disk
  is elliptical, then the base profile will precess as described in
  \S\ref{variability_timescales}). One should be cautious in
  associating the minimum spectrum with the profile of an underlying
  ``static'' disk to which all other features are added.}

\item{The less luminous systems (1E~0450.3--1817, Pictor~A, and
  PKS~0921--213) vary on shorter timescales than the more luminous
  systems. The black hole masses ($\mbh \sim 4\times10^{7} \msol$) for
  these three objects have been estimated from the stellar velocity
  dispersion \citep{le06}, and the physical timescales for the
  variability in these objects will be discussed in more detail in
  \S\ref{variability_discussion}}.

\item{If one assumes that lumps from one observation epoch can be
tracked in following observations, several general conclusions can be
drawn.}

\begin{enumerate}

\item{The lumps can persist for several years; in a few objects, some
lumps appear to persist throughout the duration of the monitoring
observations such as the red lumps in 3C~59 and PKS~1739+18C. These
lumps still do vary slightly in amplitude and shape, but they do not
dissipate completely.}

\item{When the lumps do disappear over time, they typically arise on
  rather short timescales ($\lesssim$ one year) and dissipate more
  slowly (over several years), generally becoming broader in the
  process of decaying.}

\item{Lumps can drift across the profile in either direction. We note
  that it is more common for lumps to drift from positive to negative,
  reminiscent of Arp~102B \citep{sps00}. However \citet{gezari07} do
  not observe such a trend in the longer-baseline observations of
  Arp~102B, nor in any of the other objects in their sample.}

\end{enumerate}

\end{enumerate}

1E~0450.3--1817, Pictor~A, CBS~74 and PKS~0921--213 all show striking
variability which we describe in more detail here.

\begin{description}

\item
{\it 1E 0450.3--1817.  --} This object is unique in the large number
of distinct lumps which have appeared, drifted across the profile, and
dissipated; during some observation epochs there are up to three
different lumps present at once. The lumps emerged at both small and
large (as well as negative and positive) projected velocities. Some of
the lumps are long-lived and do not vary significantly in strength,
shape, or position over a period of $\sim$ 8 years, whereas other
lumps emerge, drift, and dissipate over four years. 

\item
{\it Pictor A. --} This object
shows the most dramatic variability, in the sense that the line was
not observed to be double-peaked prior to 1994 \citep{he94,s95}. The
blue lump originated at a projected velocity of $v_{p} \sim 7500
\;\kms$, suggesting that the emitting gas is located at characteristic
radius ($r_{c}$) of less than $1500\;\rg\; \sin^2 i$, where $i$ is the
angle between the line of sight and the axis of the disk. \citet{eh03}
find that $i \sim 30{\rm^{\circ}}$ when an elliptical disk model is used, and
$i>65^{\rm {\circ}}$ for a circular disk. This lump steadily drifted to less
negative projected velocity ($\Delta v_{p} \sim 3000\;\kms$) over a
period of approximately 4 years and has been dissipating slowly
throughout the campaign. The red lump originated at an earlier time
and also drifted towards smaller projected velocity, however it
significantly decreased in intensity over the first several years.

\item
{\it CBS 74. --} The blue side of the profile as a whole
is quite variable after 1998 October, and it is difficult to describe in
terms of individual lumps of emission. A lump on the red side of the
profile steadily increased strength and then began to decrease again
over a 5 year period. This lump of emission drifted blueward at first
and then drifted slightly redward before dissipating.

\item
{\it PKS 0921--213. --} The profile of this object in the
early observations is very unusual and appears to have two
double-peaked components, one with closely separated peaks overlying
another with more widely separated shoulders. Within a year, the
closely-spaced component appears to dimish, leaving a double-peaked
profile that has peaks with an intermediate spacing. After this, the
most prominent feature is a red lump, at a projected velocity of $\sim
4500\;\kms$, suggesting a characteristic radius of less than
$4000\;\rg\;\sin^2 i$ \citep[$i > 40^{\circ}$;][]{eh03} that
strengthened over two years then drifted towards lower projected
velocity by $\sim 2000\;\kms$ and was still quite strong during the
last observation.

\end{description}

In the above descriptions, we have assumed that one can track a lump
from one epoch in subsequent observations. This does not necessarily
imply that the same parcel of gas is responsible for the emissivity
enhancement.  Indeed this {\it cannot be the case}, at least for
PKS~0921--213 and 1E~0450.3--1817, if these gas parcels are part of a
Keplerian disk. The dynamical timescales for these objects are known
to be only a few months \citep{le06} and each of these objects was
closely monitored at times (with gaps of only 1--2 months between
observations); there was clearly no orbital motion.  Thus, these lumps
of emission should not be likened to the bright spot postulated by
\citet{ne97} in the context of Arp~102B. The fact that the properties
of the lumps (i.e. the width, amplitude, and especially projected
velocity) are not randomly distributed from one observation to the
next strongly suggests some kind of physical connection between them,
but the lumps are likely to be associated with a {\it location} in the
disk, rather than any particular parcel of matter within the disk.  We
offer a possible interpretation of this phenomenon in
\S\ref{fragmented_spiral}

\subsection{Variations in Profile Parameters}\label{variability_parameter_plots}

\begin{figure}
\epsscale{0.65}
\plotone{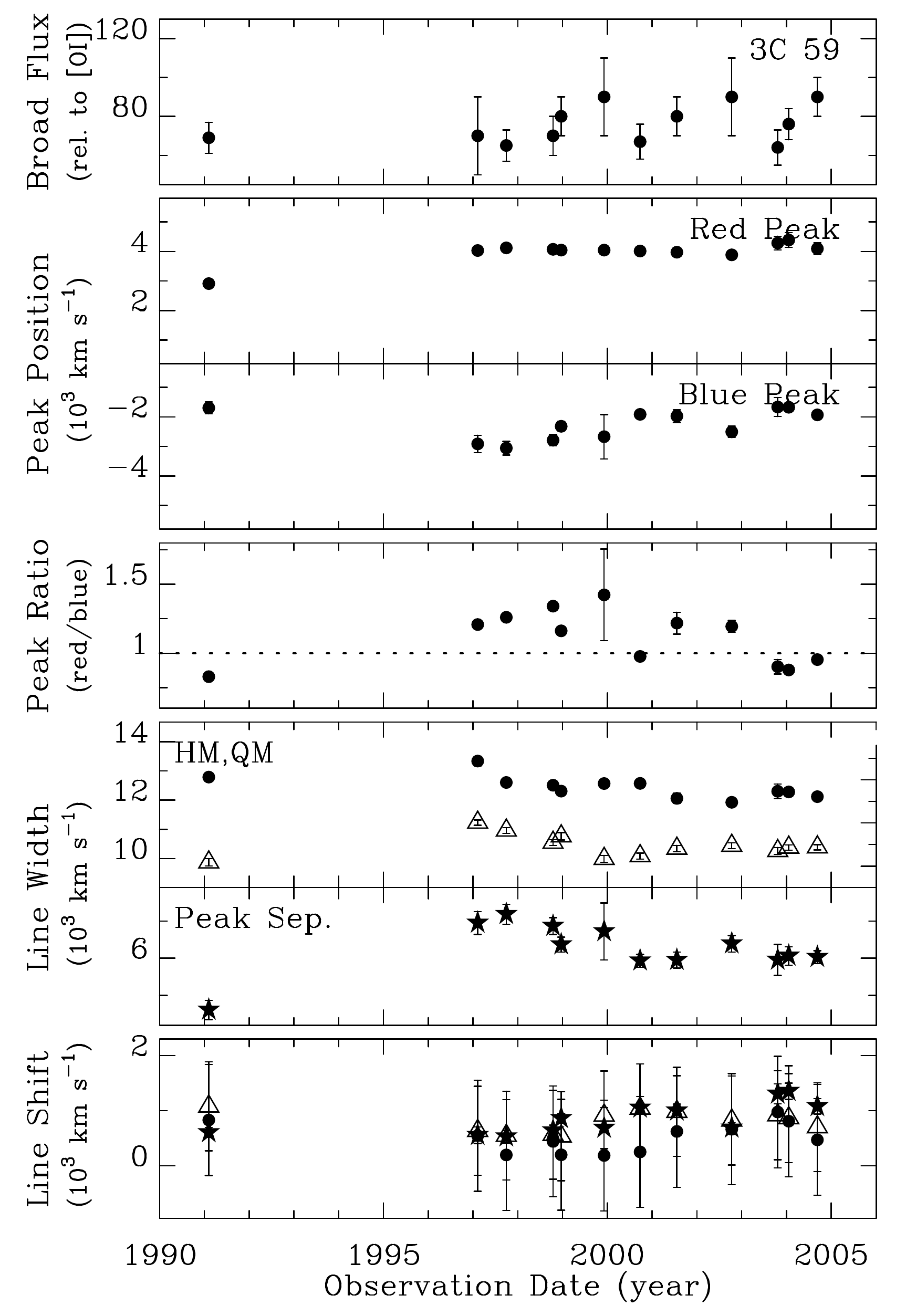}
\caption{\label{3c59_var} Variations in the profile properties or
3C~59: (1) the total broad-line flux, relative to
\oxi$\lambda6300$; (2) the velocity shift of the red and blue peaks,
(3) the ratio of the red to blue peak flux; (4) the full-widths at
half and quarter maximum and the peak separation (denoted by open
triangles, filled circles, and filled black stars, respectively); (5)
and the velocity shift of the profile at the HM, QM and the offset of
the average peak velocity with the same symbols as for the widths. All
velocities are relative to the narrow H$\alpha$ emission line. The
measurement procedure is described in
\S\ref{variability_parameter_plots}.}
\end{figure}
\begin{figure}
\epsscale{0.65}
\plotone{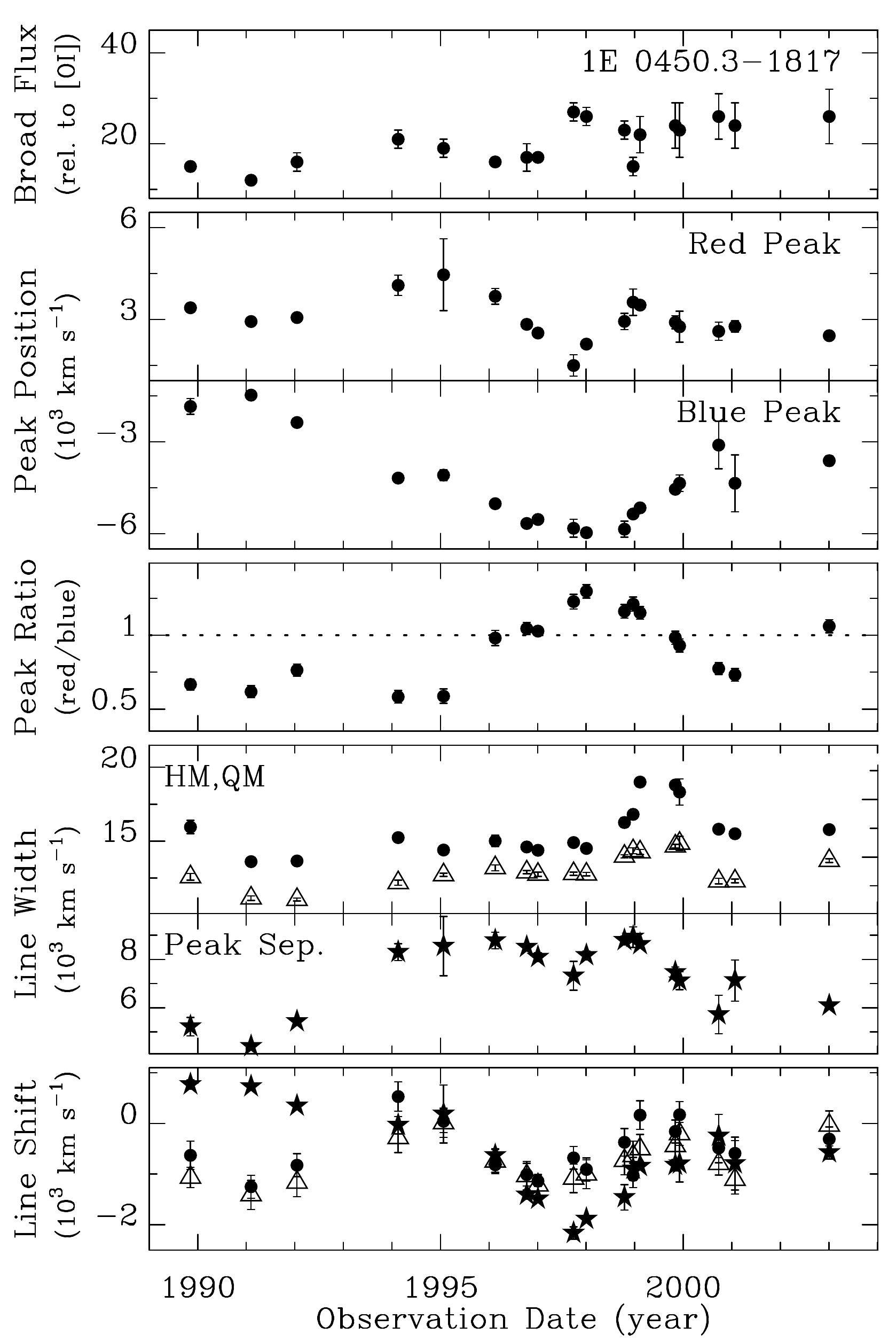}
\caption{\label{1e0450_var} Same as Figure \ref{pictora_var}, but for
1E~0450.3--1817.}
\end{figure}
\begin{figure}
\epsscale{0.65}
\plotone{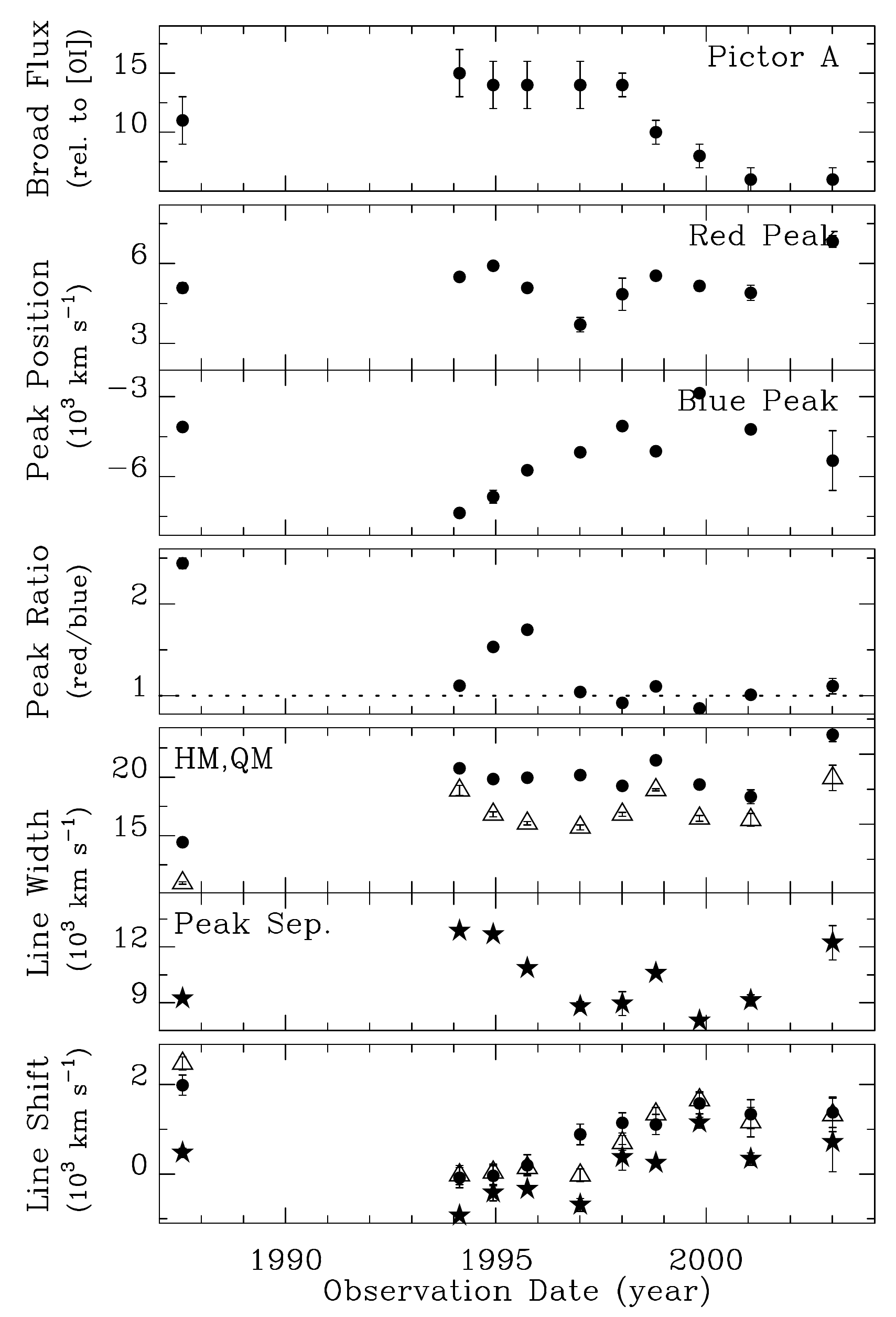}
\caption{\label{pictora_var} Same as Figure \ref{3c59_var}, but for Pictor~A.}
\end{figure}
\begin{figure}
\epsscale{0.65}
\plotone{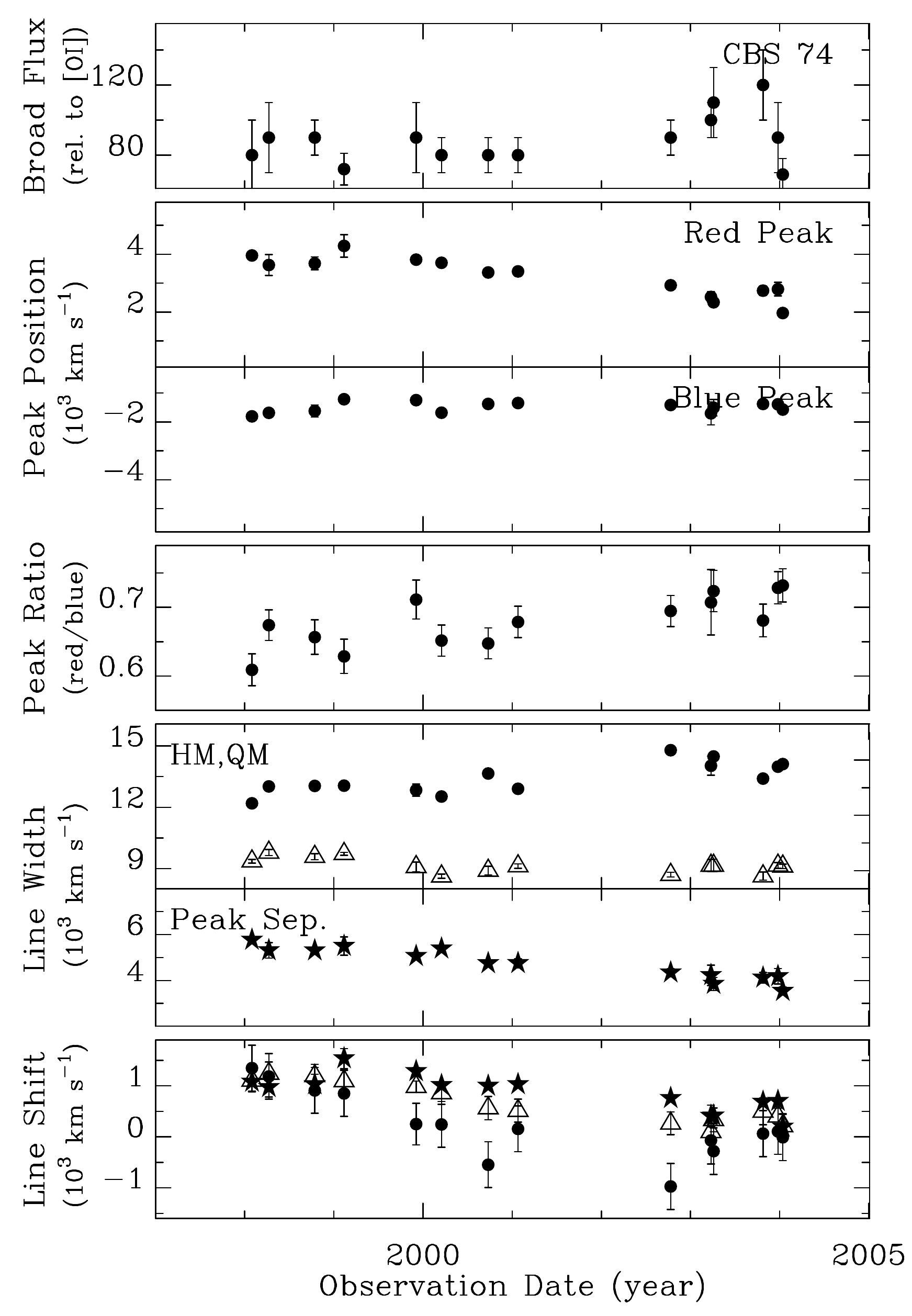}
\caption{\label{cbs74_var} Same as Figure \ref{3c59_var}, but for
CBS~74.}
\end{figure}
\begin{figure}
\epsscale{0.65}
\plotone{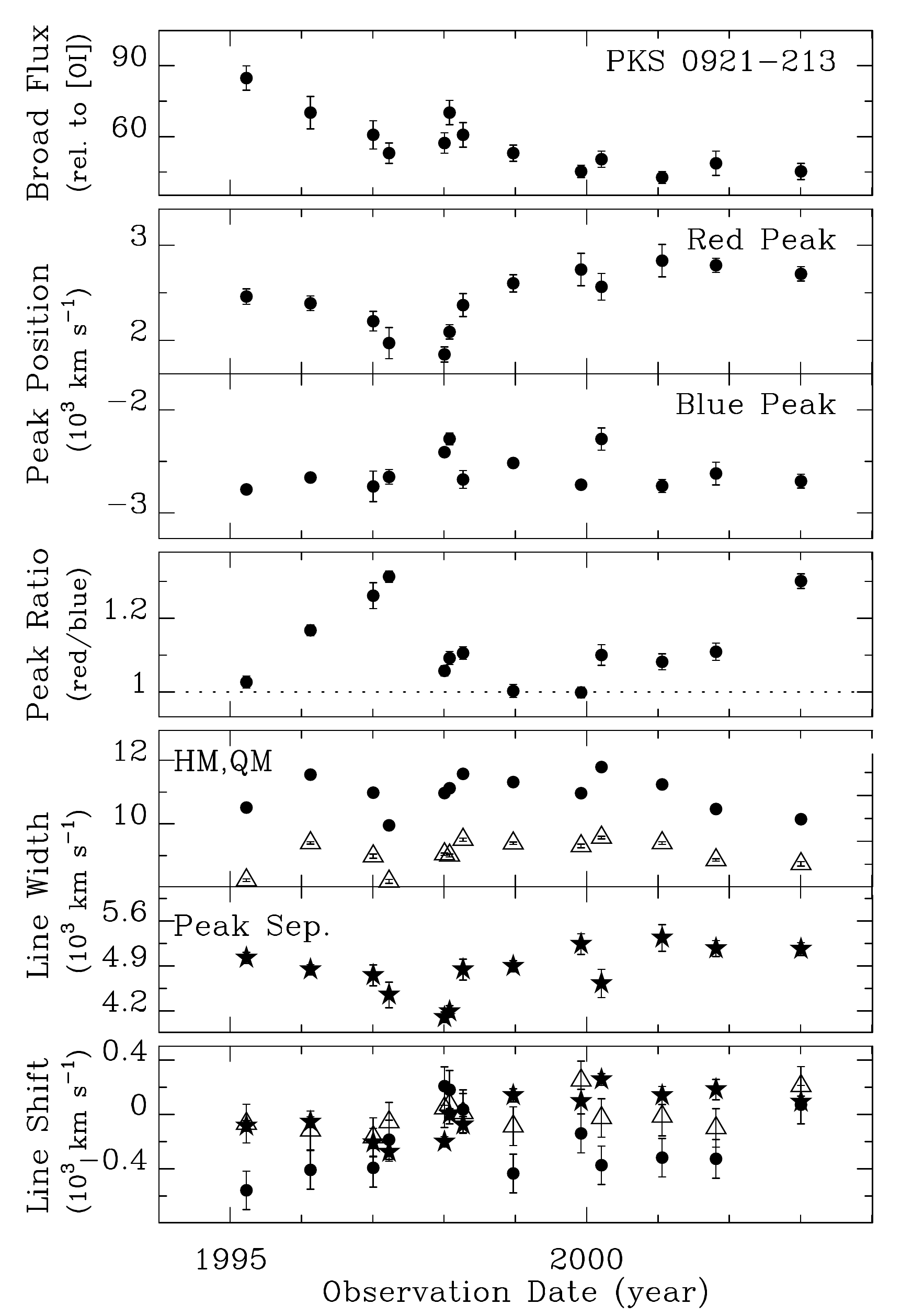}
\caption{\label{pks0921_var} Same as Figure \ref{3c59_var}, but for
PKS~0921--213.}
\end{figure}
\begin{figure}
\epsscale{0.65}
\plotone{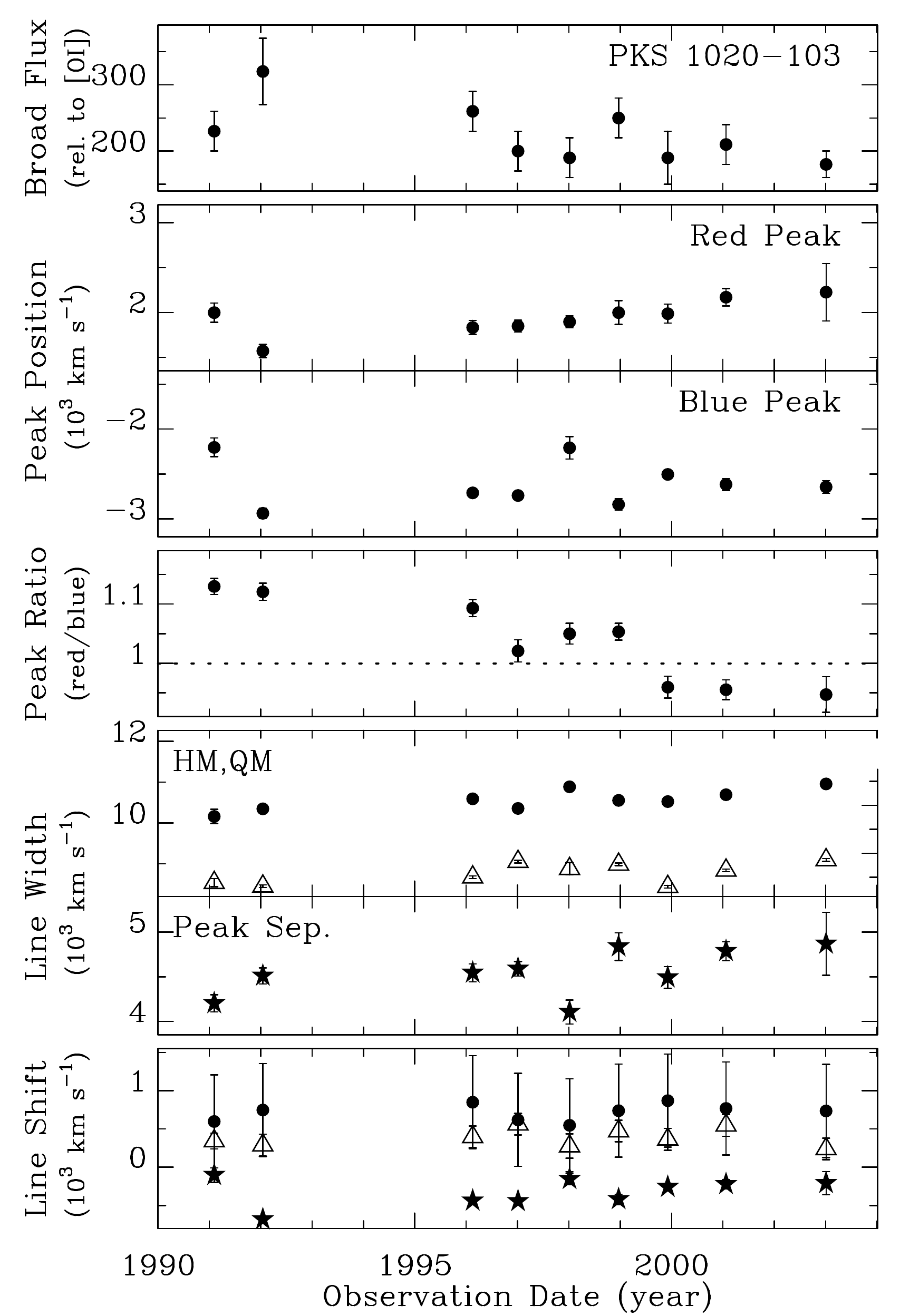}
\caption{\label{pks1020_var} Same as Figure \ref{3c59_var}, but for
PKS~1020--103.}
\end{figure}
\begin{figure}
\epsscale{0.65}
\plotone{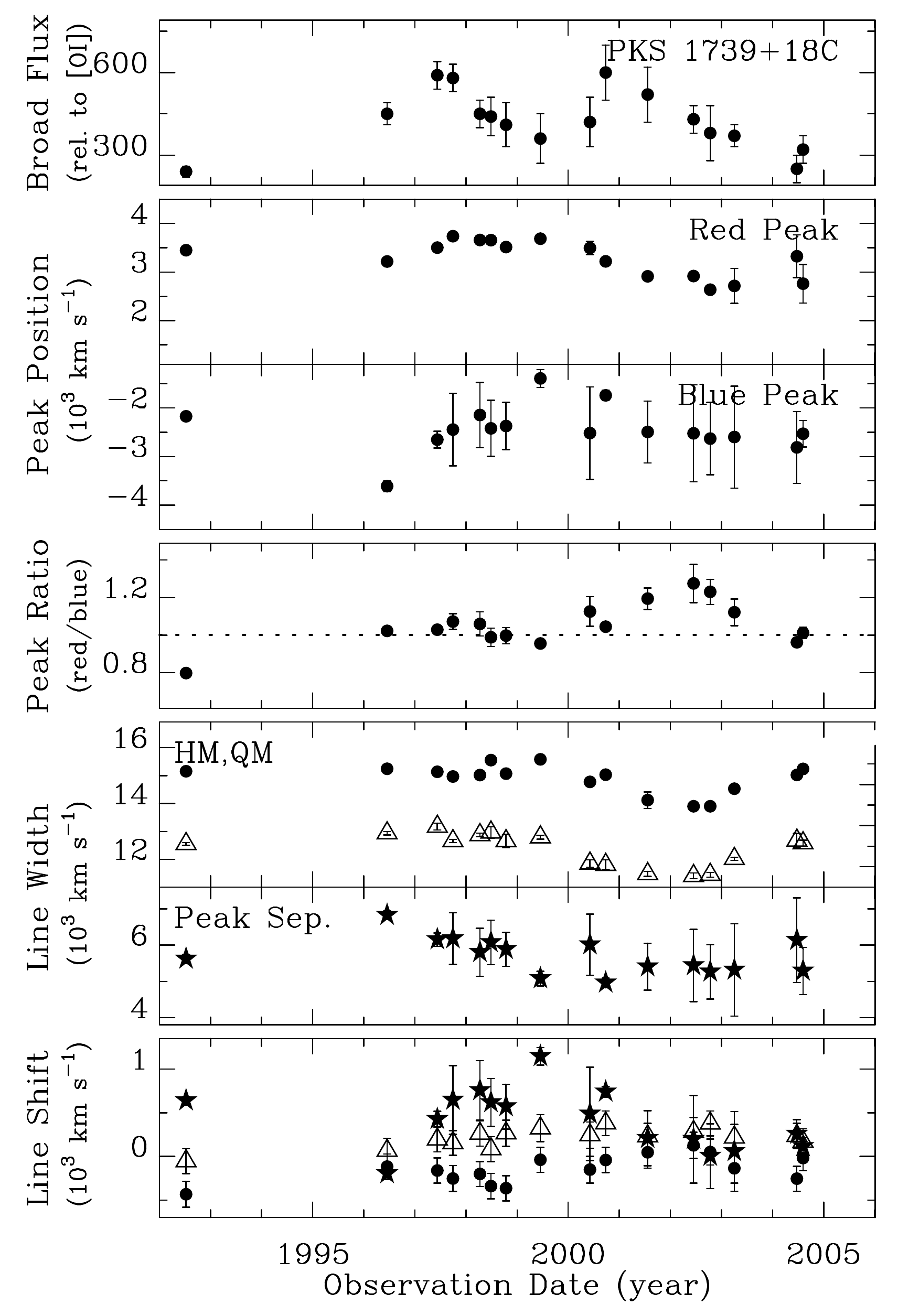}
\caption{\label{pks1739_var} Same as Figure \ref{3c59_var}, but for
PKS~1739+18C.}
\end{figure}

Although the profiles are extremely complex, to first order they can
be reduced to a small set of easily measurable quantities: the
velocity shifts of the red and blue peaks, the ratio of the red to
blue peak flux, the full-width of the profile at half-maximum and
quarter-maximum (HM and QM), the velocity separation of the two peaks,
the velocity shifts of the profile centroid at HM and QM, and the
average velocity shift of the peaks. The maximum is defined to be the
flux density of the higher peak and all velocities are measured with
respect to the narrow H$\alpha$ emission line.  The location and flux
density of the two peaks were measured by fitting a Gaussian to the
region around each peak. The resulting measurements of profile
parameters are included in Table~4 and are plotted as a
function of time in Figures~\ref{3c59_var}--\ref{pks1739_var}.

In determining the profile parameters for a given spectrum, there are
three major sources of uncertainty. First and foremost is the error in
measuring the various parameters. The velocity shifts of the red and
blue peaks and the shifts of the profile at HM and QM can be quite
uncertain. For example, when a peak is flat-topped, such as the blue
peak of PKS 1739+18C, there is considerable ambiguity in the location
of the peak. In objects where the peak fluxes are relatively low or
the profile is very broad, the profile at the HM or QM is contaminated
by the narrow lines, making it difficult to determine the width and
especially the shifts. To take into account this uncertainty, each
profile property was measured three times, the values were averaged
and the standard deviation in the measurements was assigned as an
approximate error bar. In some objects (1E~0450.3--1817, CBS~74, and
PKS~0921--213,), there appear to be multiple peaks in a small number
of spectra and in all cases the stronger peak (which also had the less
extreme projected velocity) was adopted rather than measuring the
positions of both peaks and assigning extremely large errors to the
peak position and flux ratio.

In addition, continuum subtraction introduces additional uncertainty
in the measured profile properties, particularly in the HM and QM.
It was not practical to explore the effect of continuum subtraction
for each spectrum individually. Instead, the effects of continuum
subtraction were explored in great detail for two representative
spectra, one from PKS 0921--213, which has a large starlight fraction,
and one from PKS 1739+18C, which has a negligible contribution from
starlight. For both objects, several different template galaxies,
powerlaw indices, and fitting intervals were used. For each individual
global continuum subtraction, several different local continuum 
subtractions were performed as well. Over 100 sets of measurements 
to the broad-line profile were made for each representative
spectrum, which allowed us to separately estimate the additional 
uncertainty introduced by the global and local continuum fitting 
processes. For both PKS~0921--213 and PKS~1739+18C, the fractional
uncertainty introduced by the global continuum subtraction was very 
similar; although PKS~0921--213 had a larger contribution from starlight, the
continuum fit was constrained by the many stellar absorption features.
The fractional error resulting from the global continuum fit was 
adopted for all objects.

For both PKS~0921--213 and PKS~1739+18C, the uncertainty introduced
due to the local continuum subtraction were also similar. Fitting the
local continuum is a much simpler process and we were able to use
several spectra for each of our objects to estimate the uncertainties
from the local continuum subtraction. If the error for a particular
object was larger than that found for PKS~0921--213 or PKS~1739+18C,
the error bars were increased accordingly.

Additionally, a few objects were observed multiple times
within a few days, thus it was also possible to estimate the errors
that are induced by different observing conditions. This error is
relatively small compared to those from continuum subtraction and the
measurement process, but not negligible. When multiple spectra were
obtained within one month of each other, the results were averaged
together, and in these instances it was not necessary to factor in
this final error.

The variability plots reveal a few additional trends, which were not
readily apparent from the difference spectra presented in
\S\ref{variability_diff_spec}.  

\begin{enumerate}

\item{Blueshifts in the profile at HM, QM and of the average peak
  velocity are quite common and the blueshifts can be as large as
  $2000 \;\kms$, although they are more typically
  $\sim500\;\kms$. This is in contrast to the expectations of a simple
  axisymmetric disk, in which the profile, as a whole, should be
  redshifted due to general relativistic effects. There is also
  considerable variability in these shifts. However, the errors on the
  shifts can be large at times.}

\item{With the exception of the profile widths at HM and QM, the
  various parameters rarely vary in concert. In some objects the peak
  separation varies in a similar way as the profile widths, but not in
  all instances.}

\item{It is not uncommon for the red peak to be stronger than the blue
  peak, again in contradiction with the expectations for a simple
  axisymmetric disk. With the exception of CBS~74, {\it all} of the
  objects in this sample have a red peak that is stronger than the
  blue in at least 50\% of the observations. We note that due to the
  non-uniform sampling, this does not imply that the red peak is
  stronger than the blue 50\% of the time.  More puzzling is that in
  some cases the peak reversal is extreme, with the red peak being
  20-30\% stronger than the blue. The most striking instances of a
  peak reversal are in PKS~0921-213 and Pictor~A, in which the blue
  peak has at best achieved a flux equal to that of the red peak. In
  the context of a disk model, a rather dramatic asymmetry is required
  to produce such a strong peak reversal. If this asymmetry is
  persistent, then objects which exhibit these strong red peaks
  should, at some epoch, exhibit extremely large ratios of the blue to
  red peak flux, especially considering that any asymmetry in the
  emissivity will be Doppler boosted on the blue side.  These objects
  may simply have not been observed over a long enough time period to
  observe large blue peak fluxes yet, however the lack of strong blue
  peaks is very curious.}

\end{enumerate}

The objects studied by \citet{gezari07} do not show this high
incidence of strong red peaks. Two objects show a stronger red peak in
just one observation but Mkn~668, similar to PKS~0921--213 and
Pictor~A, has a strong red peak throughout the monitoring
campaign. There were also very few objects in that study which
exhibited blueshifts at the QM, but for 3C~227 and Mkn~668 the
blueshifts were extreme ($\sim 1500\;\kms$). In the SDSS sample of
double-peaked emitters, \citet{s03} finds that it is also not uncommon
for the red peak to be stronger than the blue peak or for the profile
to have an overall blue-shift of 1000 $\kms$ or more.

We also searched for correlations between the broad H$\alpha$ flux and
the other profile parameters.  A negative correlation between the
broad-line flux and peak separation was observed in NGC~1097
\citep{sb03} and 3C~332 \citep{gezari07}. Such an anti-correlation can
be easily explained in the context of the disk model.  As the
illuminating flux increases, the portion of the disk that has the
proper ionization state to produce the Balmer lines will drift towards
larger radii. As a result, lower velocity gas makes a greater
contribution to the profile and the peaks will move closer
together. Additionally, one would expect that the line profile becomes
narrower for the same reason, as is observed in 3C~332.

A few objects (1E~0450.3--1817, CBS~74, and PKS~0921--213) show
negative correlations between the peak separation and the broad
H$\alpha$ flux for a {\it portion} of the observations, but never for
the entire set of data. Additionally, a corresponding negative
correlation between the line width (either FWHM or FWQM) and the broad
H$\alpha$ flux is not observed. Any efforts to investigate
correlations between the broad H$\alpha$ flux and other properties of
the profile are hampered by the fact that during the course of these
observations, the flux varies nearly monotonically for all objects and
the overall changes in the flux are not large. We note that if the
line emitting region is quite large, a small change in the flux will
not significantly change the inner and outer radius of the line
emitting region. Indeed NGC~1097 and 3C~332 have rather small disks
\citet{eh94,eh03} compared to many studied in this sample. It is
certainly plausible that the ionizing flux can affect the structure of
the disk on long timescales, but we do not find any evidence in these
data for a simple relationship between the ionizing flux and the
profile properties.

\section{Model Profile Characterization}\label{variability_models}

In this section we characterize the profile variability of two
families of models---an elliptical disk and a circular disk with a
spiral emissivity perturbation---which have been used in the past to
model the profile variations in some double-peaked emitters (see
\S\ref{variability_intro}.) Although these models are simple, there is
strong motivation for each as we explain further below; they are not
simply convenient ways to ``parameterize'' the observed profile
variability.

\subsection{Physical Motivation}\label{variability_model_motivation}

\begin{description}

\item{\it Elliptical Disk --} An elliptical accretion disk could form
due to the presence of a second supermassive black hole, analogous to
the formation of an elliptical disk in some CVs due to perturbations
from the companion star \citep[see the discussion of][]{e95}. Because
galaxies (and in particular the massive early-type galaxies that
typically host radio galaxies) are thought to form through mergers, it
is likely that some galaxies contain binary black holes pairs
\citep*{bbr80} and a few close pairs have been observed in NGC~6240
\citep{komossa03} and the radio galaxy 0402+379 \citep{rtz06}. Due to
the continuous perturbation by the second black hole, the accretion
disk around the primary can attain a uniform eccentricity and maintain
that eccentricity for some time. Alternatively, if the accretion disk
is only perturbed temporarily by a massive body, the inner regions of
the disk will begin to circularize due to differential precession,
while the outer regions can retain their eccentricity for $\sim
10^4$~years.

A second formation mechanism for an elliptical disk suggested by
\citet{e95},  inspired by the sudden appearance of double-peaked
emission lines in the LINER NGC 1097, is the tidal disruption of a
star by a black hole. For a black hole with a $\mbh \lesssim
10^{8}\msol$, a solar-mass star will be tidally disrupted before it is
accreted. This tidal debris initially has very eccentric orbits, and
\citet{sc92} find that the debris will form an eccentric accretion
disk within a viscous timescale.

\item{\it Spiral Arms. --} The second model we consider is that of a
circular disk with a spiral emissivity perturbation.  They are a
mechanism of angular momentum transport \citep[see, e.g,][]{ls91,m89}
thus they are an attractive process in the outer disk where other
proposed mechanisms (such as ``viscosity'' and hydromagnetic winds)
may be ineffective.  Spiral arms are directly observed in galactic
disks and indirectly CVs \citep*[see,
e.g.,][]{shh97,bap05,p05}. Spiral waves are thought to form in CVs due
to tidal interactions with the binary companion whereas in
self-gravitating galaxy disks they form due to gravitational
instabilities.

The outer portions ($r \ge 1000\;\rg$) of an AGN accretion disk are
self-gravitating \citep[e.g.][]{eh94}, thus spiral arms could be
launched from the outer regions of the disk and propagate inwards
\citep*{ars89}. The passage of a massive star cluster or a second
supermassive black hole could also trigger the formation of spiral
arms in the disk \citep[e.g.][]{cw93}. \citet{cw93} and \citet{cw94}
first suggested that spiral shocks might be a major contributor to the
profile variability of double-peaked emission lines in CVs and AGNs
(Arp~102B and 3C~390.3), respectively.

\end{description}

In the scenarios presented above, either the eccentric disk or
the spiral arm will precess on long timescales, leading to variability
in the observed line profile. The timescales over which the
variability is expected are discussed further in
\S\ref{variability_timescales}. The line profiles were calculated
following the weak-field formalism laid out by \citet{chf89} and
\citet{ch89}, which takes into account several relativistic effects,
namely: Doppler boosting, gravitational redshifting, and light
bending.  It is important to note that the weak-field approximations
made in this calculation are invalid for $r \lesssim 100 \rg$
\citep[see, e.g.,][]{f89}. The codes used to implement both the
elliptical and spiral-arm models for the line profiles are the same as
those used by \citet{sb03}.

\subsection{Model Profile Calculation}\label{variability_model_calc}

A circular accretion disk is described by the inner and outer radii
of its line-emitting region ($\xi_{1}$ and $\xi_{2}$) in units of the
gravitational radius ($\rg$), the inclination angle $i$ ($i\equiv0$
for a face-on disk), the powerlaw emissivity index ($q$,
$\epsilon(\xi) \propto \xi^{-q}$), and a broadening parameter
$\sigma$, which accounts for local broadening due to turbulent motions
in the disk. The value of $q$ lies between 1 and 3 \citep[see,
e.g.][]{ch89,dcs90a,s03,eh03}.

We use the elliptical disk model presented by \citet{e95}, where the
disk is described by the same parameters as the circular disk, except
that $\xi_{1}$ and $\xi_{2}$ represent the inner and outer pericenter
distances.  The eccentricity is described with three
parameters---$e_{1}$, $e_{2}$, and $\xi_{e}$. The disk has an
eccentricity $e_{1}$ from $\xi_{1}$ to $\xi_{e}$ and increases
linearly from $e_{1}$ to $e_{2}$ over the distance $\xi_{e}$ to $\xi_{2}$.
Thus a wide range of disks can be described, from disks with constant
eccentricity to those in which the inner regions have
circularized. Time evolution of the profiles via precession of the
elliptical disk is simulated by varying the angle that the major axis
makes to the line of sight ($\phi_{0}$).

The spiral-arm model was implemented by introducing an emissivity
perturbation with the shape of a spiral arm, as described by
\citet{g99} and \citet{sb03}.  In addition to the five circular disk
parameters, the spiral arm is described by a ``contrast'' ($A$)
relative to the underlying disk, a pitch angle ($p$), angular width
($\delta$), and inner termination radius $\xi_{sp}$ (the arm is
launched from the outer edge of the disk). Again time variability is
simulated by varying the viewing angle $\phi_{0}$.

A one-armed spiral is preferred in a self-gravitating disk that has a
Keplerian rotation curve \citep{ars89}. \citet{lk96} find that even if
a disk is linearly stable to the growth of an $m=1$ mode, the
formation of a single-armed spiral can be triggered by coupling
between higher mode perturbations (e.g. the $m=3$ and $m=4$
modes). Thus from a theoretical standpoint, a single-armed spiral
seems to be preferred in the absence of an external
perturbation. Furthermore, \citet{g99} found that the presence of
multiple spiral arms leads to a disk that is too symmetric to
reproduce the peak reversals observed in 3C 332 and 3C 390.3.

\subsection{Model Characterization}\label{variability_model_characterization}

\begin{figure}
\epsscale{0.85}
\plotone{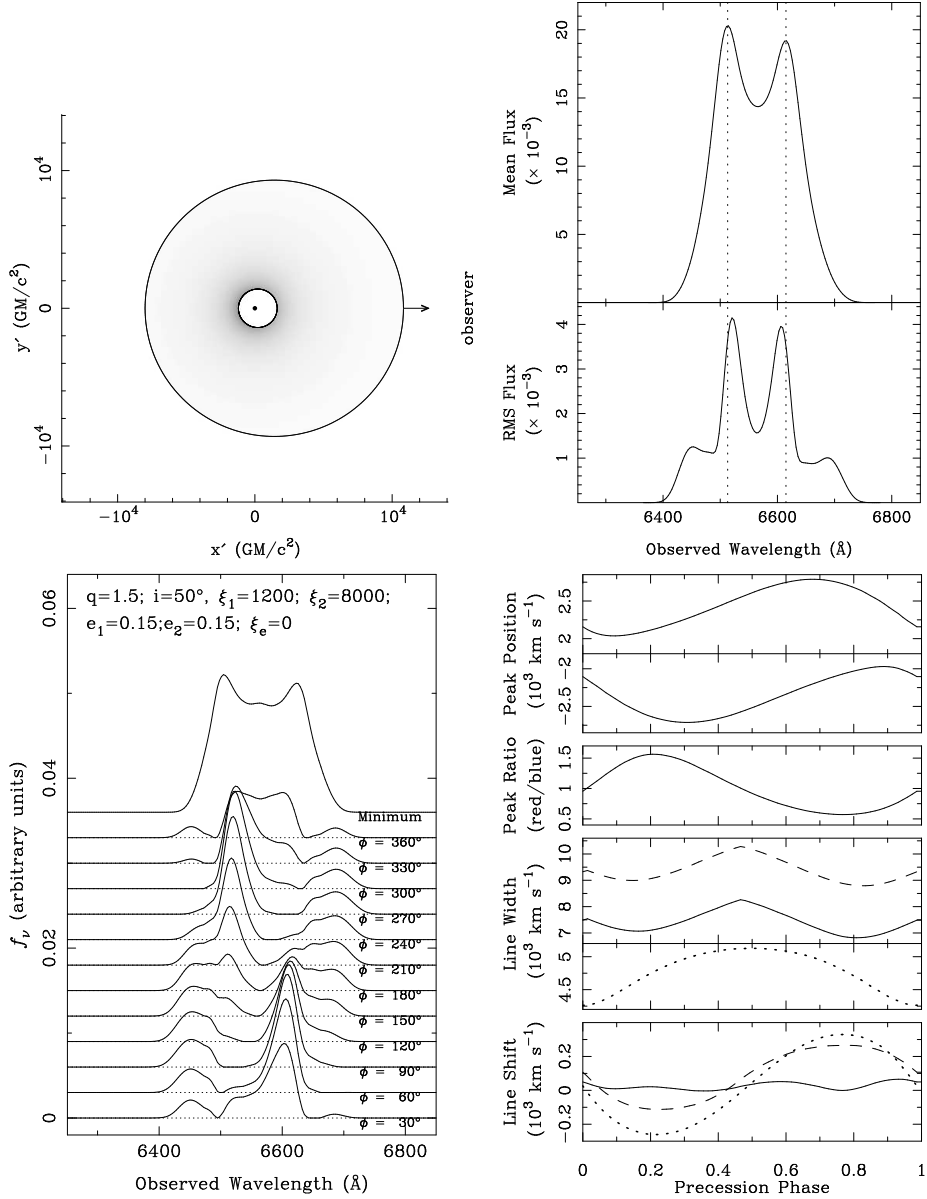}
\caption{\label{pks0921_ell} Elliptical disk results for a disk based
on the circular disk to PKS~0921--213 ($q$=1.5, $i$=$50^{\rm
\circ}$, $\xi=1200$--8000, $\sigma$=600$\; \kms$) from \citet{eh03}.
For illustration, a constant eccentricity of 0.15 was used. {\it Top
left:} grayscale image of the disk emissivity. {\it Top right:} mean
and rms profiles. {\it Bottom left: } differences from the minimum
spectrum at various precession phases. {\it Bottom right:} variations
in profile parameters with precession phase. In the lower two panels,
the velocity width and shift of the half-maximum are represented by
solid lines, those at the quarter-maximum by dahsed lines, and the
peak separation and the average peak velocity by dotted lines. See \S
\ref{variability_model_calc} for a description of the model
parameters.}
\end{figure}

\begin{figure}
\epsscale{0.85}
\plotone{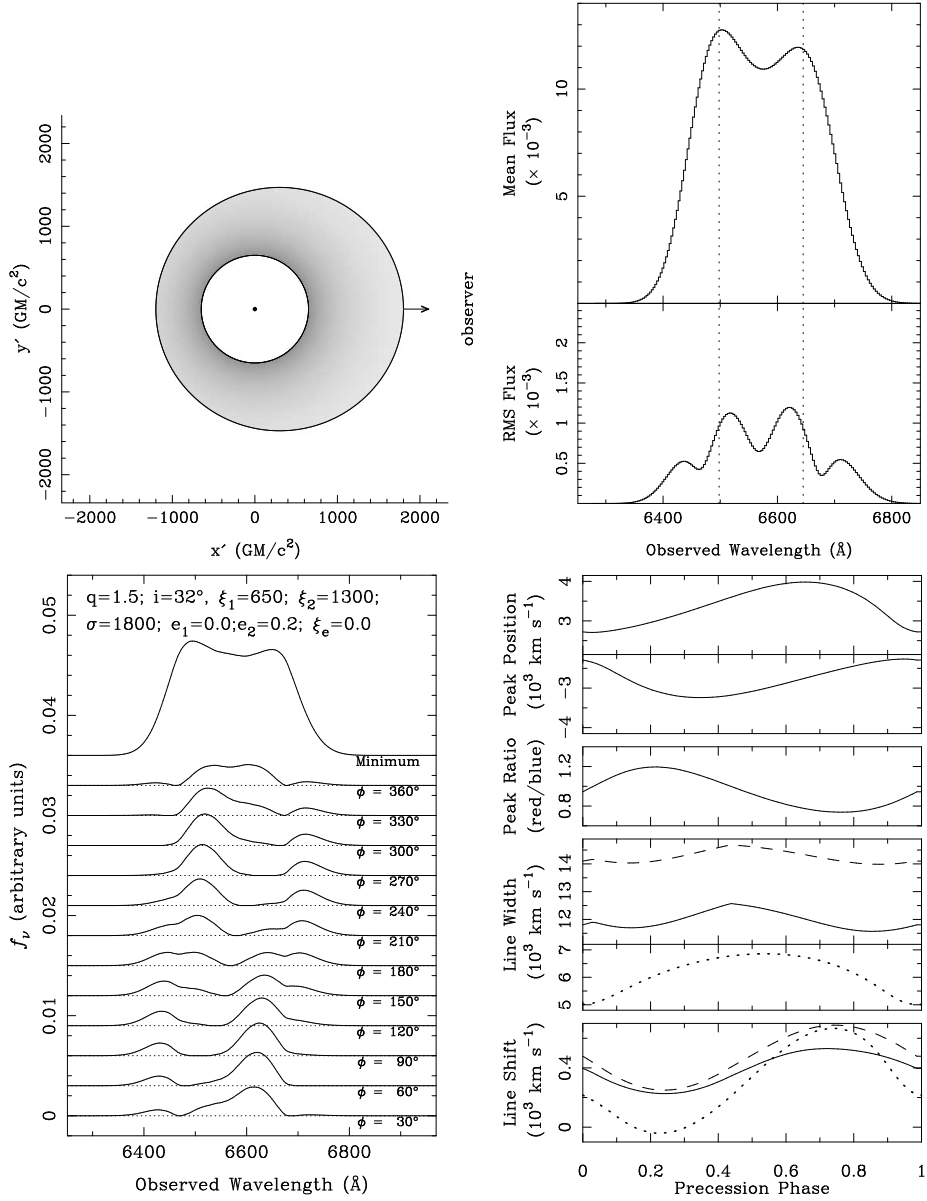}
\caption{\label{pks1739_ell} Similar to Figure \ref{pks0921_ell}, but
for an elliptical disk model based on the circular disk fit to
PKS~1739+18C ($q$=1.5,$i$=$32^{\rm \circ}$,
$\xi$=650--1300, $\sigma$=1800$\;\kms$) from \citet{eh94}.  For
illustration a eccentricity which increases linearly from $e=0$ at
$\xi$=650 to $e=0.2$ at $\xi_{e}$=1300 was used.}
\end{figure}

\begin{figure}
\epsscale{0.85}
\plotone{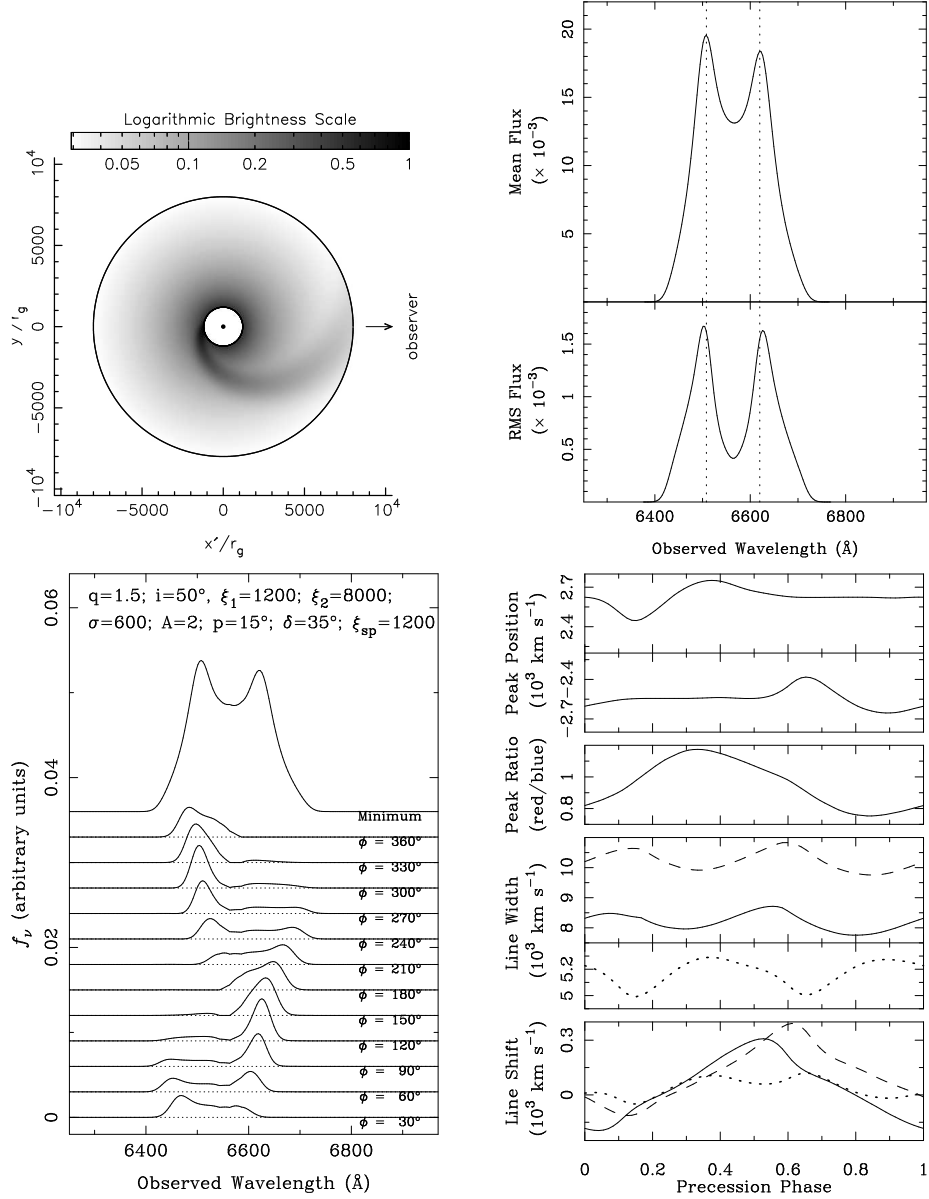}
\caption{\label{pks0921_sp1} Similar to Figure \ref{pks0921_ell}, but
for a model with a single, spiral arm. As with Figure
\ref{pks0921_ell}, the circular disk parameters are based on those
that best-fit PKS~0921--213.  For illustration, a spiral arm with
$A$=2, $p$=15$^{\rm \circ}$, $\delta$=35$^{\rm \circ}$, and
$\xi_{sp}$=1200 was used.}
\end{figure}

\begin{figure}
\epsscale{0.85}
\plotone{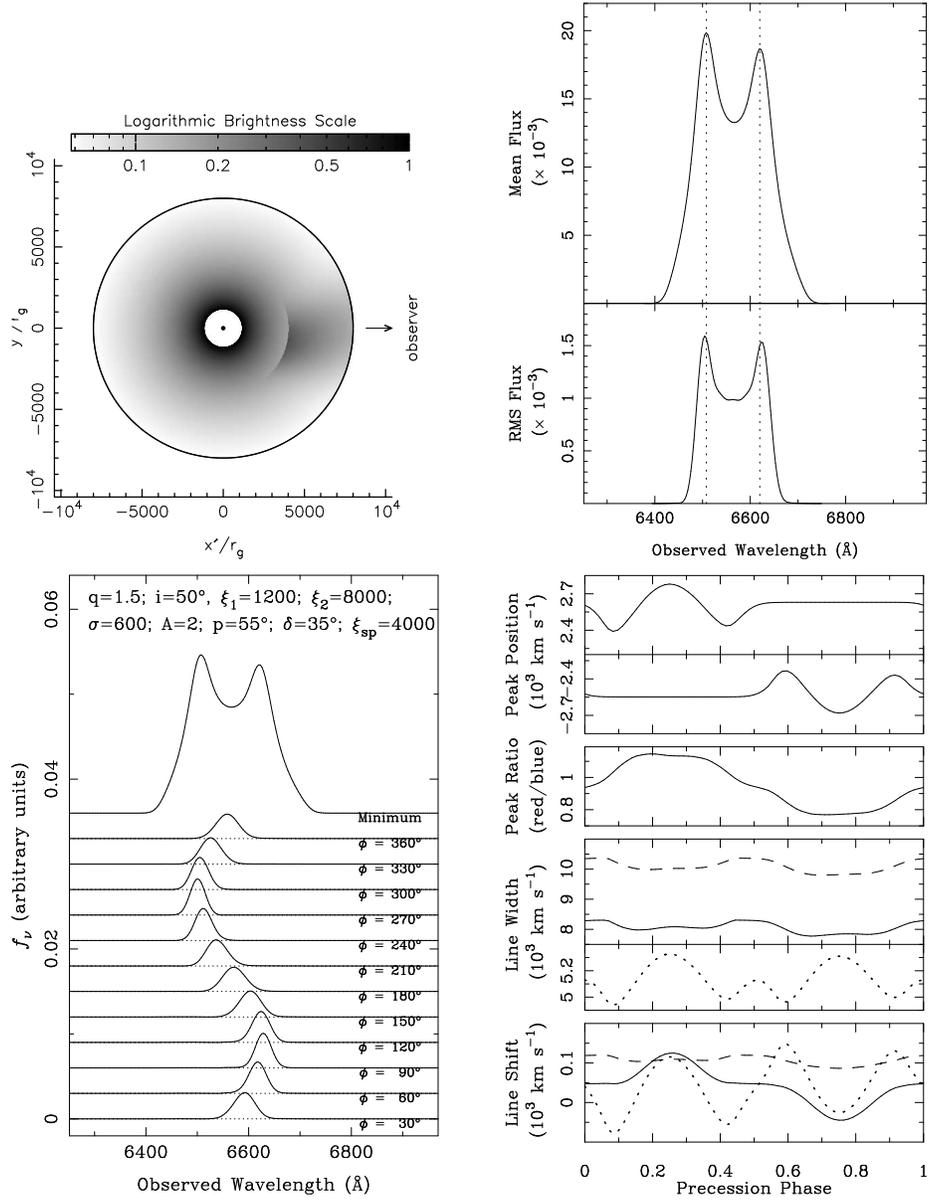}
\caption{\label{pks0921_sp2} Similar to Figure \ref{pks0921_sp1}
but for a spiral arm with $A$=2, $p$=55$^{\rm \circ}$,
$\delta$=35$^{\rm \circ}$, and $\xi_{sp}$=4000.}
\end{figure}

\begin{figure}
\epsscale{0.85}
\plotone{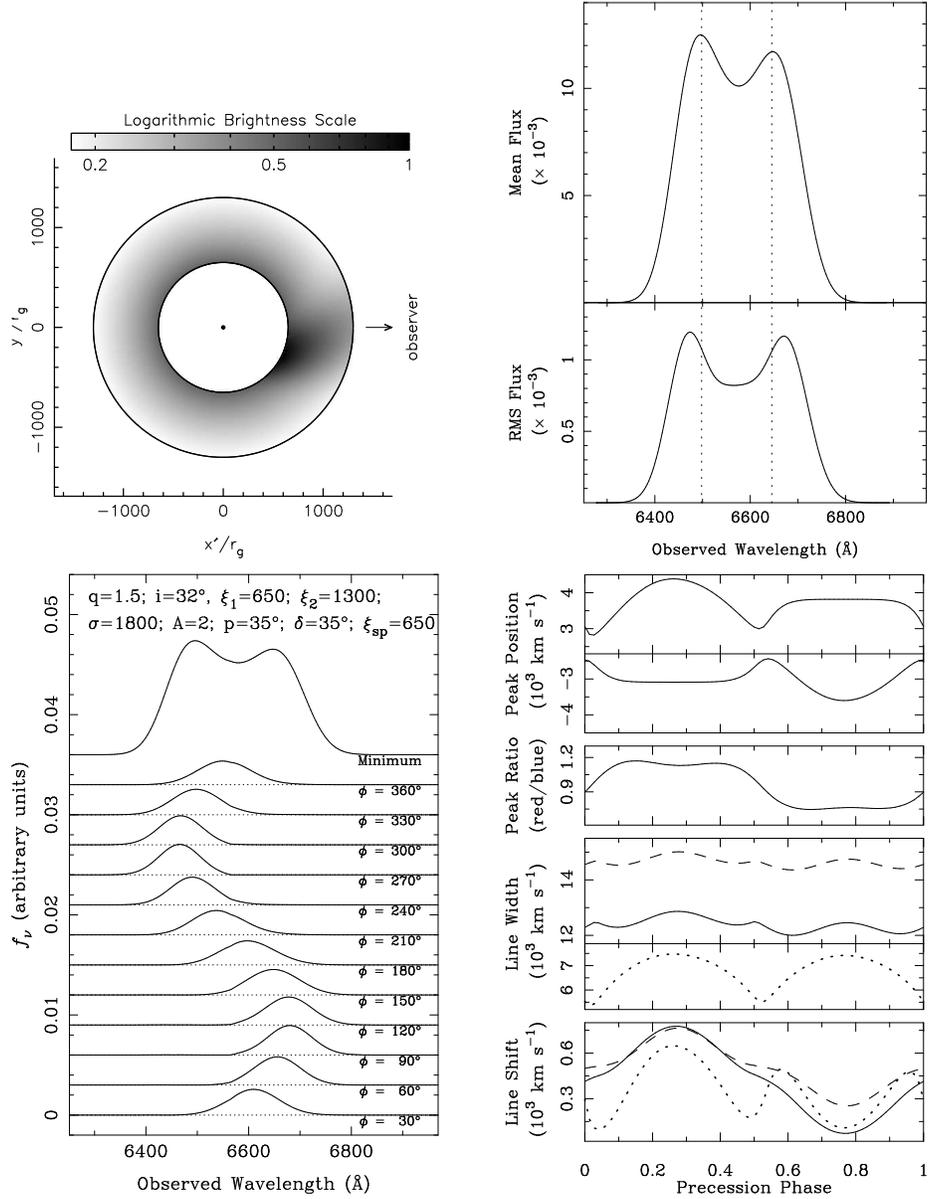}
\caption{\label{pks1739_sp1} Similar to Figure \ref{pks0921_sp1} but for
a circular disk based on the best-fit model for PKS~1739+18C (see
Figure \ref{pks1739_ell}) and a spiral arm with $A$=2, $p$=35$^{\rm
\circ}$, $\delta$=35$^{\rm \circ}$, and $\xi_{sp}$=650. }
\end{figure}

\begin{figure}
\epsscale{0.85}
\plotone{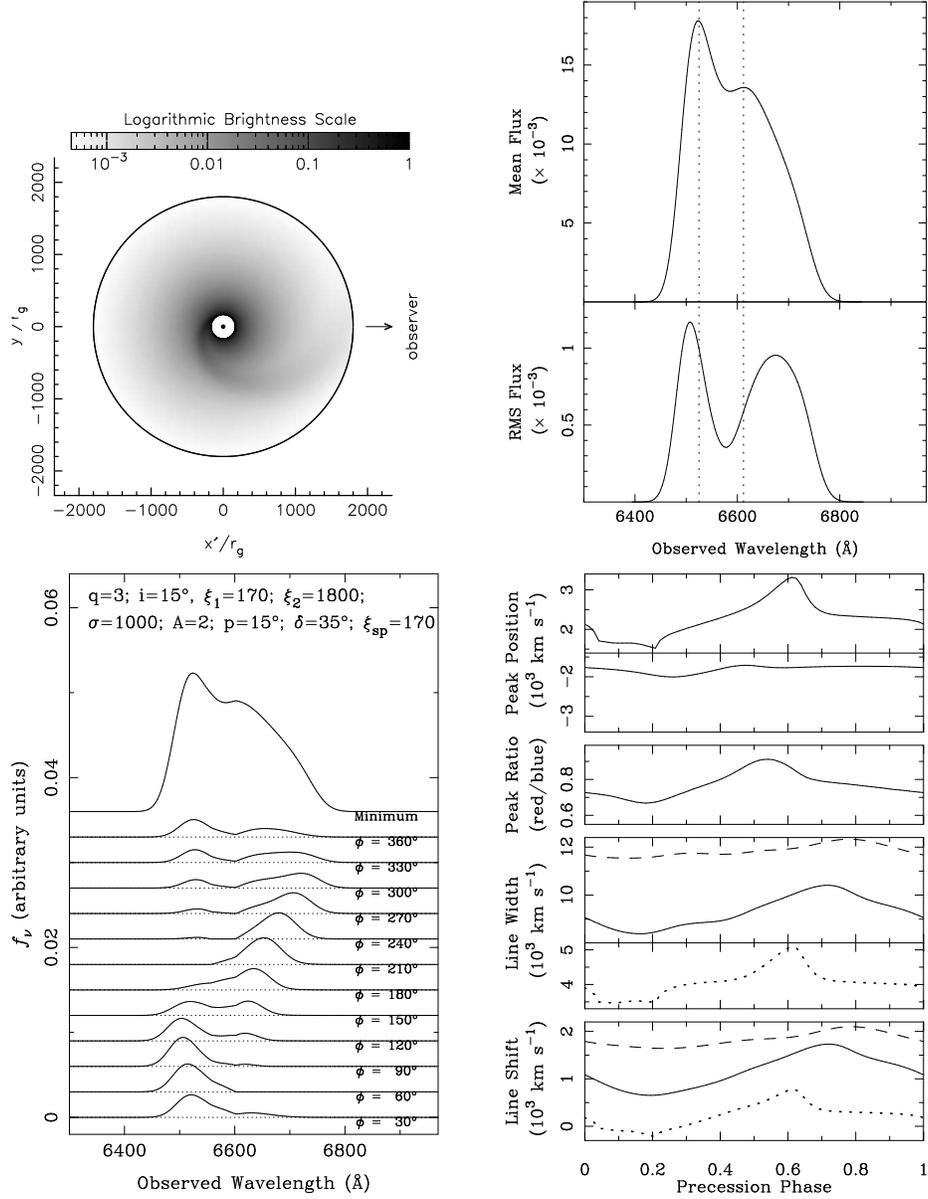}
\caption{\label{cbs74_sp} Similar to Figure \ref{pks0921_sp1} but for a
circular disk based on the best-fit model for CBS~74
($q$=3, $i$=$15^{\rm \circ}$, $\xi$=170--1800, $\sigma$=1000$\;\kms$)
from \citet{eh03} and a spiral arm with $A$=2, $p$=15$^{\rm \circ}$,
$\delta$=35$^{\rm \circ}$, and $\xi_{sp}$=170.}
\end{figure}

We have characterized extensive libraries of these model profiles in
exactly the same way as for the data, namely by constructing rms and
difference spectra and plots of the profile parameters as a function
of time.  For each set of model profiles, the primary disk parameters
($\xi_{1},\xi_{2},i,q,\sigma$) were chosen to specifically match some
of the objects being studied here to facilitate a comparison. In
Figures \ref{pks0921_ell}--\ref{cbs74_sp} we show a few representative
examples for models based on PKS~0921--213, PKS~1739+18C, and CBS~74
whose best-fit circular disk models span a range of ratios of inner to
outer disk radii and inclination \citep{eh94,eh03}. The other model
properties were chosen simply for illustration. In each of these
figures, the emissivity of the disk is shown in the top left panel.
The elliptical disk model leads to qualitatively similar variability
properties over a wide range of model parameters, although the
amplitude of the variations is certainly dependent on specific model
parameters. The spiral-arm model exhibits more diverse variability
patterns thus, a larger number of examples are shown. We note that
CBS~74, due to its skewed profile, does not always exhibit the same
trends as the other objects.

\begin{description}

\item
{\it rms spectra -- see the top right panel in
Figures \ref{pks0921_ell}--\ref{cbs74_sp}. } There is a striking
difference between the rms spectra for the elliptical and spiral arm
models. The elliptical disk models show two sets of peaks, with
primary peaks occurring at projected velocities slightly smaller than
those in the mean and a secondary set at larger projected velocities.
The overall amplitude of the rms profile increases with the
eccentricity, but the shape of the profile is primarily determined by
the eccentricity prescription (constant versus linearly increasing) with
larger primary/secondary peak ratios and more widely separated
secondary peaks for constant eccentricity disks. It is also not
uncommon for the red peaks of the rms profile to be stronger than the
blue.  The models based on CBS~74 show behavior that is fundamentally
the same as the other objects, but the peak shifts are so extreme that
that the red peak in the mean usually corresponds to a minimum in the
rms profile.

On the other hand, the spiral-arm model produces rms profiles that are
more similar to the mean profile, with just one set of peaks that are
always at larger projected velocities than in the mean.  The amplitude
of the rms variability does not depend on any one parameter.  As would
be expected, the FWHM of the rms profile is larger when the spiral arm
extends to small radii, however the peaks of the profile remain
constant over a wide range of model parameters.  The low projected
velocity portion of the profile is the most sensitive to the changes
in the model parameters (see
Figures~\ref{pks0921_sp1}--\ref{pks0921_sp2}). However, this region is
contaminated by the narrow ${\rm H\alpha}$+\nii\ lines so this is not
a useful diagnostic. Again, models based on CBS~74 are somewhat
different; the projected velocity of the red peak of the rms profile
decreases quite dramatically as the inner radius of the spiral arm
increases. Also, as with the elliptical models, the red peak in the
mean usually corresponds to a minimum in the rms profile.

\item
{\it Difference spectra -- see the bottom left panel Figures
\ref{pks0921_ell}--\ref{cbs74_sp}. } The difference spectra for the
spiral-arm models, as expected, reveal a lump of excess emission that
travels across the profile.  When the arm has a small pitch angle, at
certain phase angles the line of sight passes through the arm at two
different locations, leading to two lumps of emission at different
velocities (e.g. Fig \ref{pks0921_sp1}.  However, there is typically
a single lump of emission and the difference spectra are quite
simple. This lump of emission changes shape (broadening and
sharpening) depending on whether the line of sight passes across or
along the spiral arm. The difference spectra for the elliptical disk
models are comparatively complex.  There are always two main lumps of
emission, one blue-shifted and another red-shifted, which alternate in
strength, and the stronger lump is generally narrower than the weaker
lump.

\item
{\it Variability plots --see the bottom right panel Figures
\ref{pks0921_ell}--\ref{cbs74_sp}. } When considering the variability
plots, it is the elliptical disk model which yields extremely simple
variability patterns.  All of the profile parameters, with the
exception of the peak separation, vary nearly in concert (i.e. the
minima and maxima co-incide). Every parameter, except the profile
width at HM and QM, has exactly one minimum and one maximum per
precession cycle, and the profile variations occur smoothly and
symmetrically. The profile widths and shifts at HM and QM often
undergo very little change in comparison to the peak separation and
the shift in the average velocity of the peaks. When the eccentricity
is increased, or made constant throughout the disk, the only effect is
to increase the amplitude of the variations.

On the other hand, the profile properties of the spiral-arm model vary
in a non-uniform manner. A parameter can remain nearly constant and
then undergo a significant change over a small fraction of the
precession period; this is best exemplified by the velocities of the
red and blue peaks. This is because the spiral arm is localized in
projected velocity space. The FWHM and FWQM are the only parameters
that vary at similar times and by the same magnitude. Although the
shifts in the profile centroids at HM and QM tend to vary at roughly
the same time, they can vary by different magnitudes. All of the
models shown are for spiral arms with $A=2$ to facilitate comparison;
when the amplitude is increased, the variability patterns are not
qualitatively different, but the amplitude of the variations
increases.

It must be noted that there are certainly instances, particularly when
the pitch angle and/or arm width is large, that the spiral-arm model
predicts smoothly varying profile parameters that resemble those seen
from the elliptical disk models; smooth variability does not preclude
a spiral arm but non-uniform variability does eliminate an elliptical
disk unless additional perturbations are added.

\end{description}

This characterization has shown that the spiral arm and elliptical
disk models are (in most instances) clearly
distinguishable. Furthermore, in the case of the spiral-arm model, the
variability plots are very sensitive to the input model parameters;
therefore plots of this type could be extremely useful in
quickly pinpointing which spiral arm parameters are most likely to
reproduce a sequence of observed profiles.

\section{DISCUSSION AND INTERPRETATION}\label{variability_discussion}

\subsection{Comparison of Observed Profile Variability With Models}\label{compare_observed}

The greatest difficulty in comparing the observed variability patterns
with those predicted by the models is that the observed data need not
correspond to a full precession cycle. Although this has no impact on
the variability plots, it does affect the characterization of the rms
and difference spectra. Therefore, for a subset of the models in our
library, we constructed the rms profiles and the two comparison
profiles (minimum and average) used for the difference spectra using
only a quarter or half of the cycle, starting at various initial phase
angles.

For the elliptical disk model, we find that the rms profile has the
same basic characteristics (two sets of peaks) even when just a
quarter cycle is used. On the other hand the minimum and average
profiles are very different; even when using a half cycle and the
difference spectra are greatly affected. However, we note that the
difference spectra still have lumps of emission at both negative and
positive projected velocity.  For the spiral-arm model, the minimum
spectrum is very robust. The minimum profile for a half-cycle is
indistinguishable from that for the full cycle, and when a quarter
cycle is used, the relative heights of the two peaks are affected
slightly but the profile is still quite similar. The average profile
is affected more than the minimum by partial cycles, but again the
profiles are still quite similar to those from a full cycle. Thus for
the spiral-arm model, the difference spectra presented in Figure
\ref{pks0921_sp1}--\ref{cbs74_sp} remain qualitatively the same even
if a fraction of a cycle is used. However, the rms plot changes
considerably with one of the two peaks often disappearing.  Thus for
the elliptical model, comparisons between the observed and predicted
difference spectra should be made with extreme caution, whereas for
the spiral-arm model it is the rms profiles that are subject to the
greatest uncertainty. With this in mind, we now compare the most
prominent observed variability trends with those predicted by the
models.

\begin{description}

\item {\it rms Variability. --} Neither model is very successful in
  reproducing the rms profiles observed, which often have three
  distinct features.  The observed rms profiles do not have four
  peaks, as predicted by the elliptical disk model even for partial
  cycles. In some objects with peaks close to the narrow
  H$\alpha$+\nii\, the inner pair of peaks could be masked by the
  narrow lines, but this is not the case for most objects. The spiral
  arm model, on the other hand, leads to rms profiles with at most two
  peaks (depending on what fraction of the cycle is used).  Both
  models can account for the unusual rms profile of CBS~74, which has
  a red peak in the rms profile that is at a significantly larger
  projected velocity than the peak in the mean profile as well as a
  minimum in the rms profile at the velocity of the red peak in the
  mean profile. The behavior of PKS~1739+18C (which has a minimum in
  the rms profile coinciding with a peak in the mean profile) is not
  reproduced by either model.

\item 
{\it Difference spectra. --} The observed difference spectra show
multiple lumps of excess emission, both at positive and negative
projected velocity, which is consistent with the elliptical disk but
disfavors a simple spiral-arm model.  The spiral-arm model can only
produce multiple lumps that are both red- and blueshifted at specific
lines of sight through the spiral arm. However, the lumps exhibit
large variations in projected velocity which is not seen in the
elliptical disk models, where the two primary lumps of emission simply
trade off in amplitude without the lumps drifting significantly in
projected velocity. Considering that the difference spectra for the
elliptical disk models are sensitive to partial cycles, the observed
difference spectra could be qualitatively consistent with those
predicted by the elliptical disk model.

\item
{\it Profile parameter variability. --} Both models naturally lead to
reversals in the peak flux ratio without fine-tuning, however the
elliptical disk model can lead to more dramatic peak reversal than the
spiral-arm model. Neither model can reproduce the large number of
spectra where the red peak is stronger than the blue, and the red peak
is never stronger than the blue for more than half of the
cycle. Unless all of the objects were preferentially observed at a
time when the red peak was stronger, neither model can explain the
large fraction of observations with the red peak stronger than the
blue.  Both models can produce profiles that are blueshifted at times
but blueshifts are more easily produced by the elliptical disk model
because the variations are always symmetric. Within the parameter
space explored here, extreme blue shifts (on the order of 1000 $\kms$)
are not easily produced. \citet{s03} explored a much wider range of
parameter space for the elliptical disk model and find that such
blue-shifts are completely consistent with the elliptical disk model
at least. The spiral-arm model has a much larger parameter space and
it has not been explored as extensively. The most distinctive
characteristic of the observed profile parameter plots is the fact
that the parameters do not vary in concert and the variations,
although systematic, are not smooth. This disfavors the elliptical
disk model, but is consistent with the spiral-arm model.

\end{description}

\subsection{Variability Timescales}\label{variability_timescales}
Besides the various differences in profile variability, another
significant difference between these two models is the pattern
precession timescale.  If a spiral perturbation is triggered by the
self-gravity of the disk, the resulting pattern can be expected to
roughly co-rotate with the disk. Thus the variability would occur on
nearly the dynamical timescale.

\begin{equation}\label{dynamical}
\tau_{\rm dyn} = 6\;M_{8}\; \xi_{3}^{3/2}~ {\rm months}
\end{equation}

\noindent where $M_{8}$ is the mass of the black hole in units of
$10^{8}\; \msol$, $\xi_{3}$ is the radius from the center of the black
hole, in units of $10^{3}\; \rg$.

Simulations indicate that when the disk is less massive than the
central object, the disk is stable to the development of a
single-armed ($m$=1) mode, however a single-armed spiral can result from
the interaction between higher order perturbations.\citet[][and
  references therein]{lk96}. These authors find that the pattern speed
will be similar to the beating frequency of these two modes; in the
case of an interaction between the m=3 and m=4 modes, the expected
pattern speed is about an order of magnitude longer than the dynamical
timescale. Finally, a spiral arm that is externally triggered (e.g. by
a binary companion) will precess on characteristic time scale
associated with the perturbing object.

As described in \citet{e95}, an elliptical disk can precess due to two
effects, the precession of the pericenter due to relativistic effects,
which occurs on a timescale of

\begin{equation}\label{elliptical_gr}
P_{\rm GR} \sim 10^{3}\; M_{8}\; \tilde{\xi}_{3}^{5/2}~ {\rm yr}
\end{equation}

\noindent or in the case of a binary, also due to the tidal forces of
the secondary

\begin{equation}
\tau_{\rm tidal} \sim 500 \left(\frac{q_4^3}{1+q_4}\right)
\; a_{17}^{3/2}\; M_8^{-1/2} ~{\rm yr}
\end{equation}

\noindent where $\tilde{\xi}_{3}$ is the pericenter distance in units
of $10^{3} \rg$, $q_4$ is the mass ratio divided by four, and $a_{17}$
is the binary separation in units of $10^{17}$~cm. Equation
\ref{elliptical_gr} is only applicable to an accretion ring (i.e.\
when the ratio of the outer to inner radius is small). In a
large disk (and in the absence of a constant perturbing force) the
inner disk will circularize before the outer disk has had an
opportunity to precess significantly.

For three objects in this study, Pictor~A, PKS~0921-213, and
1E~0450.3-1817, the black hole masses are known \citep[$\mbh \sim
  4\times10^{7} \msol$;][]{le06}, giving dynamical timescales of
$\sim$ 1--4 months at r=1000 $\rg$\ (where the spread is the result of a
50\% uncertainty in the black hole mass). If the disk is unstable to
the formation of a single-armed mode, the precession timescale could
be as small as a few months. However in the more likely case of
interactions between higher order modes or external perturbation, the
precession timescale for a spiral arm is several years or longer. On
the other hand, the expected precession timescale for the elliptical
disk is 400 years!  The fact that clear variability is observed within
one decade in these objects strongly suggests that the elliptical disk
model is untenable. Although black hole masses have not been obtained
for the other objects in this study, these objects are unlikely to
have black hole masses that are significantly smaller than these. The
X-ray luminosities of the remaining objects are all on the order of
$10^{44}\;\ergs$; if the black hole masses were less than $10^{7}\;
\msol$, the X-ray luminosity alone would exceed the Eddington
luminosity ($1.3\times 10^{38} \;\mbh/\msol\; \ergs$). Thus it appears
that the elliptical disk model may not be tenable for any objects in
this sample.

\subsection{Refinements to the Spiral-Arm Model}\label{fragmented_spiral}

Like \citet{gezari07} we find that neither of these simple disk models
can explain the observed profile variability behavior in detail,
although both account for some of the common trends. However, we must
keep in mind that these models are the simplest extensions to the
circular disk. There is much room for improvement and it is possible
that alternatives or modifications to the above models might yield
better agreement with the observed profile variability. Here, we focus
on the spiral-arm model, which at least based on the variability
timescale, is a viable scenario.  One simple modification which has
already been used successfully in NGC 1097 \citep{sb03} is to allow
the emissivity power-law index to vary with both radius and
time. However, as noted in \S\ref{rms_corr} and
\S\ref{variability_parameter_plots}, no unambiguous trends between the
profile properties and broad H$\alpha$ flux were found thus, such a
scenario is not justified for the objects studied here.

In the detailed descriptions of the observed difference spectra, we
noted that while it was implausible that the lumps of emission were
due to persistent bright spots orbiting in the disk, the non-random
distribution of lumps in projected velocity strongly suggested that
the various lumps were somehow physically associated with each other.
There appear to be two aspects of the variability; the projected
velocity of the lumps changes on long timescales, whereas the
properties of the lumps can change on much shorter timescales.  These
two aspects of the variability might be naturally explained if the
spiral arm is fragmented. 

\citet{cw93} modeled spiral arms in AGN disks triggered by the close
passage of another body and find that the arm does fragment and
reform.  Their primary motivation was to explore whether such a
mechanism could explain the flux variability of AGN, and indeed they
found that as an arm forms, the flux increases on timescales of a year
to several years and then small 1\% variations in the flux occur on
timescales of several months as the arm fragments and reforms.
However the simulations were limited to mass ratios of $>10^{-3}$ and
it is unclear what effect the close passage of a more common stellar
mass object would have or how long-lived the induced spiral arm would
be.  However, a molecular cloud or star cluster may be massive enough
to trigger an instability.

Alternatively, the disk itself might be non-uniform with the
sub-structure of the disk changing on short timescales, comparable to
the dynamical timescale. For example, if the emissivity of the arm is
dominated by isolated clumps that happen to be passing through the
arm, this could lead to the observed variability in the amplitude and
shape of the lumps while preserving the slow changes in the projected
velocity of the lumps.

\citet{fe08} performed simulations to investigate the applicability of
disks with stochastically changing bright regions to the variability of
Arp~102B and 3C~390.3. They consider several mechanisms to create
emissivity enhancements in the disk: self-gravitating clumps
\citep[e.g.][]{rla05b}, hot spots generated in a collision between the
disk and stars \citep{zurek94}, or baroclinic vortices
\citep{petersen07} that result from the combined effects of the
temperature gradients in the disk and the differential rotation. 

\citet{gt04} present estimates for the radius at which an AGN
accretion disk will become marginally stable to self-gravity. Using
these results, and the range of Eddington luminosities and black hole
mass for Pictor~A, 1E~0450.3--1817, and PKS~0921--213 presented in
\citet{le06}, we calculate the range of radii at which the disks in
these objects might become unstable, with the large range resulting
primarily from uncertainty in the black hole mass. The disk of
1E~0450.3--1817 is marginally stable at radii larger than
1700--3300~$\rg$ whereas the disks of Pictor~A and PKS~0921--213 are
maringally stable at radii larger than 3100--4400~$\rg$. For both
1E~0450.3--1817 and PKS~0921--213 the best-fit disk parameters
\citep{eh94,eh03} indicate that the region of instability is well
within the line-emitting portion of the disk, thus the formation of
self-gravitating clumps is a possibility for these objects. However,
Pictor~A has a much smaller disk and based on these estimates, the
disk should be stable to self-gravity.

In all of the scenarios considered by \citet{fe08}, the emissivity
enhancements are fairly long-lived, thus one would expect such
emissivity enhancements to rotate through the disk on a dynamical
timescale.  We do not observe such behavior in the few objects that
were closely monitored. This suggests that the decay time of bright
spots is shorter than the dynamical time. Alternatively, the
bright spots are generated when pre-existing clumps are compressed
further as they pass through the spiral arm.

With these considerations, a model in which the arm itself is
undergoing fragmentation seems much more attractive as the changing
emissivity pattern would naturally travel with the arm as it precesses
through the disk. 

The situation of having more than one lump of emission, on both sides
of the profile might also be accommodated with more complex spiral arm
models. Although \citet{g99} found that multiple spiral arms led to a
disk that was too axisymmetric to reproduce the observed peak
reversals, this was for a {\it uniform} spiral perturbation, not one
in which isolated clumps or fragments dominated the emissivity. In
this scenario, there may be no {\it observational} obstacle to the
presence of multiple spiral arms. There are some significant problems
that this clumpy spiral-arm model does not remedy, namely the large
blueshifts that are observed at times and the fact that most objects
have red peaks stronger than the blue in at least 50\% of the spectra.

\subsection{Mass in the Lumps of Emission}

A final consideration is how much mass is involved in the excess lumps
of emission identified in the difference spectra. The lumps are
typically 5\%--10\% of the {\it total} broad H$\alpha$ flux and the
luminosity of the lumps ranges from $4\times10^{40}$ to
$\;2\times10^{42} \ergs$, with an average luminosity of
$5\times10^{41} \ergs$, although as noted previously, the absolute
luminosities are uncertain by up to a factor of 2.  Assuming that
the excess luminosity is due to Hydrogen recombination, one can
roughly estimate the mass of Hydrogen involved, assuming constant
density and temperature:

\begin{equation}\label{lump_mass_eqn}
\frac{L_{\rm H\alpha}}{M_{\rm H\alpha}} = \frac{ n_{e}\;{\rm h\nu_{H\alpha}}\;\alpha_{\rm eff, H\alpha}}{\rm m_{H}}
\end{equation}

\noindent where $L_{\rm H\alpha}$ is the H$\alpha$ luminosity, $M_{\rm
  H\alpha}$ is the mass of hydrogen gas participating in the emission,
$n_{e}$ is the electron density in ${\rm cm^{-3}}$, $\rm{m_{{\rm H}}}$
is the mass of the hydrogen atom, and $\alpha_{\rm eff, H\alpha}$ is
the recombination coefficient, which we assume to be $\sim
10^{-13}\;{\rm cm^3s^{-1}}$, appropriate for gas with temperatures
between 5000-10,000$\;$K \citet[][Table 2.1]{osterbrock}.  We note
that $M_{\rm H\alpha}$ is a lower limit to the total mass of the gas,
since it is expected that only a photo-ionized skin of the matter
involved will emit H$\alpha$.  For an electron density of $10^{11}\;
{\rm cm^{-3}}$, typical for the broad-line region \citep[BLR;
  e.g.,][]{petersonagn}, $M_{\rm H\alpha}$ in the average lump is
$\sim 0.05\;\rm{M_{\odot}}$, but the electron densities inferred for
the accretion disk is $10^{13-15}\; {\rm cm^{-3}}$
\citep[e.g.,][]{cs87}, so the emitting mass is probably of order
$10^{-4}\; \msol$ or less. Based on comparison with the model
profiles, the amplitude of the spiral emissivity pattern is probably
$A\ls 1.5$ and only a 20\% perturbation in density would be
required. In general, we expect that very little mass is needed to
produce the observed lumps of excess emission in the H$\alpha$ line
profiles.

\section{Summary and Conclusions}\label{variability_conclusions}

In this paper we have characterized, in a model-independent way, the
variability of the broad, double-peaked H$\alpha$ emission lines in
seven objects (Pictor A, PKS 0921--213, 1E 0450.3--1817, CBS 74, 3C
59, PKS 1739+18, and PKS 1020--103) which have been monitored over the
past decade. As indicated by the rms profiles, the greatest
variability occurs at larger projected velocities, and in particular
many objects show a peak in the rms profile at large negative
projected velocities. Difference spectra showed that the variability
is caused primarily by discrete lumps of excess emission that change in
morphology and amplitude on timescales of a few years and drift in
projected velocity on longer timescales.  There are often multiple
lumps of emission observed at a single epoch and they are generally
located at both positive and negative projected velocities.  For some
objects the dynamical timescale is known to be only a few months, thus
these lumps cannot be orbiting bright spots such as those used to
model Arp~102B \citep{ne97,sps00}.

The most striking profile variations observed are changes in the ratio
of the red to blue peak flux, thus this trend which has long been
observed in some of the better-studied double-peaked emitters is very
common indeed. In fact, with the exception of CBS~74, all of the
objects in this study have a red peak that is stronger than the blue
in at least 50\% of the observations. Some objects in this sample are
very extreme, most notably PKS 0921--213 and Pictor~A, in that the
blue peaks, which are supposed to be boosted due to relativistic
effects, are rarely observed to be stronger than the red peak. We also
noted that many objects had profiles that were blue-shifted by up to
1000$\;\kms $, again contrary to the expectations of the simple
circular disk model. Further observation of these objects is
important; should they continue to show strong red peaks and overall
blueshifts this will place important constraints upon models for the
broad-line region in these objects. 

We compared these variability trends with those expected from two
simple models, an elliptical accretion disk and a circular disk with a
single-armed spiral emissivity perturbation. In general, neither of
these models reproduces the observed variability trends in detail;
most importantly spiral-arm models do not predict the presence of
multiple lumps of emission at a single epoch and the elliptical disk
model predicts profile parameters variations that are extremely
smooth, uniform, and symmetric and also rms profiles with four peaks,
neither of which is observed. From a consideration of physical
timescales, at least for the three objects with a known black hole
mass (Pictor A, PKS 0921--213, and 1E 0450.3--1817), the spiral arm
models is able to produce variability on a reasonable timescale, while
the eccentric disk model appears to be untenable.

Thus we propose an extension to the spiral-arm model in which one or
more {\it fragmented} spiral arms are present in the accretion disk.
This model retains many of the general theoretical characteristics of
the simple, uniform spiral arm but is likely to produce the observed
variability more successfully. This model is inherently stochastic and
it will be necessary to perform simulations to determine whether,
statistically, such a model can reproduce the types of behavior
observed. Such work is already being undertaken, in the context of a
clumpy disk by \citet{fe08}, but has not yet been extended to the
scenario proposed here. We reiterate that the models considered here
are the simplest extensions to a circular disk and more sophisticated
models will be required. In particular it may be important to consider
an accretion disk wind or other outflow to explain the full range of
behaviors that are observed.

Finally, we note that significant variability appears to be occurring
on timescales of less than a year in some objects. In particular lumps
of emission were observed to change significantly in shape and/or
amplitude within one year, and it is quite possible that some rapid
variations are being missed by the current observing strategy. It
would be very useful to monitor all objects at least twice per year,
and to intersperse periods of intense monitoring (perhaps as often as
every few weeks) for a few of the more variable objects such as
1E~0450.3--1817 or PKS~0921--213. In particular, to test the
fragmented spiral-arm model, it is necessary to determine the lower
limit for the timescale over which the individual lumps of emission
change in amplitude and morphology, which would provide an estimate
of the fragmentation timescale.

\acknowledgments

K.T.L. was funded by the NASA through the Graduate Research Fellows
Program (NGT5-50387), the NASA Postdoctoral Program Fellowship (NNH
06CC03B), and the American Astronomical Society's Small Research Grant
Program. Support was also provided by the Dickinson College Research
\& Development Committee. A substantial part of this work was carried
out between 1995 September and 1998 August, while M.E. was a Hubble
Fellow at the University of California, Berkeley.  That part of the
work was supported by NASA through the Hubble Fellowship grant
HF-01068.01-94A awarded by the Space Telescope Science Institute,
which is operated by the Association for the Universities for Research
in Astronomy, Inc.for NASA, under contract NAS 5-26555. This award
covered a significant fraction of the observing-related costs from
1995 September to 1998 August.

We thank S. Sigurdsson for helpful discussions, and A. Filippenko,
S. Simkin, and R. Becker for providing spectra of Pictor~A. We are
especially grateful to the CTIO and KPNO staff for their expert help
and hospitality during the course of this observing campaign.  Above
all we are indebted to Jules Halpern for his help in planning and
carrying out many of the observations for this program, for his
invaluable advise on aspects of the project, and for his enthusiastic
support over the past 20 years.

The National Optical Astronomy Observatory, which operates the Kitt
Peak National Observatory and the Cerro-Tololo Interamerican
Observatory, is operated by AURA, under a cooperative agreement with
the National Science Foundation. 

The Hobby-Eberly Telescope (HET) is a joint project of the University
of Texas at Austin, the Pennsylvania State University, Stanford
University, Ludwig-Maximilians-Universit${\rm \ddot{a}}$t M${\rm
\ddot{u}}$nchen, and Georg-August-Universit${\rm \ddot{a}}$t G${\rm
\ddot{o}}$ttingen. The HET is named in honor of its principal
benefactors, William P. Hobby and Robert E. Eberly.

The Marcario Low Resolution Spectrograph is named for Mike Marcario of
High Lonesome Optics who fabricated several optics for the instrument
but died before its completion. The LRS is a joint project of the
Hobby-Eberly Telescope partnership and the Instituto de Astronom${\rm
\acute{\i}}$a de la Universidad Nacional Autonoma de M${\rm
\acute{e}}$xico.

Many of the spectra from the 3m Shane reflector at Lick Observatory
were taken with the Kast double spectrograph, which was made possible
by a generous gift from William and Marina Kast.


\begin{deluxetable}{lcccccc}
\tablecaption{Galaxy Properties\label{variability_objects}}
\tablewidth{0pt}
\tablehead{
  \colhead{Object} & \colhead{}        & \colhead{}                     & \colhead{}                        & \colhead{Starlight\tablenotemark{c}} & \colhead{$L_{X} {\rm (erg\;s^{-1})}$}          & \colhead{X-ray}\\
  \colhead{Name}   & \colhead{$m_{V}$} & \colhead{$z$\tablenotemark{a}} & \colhead{E(B-V)\tablenotemark{b}} & \colhead{Fraction}                   & \colhead{(0.1--2.4 keV obs.)}  & \colhead{Ref.}
          }
\startdata
3C 59          & 16.0 & 0.1096 & 0.064 & 20\%--30\% & $1.6\times10^{44}$ & 1 \\ 
1E 0450.3-1817 & 17.8 & 0.0616 & 0.043 & 40\%--50\% & $5.4\times10^{42}$ & 2 \\
Pictor A       & 16.2 & 0.0350 & 0.043 & 10\%     & $3.9\times10^{43}$ & 3 \\
CBS 74         & 16.0 & 0.0919 & 0.036 & 10\%     & $2.6\times10^{44}$ & 4 \\
PKS 0921--213  & 16.5 & 0.0531 & 0.060 & 30\%--40\% & $3.3\times10^{43}$ & 5 \\
PKS 1020--103  & 16.1 & 0.1965 & 0.046 & 10\%     & $8.4\times10^{44}$ & 3 \\
PKS 1739+18C   & 17.5 & 0.1859 & 0.062 & 10\%     & $5.7\times10^{44}$ & 3 \\
\enddata  
\tablenotetext{a}{Redshifts from \citet{eh04}}
\tablenotetext{b}{Reddening obtained from \citet*{sfd98}}
\tablenotetext{c}{Typical starlight fraction within the spectroscopic aperture used for the 
observations over a rest wavelength range of 5500--7000${\;\rm \AA}$. It may vary due to seeing and/or 
the aperture size.}
\tablenotetext{d}{{\it  References for X-ray luminosity} -- 
observed X-ray fluxes for 0.1--2.4 keV band taken from the {\it Rosat} All-Sky Survey in an observed band of
0.1--2.4 keV, except 1E~0450.3-1817 ({\it Einstein})
and PKS~0921-213 ({\it XMM-Newton}). The fluxes were not corrected for absorption.
(1) \citet{brinkmann95}.
(2) \citet{stocke83}; 
(3) \citet*{brinkmann94}; 
(4) \citet{bade98}; 
(5) \citet{l10}; 
}
\end{deluxetable}

\clearpage

\begin{deluxetable}{lccccc}
\tablecaption{Instrumental Configurations\label{variability_instruments}} 
\tabletypesize{\small}
\tablewidth{0pt}
\tablehead{ 
\colhead{Code} & \colhead{Telescope} & \colhead{Spectrograph} & \colhead{Grating} & \colhead{Slit Width
($\arcsec$)} & \colhead{Spec. Res. (\AA)}
 } 
\startdata 
1a & KPNO 2.1m & GoldCam & 240    & 1.8--1.9 & 4.5--5.5\\ 
1b & KPNO 2.1m & GoldCam & 35     & 1.8--2.0 & 3.5--4.4\\ 
2a & KPNO 4m   & RC      & BL 181 & 1.7      & 6.0 \\
3a & CTIO 1.5m & Cass    & 35     & 1.8      & 3.0--4.0\\
3b & CTIO 1.5m & Cass    & 26     & 1.8      & 3.5--4.0\\ 
3c & CTIO 1.5m & Cass    & 9      & 1.3      & 8.0\\ 
3d & CTIO 1.5m & Cass    & 32     & 1.8      & 6.0--7.0\\ 
4a & CTIO 4m   & RC      & G510   & 1.5--2.0 & 6.0--8.0\\
4b & CTIO 4m   & RC      & KPGL2  & 1.5      & 7.0--8.0\\ 
4c & CTIO 4m   & RC      & BL 181 & 1.5      & 1.0--2.0\\ 
4d & CTIO 4m   & RC      & G250   & 1.5      & 8.0 \\ 
5a & Lick 3m   & Kast    & 2      & 2.0      & 6.3 \\ 
6a & MDM 1.3m  & MK III  & 600 l/mm & 1.5--2.0 & 5.0--6.0\\ 
7a & MDM 2.4m  & Modspec & 600 l/mm & 1.25--1.5 & 4--5\\ 
7b & MDM 2.4m  & Modspec & 830 l/mm & 1.25   & 2.9 \\ 
8a & HET 9.2m  & LRS     & Grism 3  & 1.5    & 4.8\\ 
8b & HET 9.2m  & LRS     & Grism 1  & 1.0    & 9.3\\ 
9a & du Pont 2.5m & Boller \& Chivens & BL 1200 l/mm & 2.0 & 5.0 \\ 
\enddata
\end{deluxetable}
\clearpage

\begin{deluxetable}{lrc}
\tablecaption{Observation Log\label{all_obs}} 
\tabletypesize{\small}
\tablewidth{0in}
\tablehead{
\colhead{}                                   & \colhead{Exposure} & {Instr.}\\
\colhead{\hbox to 0.75truein{UT Date\hfil }} & \colhead{Time (s)} & \colhead{Code \tablenotemark{a}}
}
\startdata
\multicolumn{3}{c}{3C 59} \\
\noalign{\vskip 3pt\hrule \vskip -6pt}\\
1991 Feb 04 & 1800 & 2a \\ 
1991 Feb 05 & 1500 & 3a \\
1997 Feb 07 & 3000 & 1a \\
1997 Sep 28 & 3600 & 3a \\
1998 Oct 13 & 3600 & 3a \\
1998 Oct 15 & 3600 & 1a \\
1998 Dec 20 & 2400 & 7a \\
1999 Dec 04 & 3600 & 1b \\
2000 Sep 24 & 3600 & 1b	\\
2001 Jul 21 & 2700 & 1b	\\
2002 Oct 11 & 3600 & 1b	\\
2003 Oct 21 &  600 & 8a	\\
2004 Jan 19 &  600 & 8a	\\
2004 Sep 09 &  600 & 8a	\\
\noalign{\vskip 3pt\hrule\vskip -6pt}\\
\multicolumn{3}{c}{1E~0450.3--1817} \\
\noalign{\vskip 3pt\hrule\vskip -6pt}\\
1989 Nov 06 & 3600 & 6a \\
1989 Nov 07 & 5400 & 6a \\
1991 Feb 05 & 1800 & 2a \\
1992 Jan 16 & 7200 & 4a \\
1994 Feb 16 & 6000 & 4a \\
1995 Jan 23 & 5447 & 1b \\
1996 Feb 15 & 3600 & 1a \\
1996 Feb 16 & 5400 & 1b \\
1996 Oct 10 & 5400 & 5a \\
1997 Jan 02 & 3600 & 3a \\
1997 Jan 04 & 3600 & 3a \\
1997 Sep 27 & 5700 & 1a \\
1998 Jan 02 & 7200 & 3a \\
1998 Oct 15 & 7200 & 1a \\
1998 Dec 19 & 6000 & 7a \\
1999 Feb 09 & 3600 & 7a \\
1999 Nov 01 & 6600 & 3a \\
1999 Dec 02 & 3000 & 1a \\
1999 Dec 03 & 3000 & 1a \\
1999 Dec 04 & 3600 & 1b \\
2000 Sep 23 & 3600 & 1a \\
2000 Sep 24 & 3000 & 1a \\
2001 Jan 24 & 8100 & 3c \\
2003 Jan 02 & 7200 & 3a \\
\noalign{\vskip 3pt\hrule\vskip -6pt}\\
\multicolumn{3}{c}{Pictor A} \\
\noalign{\vskip 3pt\hrule\vskip -6pt}\\
1983 Aug 11\tablenotemark{{\rm b,c}} & 4200   & 9  \\
1987 Jul 21\tablenotemark{d}   & 140000 & 4c \\
1994 Feb 18\tablenotemark{b}   & 3000   & 4b \\
1994 Dec 08\tablenotemark{{\rm b,e}} & 300    & 4  \\
1995 Sep 29\tablenotemark{{\rm b,f}} & 1800   & 4a \\
1997 Jan 02\tablenotemark{d}   & 3600   & 3a \\
1998 Jan 02 & 3600 & 3a \\
1998 Oct 20 & 5400 & 3a \\
1999 Nov 01 & 3000 & 3a \\
1999 Nov 03 & 1800 & 3c \\
2001 Jan 21 & 3600 & 3a \\
2003 Jan 04 & 3600 & 3a \\
\noalign{\vskip 3pt\hrule\vskip -6pt}\\
\multicolumn{3}{c}{CBS 74} \\
\noalign{\vskip 3pt\hrule\vskip -6pt}\\
1998 Jan 30 & 3600 & 1b \\ 
1998 Apr 09 & 1200 & 6a \\ 
1998 Oct 14 & 3600 & 1a \\ 
1999 Feb 11 & 3600 & 6a \\ 
1999 Dec 02 & 3000 & 1a \\ 
1999 Dec 04 & 3000 & 1b \\ 
2000 Mar 14 & 6000 & 6a \\ 
2000 Sep 24 & 3000 & 1b \\ 
2001 Jan 24 & 1800 & 7a \\
2002 Oct 11 & 3600 & 1b \\
2003 Mar 25 &  300 & 8a \\
2003 Apr 05 &  300 & 8a \\
2003 Oct 24 & 1800 & 7a \\
2003 Dec 24 & 1200 & 8a \\
2004 Jan 13 &  300 & 8a \\
\noalign{\vskip 3pt\hrule\vskip -6pt}\\
\multicolumn{3}{c}{PKS 0921--213} \\
\noalign{\vskip 3pt\hrule\vskip -6pt}\\
1995 Mar 24 & 3000 & 1a \\ 
1995 Mar 25 & 3600 & 1a \\ 
1996 Feb 15 & 7200 & 1a \\ 
1996 Feb 16 & 3600 & 1b \\ 
1997 Jan 02 & 3600 & 3a \\ 
1997 Jan 04 & 3000 & 3a \\ 
1997 Mar 24 & 1700 & 6a \\ 
1998 Jan 02 & 4800 & 3a \\ 
1998 Jan 28 & 3600 & 1a \\ 
1998 Apr 07 & 1800 & 7a \\
1998 Apr 08 & 1800 & 7a \\
1998 Dec 20 & 3000 & 7a \\
1999 Dec 02 & 3000 & 1a \\
1999 Dec 03 & 3000 & 1a \\
2000 Mar 16 & 6000 & 6a \\
2001 Jan 21 & 6000 & 3a \\
2001 Oct 24 & 2000 & 1a \\
2003 Jan 02 & 7200 & 3a \\
\noalign{\vskip 3pt\hrule\vskip -6pt}\\
\multicolumn{3}{c}{PKS 1020--103} \\
\noalign{\vskip 3pt\hrule\vskip -6pt}\\
1991 Feb 04 &  600 & 2a \\ 
1991 Feb 05 & 2400 & 2a \\ 
1992 Jan 16 & 1800 & 4a \\ 
1996 Feb 16 & 3600 & 1a \\ 
1997 Jan 04 & 3600 & 3a \\ 
1998 Jan 05 & 3600 & 3a \\
1998 Dec 20 & 2400 & 7a \\
1999 Dec 04 & 3600 & 1a \\
2001 Jan 22 & 7200 & 3d \\
2003 Jan 03 & 6000 & 3d \\
\noalign{\vskip 3pt\hrule\vskip -6pt}\\
\multicolumn{3}{c}{PKS 1739+18C} \\
\noalign{\vskip 3pt\hrule\vskip -6pt}\\
1992 Jul 09 & 1800 & 2a \\
1996 Jun 15 & 3600 & 1b \\ 
1997 Jun 09 & 3600 & 1b \\ 
1997 Sep 28 & 3600 & 1b \\ 
1998 Apr 09 & 3000 & 7b \\ 
1998 Jun 27 & 3000 & 7a \\ 
1998 Oct 13 & 3600 & 1b \\ 
1999 Jun 15 & 3000 & 1b \\
2000 Jun 04 & 2400 & 1b \\
2000 Sep 24 & 2400 & 1b \\
2001 Jul 21 & 3000 & 1b \\
2002 Jun 14 & 2700 & 1b \\
2002 Oct 11 & 3600 & 1b \\
2003 Apr 02 &  300 & 8a \\
2004 Jun 22 &  300 & 8a \\
2004 Aug 05 &  300 & 8a 
\enddata
\tablenotetext{a}{See Table~\ref{variability_instruments} for configurations.}
\tablenotetext{b}{Originally presented by \citet{eh98}}
\tablenotetext{c}{Provided in reduced form by A. V. Filippenko, see \citet{f85}}
\tablenotetext{d}{Sum of seven exposures over three nights, 1987 Jul 20-22, provided by S. Simkin\citep[see also,][]{s95}.}
\tablenotetext{e}{Originally presented by \citet{he94}}
\tablenotetext{f}{Provided in reduced form by R.~H. Becker.}
\end{deluxetable}

\clearpage

\begin{landscape}
\begin{deluxetable}{lcccccccccc}
\tablecaption{Measured Profile Parameters \label{param_tab}} 
\tabletypesize{\scriptsize}
\tablewidth{0in}
\tablehead{
\colhead{Date}        & \colhead{Flux} & \colhead{V(R)}  & \colhead{V(B)} & \colhead{F(R)}  & \colhead{F(B)} & \colhead{F(R)/F(B)} & \colhead{FWHM}     & \colhead{FWQM}     & \colhead{V(FWHM)}  & \colhead{V(FWQM)}\\
\colhead{(1)} & \colhead{(2)}  & \colhead{(3)} & \colhead{(4)} & \colhead{(5)} & \colhead{(6)}  & \colhead{(7)} & \colhead{(8)} & \colhead{(9)} & \colhead{(10)} & \colhead{(11)}
}
\startdata
\multicolumn{11}{c}{3C 59} \\
\noalign{\vskip 3pt\hrule \vskip -6pt}\\
1991 Feb \hfill 04 & 69$\;\pm\;$ 8 &  2.9$\;\pm\;$0.1 & --1.7$\;\pm\;$0.2 &  0.35$\;\pm\;$0.03 &  0.42$\;\pm\;$0.04 &  0.83$\;\pm\;$0.02 & 10.0$\;\pm\;$0.1 & 13.4$\;\pm\;$0.2 & \phn 1.1$\;\pm\;$0.8 & \phn 0.8$\;\pm\;$1.0 \\
1997 Feb \hfill 07 & 70$\;\pm\;$20 &  4.0$\;\pm\;$0.1 & --2.9$\;\pm\;$0.3 &  0.36$\;\pm\;$0.09 &  0.29$\;\pm\;$0.07 &  1.21$\;\pm\;$0.03 & 11.6$\;\pm\;$0.1 & 14.0$\;\pm\;$0.2 & \phn 0.6$\;\pm\;$0.8 & \phn 0.5$\;\pm\;$1.0 \\
1997 Sep  \hfill 28 & 65$\;\pm\;$ 8 &  4.1$\;\pm\;$0.1 & --3.1$\;\pm\;$0.2 &  0.37$\;\pm\;$0.04 &  0.29$\;\pm\;$0.03 &  1.26$\;\pm\;$0.02 & 11.3$\;\pm\;$0.1 & 13.2$\;\pm\;$0.2 & \phn 0.5$\;\pm\;$0.8 & \phn 0.2$\;\pm\;$1.0 \\
1998 Oct  \hfill 14 & 70$\;\pm\;$10 &  4.1$\;\pm\;$0.1 & --2.8$\;\pm\;$0.2 &  0.43$\;\pm\;$0.07 &  0.32$\;\pm\;$0.06 &  1.34$\;\pm\;$0.02 & 10.8$\;\pm\;$0.1 & 13.1$\;\pm\;$0.2 & \phn 0.6$\;\pm\;$0.8 & \phn 0.4$\;\pm\;$1.0 \\
1998 Dec  \hfill 20 & 80$\;\pm\;$10 &  4.0$\;\pm\;$0.1 & --2.3$\;\pm\;$0.2 &  0.42$\;\pm\;$0.06 &  0.36$\;\pm\;$0.05 &  1.16$\;\pm\;$0.02 & 11.1$\;\pm\;$0.1 & 12.8$\;\pm\;$0.2 & \phn 0.5$\;\pm\;$0.8 & \phn 0.2$\;\pm\;$1.0 \\
1999 Dec \hfill 04 & 90$\;\pm\;$20 &  4.0$\;\pm\;$0.1 & --2.7$\;\pm\;$0.8 &  0.48$\;\pm\;$0.08 &  0.35$\;\pm\;$0.09 &  1.4\phn$\;\pm\;$0.3\phn & 10.2$\;\pm\;$0.1 & 13.1$\;\pm\;$0.2 & \phn 0.9$\;\pm\;$0.8 & \phn 0.2$\;\pm\;$1.0 \\
2000 Sep  \hfill 24 & 67$\;\pm\;$ 9 &  4.0$\;\pm\;$0.1 & --1.9$\;\pm\;$0.1 &  0.36$\;\pm\;$0.04 &  0.37$\;\pm\;$0.04 &  0.98$\;\pm\;$0.01 & 10.3$\;\pm\;$0.1 & 13.2$\;\pm\;$0.2 & \phn 1.0$\;\pm\;$0.8 & \phn 0.3$\;\pm\;$1.0 \\
2001 Jul  \hfill 21 & 80$\;\pm\;$10 &  4.0$\;\pm\;$0.1 & --2.0$\;\pm\;$0.2 &  0.42$\;\pm\;$0.08 &  0.35$\;\pm\;$0.07 &  1.22$\;\pm\;$0.08 & 10.6$\;\pm\;$0.1 & 12.6$\;\pm\;$0.2 & \phn 1.0$\;\pm\;$0.8 & \phn 0.6$\;\pm\;$1.0 \\
2002 Oct  \hfill 11 & 90$\;\pm\;$20 &  3.9$\;\pm\;$0.1 & --2.5$\;\pm\;$0.2 &  0.5\phn$\;\pm\;$0.1\phn &  0.45$\;\pm\;$0.10 &  1.19$\;\pm\;$0.04 & 10.7$\;\pm\;$0.1 & 12.4$\;\pm\;$0.2 & \phn 0.8$\;\pm\;$0.8 & \phn 0.7$\;\pm\;$1.0 \\
2003 Oct  \hfill 21 & 64$\;\pm\;$ 9 &  4.3$\;\pm\;$0.2 & --1.7$\;\pm\;$0.3 &  0.33$\;\pm\;$0.04 &  0.37$\;\pm\;$0.05 &  0.90$\;\pm\;$0.05 & 10.5$\;\pm\;$0.1 & 12.8$\;\pm\;$0.3 & \phn 0.9$\;\pm\;$0.8 & \phn 1.0$\;\pm\;$1.0 \\
2004 Jan  \hfill 19 & 76$\;\pm\;$ 8 &  4.4$\;\pm\;$0.2 & --1.7$\;\pm\;$0.1 &  0.38$\;\pm\;$0.04 &  0.44$\;\pm\;$0.04 &  0.88$\;\pm\;$0.02 & 10.6$\;\pm\;$0.1 & 12.8$\;\pm\;$0.2 & \phn 0.9$\;\pm\;$0.8 & \phn 0.8$\;\pm\;$1.0 \\
2004 Sep \hfill 09 & 90$\;\pm\;$10 &  4.1$\;\pm\;$0.2 & --1.9$\;\pm\;$0.1 &  0.48$\;\pm\;$0.06 &  0.51$\;\pm\;$0.06 &  0.95$\;\pm\;$0.03 & 10.6$\;\pm\;$0.1 & 12.6$\;\pm\;$0.2 & \phn 0.7$\;\pm\;$0.8 & \phn 0.5$\;\pm\;$1.0 \\
\noalign{\vskip 3pt\hrule\vskip -6pt}\\
\multicolumn{11}{c}{1E~0450.3--1817} \\
\noalign{\vskip 3pt\hrule\vskip -6pt}\\
1989 Nov \hfill 06 & 15$\;\pm\;$ 1 &  3.4\phn$\;\pm\;$0.1\phn & --1.8\phn$\;\pm\;$0.3\phn & 0.05\phn $\;\pm\;$0.02\phn &0.07\phn $\;\pm\;$0.03\phn & 0.67$\;\pm\;$0.04 & 12.9$\;\pm\;$0.3 & 16.7$\;\pm\;$0.5 & --1.1$\;\pm\;$0.2 & --0.6$\;\pm\;$0.3 \\
1991 Feb \hfill 05 & 12$\;\pm\;$ 1 &  2.93$\;\pm\;$0.08 & --1.47$\;\pm\;$0.09 & 0.05\phn $\;\pm\;$0.02\phn &0.07\phn $\;\pm\;$0.02\phn & 0.62$\;\pm\;$0.04 & 11.3$\;\pm\;$0.2 & 14.1$\;\pm\;$0.2 & --1.4$\;\pm\;$0.3 & --1.2$\;\pm\;$0.2 \\
1992 Jan  \hfill 16 & 16$\;\pm\;$ 2 &  3.06$\;\pm\;$0.08 & --2.37$\;\pm\;$0.07 & 0.07\phn $\;\pm\;$0.01\phn &0.09\phn $\;\pm\;$0.02\phn & 0.76$\;\pm\;$0.04 & 11.2$\;\pm\;$0.1 & 14.1$\;\pm\;$0.2 & --1.2$\;\pm\;$0.3 & --0.8$\;\pm\;$0.2 \\
1994 Feb  \hfill 16 & 21$\;\pm\;$ 2 &  4.1\phn$\;\pm\;$0.3\phn & --4.18$\;\pm\;$0.09 & 0.07\phn $\;\pm\;$0.01\phn &0.12\phn $\;\pm\;$0.01\phn & 0.58$\;\pm\;$0.04 & 12.5$\;\pm\;$0.2 & 15.9$\;\pm\;$0.4 & --0.3$\;\pm\;$0.3 & \phn 0.5$\;\pm\;$0.3 \\
1995 Jan  \hfill 23 & 19$\;\pm\;$ 2 &  4.5\phn$\;\pm\;$1.2\phn & --4.1\phn$\;\pm\;$0.2\phn & 0.06\phn $\;\pm\;$0.02\phn &0.11\phn $\;\pm\;$0.02\phn & 0.59$\;\pm\;$0.05 & 13.1$\;\pm\;$0.1 & 15.0$\;\pm\;$0.2 & \phn 0.0$\;\pm\;$0.3 & \phn 0.0$\;\pm\;$0.2 \\
1996 Feb  \hfill 15 & 16$\;\pm\;$ 1 &  3.8\phn$\;\pm\;$0.3\phn & --5.0\phn$\;\pm\;$0.1\phn & 0.08\phn $\;\pm\;$0.03\phn &0.08\phn $\;\pm\;$0.03\phn & 0.98$\;\pm\;$0.05 & 13.6$\;\pm\;$0.2 & 15.7$\;\pm\;$0.4 & --0.7$\;\pm\;$0.2 & --0.8$\;\pm\;$0.1 \\
1996 Oct  \hfill 11 & 17$\;\pm\;$ 3 &  2.8\phn$\;\pm\;$0.1\phn & --5.67$\;\pm\;$0.06 & 0.09\phn $\;\pm\;$0.02\phn &0.08\phn $\;\pm\;$0.02\phn & 1.05$\;\pm\;$0.04 & 13.3$\;\pm\;$0.1 & 15.2$\;\pm\;$0.2 & --1.0$\;\pm\;$0.3 & --1.0$\;\pm\;$0.2 \\
1997 Jan \hfill 03 & 17$\;\pm\;$ 1 &  2.6\phn$\;\pm\;$0.1\phn & --5.54$\;\pm\;$0.08 &  0.099$\;\pm\;$0.007 &  0.096$\;\pm\;$0.005 &  1.03$\;\pm\;$0.03 & 13.1$\;\pm\;$0.2 & 14.9$\;\pm\;$0.3 & --1.2$\;\pm\;$0.2 & --1.1$\;\pm\;$0.1 \\
1997 Sep  \hfill 27 & 27$\;\pm\;$ 2 &  1.5\phn$\;\pm\;$0.4\phn & --5.8\phn$\;\pm\;$0.3\phn & 0.14\phn $\;\pm\;$0.01\phn &0.11\phn $\;\pm\;$0.01\phn & 1.23$\;\pm\;$0.05 & 13.2$\;\pm\;$0.1 & 15.5$\;\pm\;$0.2 & --1.1$\;\pm\;$0.3 & --0.7$\;\pm\;$0.2 \\
1998 Jan \hfill 02 & 26$\;\pm\;$ 2 &  2.20$\;\pm\;$0.08 & --6.0\phn$\;\pm\;$0.1\phn & 0.14\phn $\;\pm\;$0.02\phn &0.11\phn $\;\pm\;$0.02\phn & 1.30$\;\pm\;$0.05 & 13.1$\;\pm\;$0.1 & 15.1$\;\pm\;$0.2 & --1.0$\;\pm\;$0.3 & --0.9$\;\pm\;$0.2 \\
1998 Oct  \hfill 15 & 23$\;\pm\;$ 2 &  2.9\phn$\;\pm\;$0.3\phn & --5.9\phn$\;\pm\;$0.3\phn & 0.08\phn $\;\pm\;$0.02\phn &0.07\phn $\;\pm\;$0.02\phn & 1.16$\;\pm\;$0.05 & 14.5$\;\pm\;$0.1 & 17.1$\;\pm\;$0.4 & --0.7$\;\pm\;$0.3 & --0.4$\;\pm\;$0.3 \\
1998 Dec  \hfill 19 & 15$\;\pm\;$ 2 &  3.6\phn$\;\pm\;$0.4\phn & --5.36$\;\pm\;$0.07 & 0.08\phn $\;\pm\;$0.02\phn &0.06\phn $\;\pm\;$0.02\phn & 1.21$\;\pm\;$0.05 & 14.9$\;\pm\;$0.2 & 17.7$\;\pm\;$0.2 & --0.6$\;\pm\;$0.3 & --1.0$\;\pm\;$0.2 \\
1999 Feb \hfill 09 & 22$\;\pm\;$ 4 &  3.5\phn$\;\pm\;$0.1\phn & --5.16$\;\pm\;$0.09 & 0.08\phn $\;\pm\;$0.02\phn &0.07\phn $\;\pm\;$0.02\phn & 1.15$\;\pm\;$0.04 & 14.8$\;\pm\;$0.2 & 20.2$\;\pm\;$0.4 & --0.5$\;\pm\;$0.3 & \phn 0.2$\;\pm\;$0.3 \\
1999 Nov \hfill 01 & 24$\;\pm\;$ 5 &  2.9\phn$\;\pm\;$0.2\phn & --4.55$\;\pm\;$0.07 & 0.11\phn $\;\pm\;$0.03\phn &0.11\phn $\;\pm\;$0.03\phn & 0.98$\;\pm\;$0.04 & 15.3$\;\pm\;$0.1 & 20.0$\;\pm\;$0.2 & --0.4$\;\pm\;$0.3 & --0.2$\;\pm\;$0.2 \\
1999 Dec \hfill 03 & 23$\;\pm\;$ 6 &  2.8\phn$\;\pm\;$0.5\phn & --4.4\phn$\;\pm\;$0.3\phn & 0.11\phn $\;\pm\;$0.04\phn &0.12\phn $\;\pm\;$0.04\phn & 0.93$\;\pm\;$0.04 & 15.5$\;\pm\;$0.5 & 19.4$\;\pm\;$1.0 & --0.2$\;\pm\;$0.2 & \phn 0.2$\;\pm\;$0.3 \\
2000 Sep  \hfill 24 & 26$\;\pm\;$ 5 &  2.6\phn$\;\pm\;$0.3\phn & --3.1\phn$\;\pm\;$0.8\phn & 0.10\phn $\;\pm\;$0.02\phn &0.13\phn $\;\pm\;$0.03\phn & 0.77$\;\pm\;$0.04 & 12.6$\;\pm\;$0.2 & 16.6$\;\pm\;$0.2 & --0.8$\;\pm\;$0.2 & --0.5$\;\pm\;$0.3 \\
2001 Jan  \hfill 24 & 24$\;\pm\;$ 5 &  2.8\phn$\;\pm\;$0.2\phn & --4.4\phn$\;\pm\;$0.9\phn & 0.08\phn $\;\pm\;$0.02\phn &0.11\phn $\;\pm\;$0.02\phn & 0.73$\;\pm\;$0.04 & 12.6$\;\pm\;$0.1 & 16.2$\;\pm\;$0.2 & --1.1$\;\pm\;$0.3 & --0.6$\;\pm\;$0.3 \\
2003 Jan \hfill 02 & 26$\;\pm\;$ 6 &  2.5\phn$\;\pm\;$0.1\phn & --3.6\phn$\;\pm\;$0.2\phn & 0.11\phn $\;\pm\;$0.03\phn &0.10\phn $\;\pm\;$0.03\phn & 1.06$\;\pm\;$0.04 & 14.2$\;\pm\;$0.1 & 16.5$\;\pm\;$0.3 & \phn 0.0$\;\pm\;$0.3 & --0.3$\;\pm\;$0.2 \\
\noalign{\vskip 3pt\hrule\vskip -6pt}\\
\multicolumn{11}{c}{Pictor A} \\
\noalign{\vskip 3pt\hrule\vskip -6pt}\\
1987 Jul  \hfill 21 & 11$\;\pm\;$ 2 &  5.1$\;\pm\;$0.2 & --4.14$\;\pm\;$0.08 &  0.041$\;\pm\;$0.008 &  0.017$\;\pm\;$0.004 &  2.44$\;\pm\;$0.06 & 10.88$\;\pm\;$0.09 & 14.06$\;\pm\;$0.09 & \phn 2.5$\;\pm\;$0.1 & \phn 2.0$\;\pm\;$0.2 \\
1994 Feb  \hfill 18 & 15$\;\pm\;$ 2 &  5.5$\;\pm\;$0.1 & --7.36$\;\pm\;$0.07 &  0.048$\;\pm\;$0.007 &  0.043$\;\pm\;$0.007 &  1.11$\;\pm\;$0.04 & 18.1\phn$\;\pm\;$0.4\phn & 19.84$\;\pm\;$0.09 & \phn 0.0$\;\pm\;$0.2 & --0.1$\;\pm\;$0.2 \\
1994 Dec \hfill 08 & 14$\;\pm\;$ 2 &  5.9$\;\pm\;$0.2 & --6.8\phn$\;\pm\;$0.2\phn &  0.056$\;\pm\;$0.006 &  0.037$\;\pm\;$0.004 &  1.53$\;\pm\;$0.04 & 16.2\phn$\;\pm\;$0.2\phn & 19.0\phn$\;\pm\;$0.2\phn & \phn 0.0$\;\pm\;$0.2 & \phn 0.0$\;\pm\;$0.2 \\
1995 Sep  \hfill 29 & 14$\;\pm\;$ 2 &  5.1$\;\pm\;$0.2 & --5.76$\;\pm\;$0.06 &  0.049$\;\pm\;$0.005 &  0.029$\;\pm\;$0.003 &  1.72$\;\pm\;$0.04 & 15.5\phn$\;\pm\;$0.1\phn & 19.10$\;\pm\;$0.08 & \phn 0.1$\;\pm\;$0.1 & \phn 0.2$\;\pm\;$0.2 \\
1997 Jan \hfill 02 & 14$\;\pm\;$ 2 &  3.7$\;\pm\;$0.3 & --5.09$\;\pm\;$0.05 &  0.039$\;\pm\;$0.005 &  0.038$\;\pm\;$0.005 &  1.04$\;\pm\;$0.04 & 15.2\phn$\;\pm\;$0.2\phn & 19.3\phn$\;\pm\;$0.2\phn & \phn 0.0$\;\pm\;$0.1 & \phn 0.9$\;\pm\;$0.2 \\
1998 Jan \hfill 02 & 14$\;\pm\;$ 1 &  4.8$\;\pm\;$0.6 & --4.10$\;\pm\;$0.06 &  0.039$\;\pm\;$0.004 &  0.042$\;\pm\;$0.004 &  0.92$\;\pm\;$0.05 & 16.2\phn$\;\pm\;$0.2\phn & 18.47$\;\pm\;$0.08 & \phn 0.7$\;\pm\;$0.2 & \phn 1.1$\;\pm\;$0.2 \\
1998 Oct  \hfill 20 & 10$\;\pm\;$ 1 &  5.5$\;\pm\;$0.1 & --5.05$\;\pm\;$0.06 &  0.027$\;\pm\;$0.002 &  0.025$\;\pm\;$0.001 &  1.10$\;\pm\;$0.04 & 18.15$\;\pm\;$0.08 & 20.47$\;\pm\;$0.04 & \phn 1.3$\;\pm\;$0.1 & \phn 1.1$\;\pm\;$0.2 \\
1999 Nov \hfill 03 & \phn 8$\;\pm\;$ 1 &  5.2$\;\pm\;$0.2 & --2.87$\;\pm\;$0.09 &  0.019$\;\pm\;$0.002 &  0.022$\;\pm\;$0.002 &  0.86$\;\pm\;$0.04 & 15.9\phn$\;\pm\;$0.2\phn & 18.6\phn$\;\pm\;$0.2\phn & \phn 1.7$\;\pm\;$0.2 & \phn 1.6$\;\pm\;$0.2 \\
2001 Jan  \hfill 21 & \phn 6$\;\pm\;$ 1 &  4.9$\;\pm\;$0.3 & --4.22$\;\pm\;$0.05 &  0.015$\;\pm\;$0.003 &  0.015$\;\pm\;$0.003 &  1.01$\;\pm\;$0.04 & 15.8\phn$\;\pm\;$0.5\phn & 17.6\phn$\;\pm\;$0.5\phn & \phn 1.2$\;\pm\;$0.3 & \phn 1.3$\;\pm\;$0.3 \\
2003 Jan \hfill 02 & \phn 6$\;\pm\;$ 1 &  6.8$\;\pm\;$0.2 & --5\phm{.}\phn\phn$\;\pm\;$1\phm{.}\phn\phn &  0.014$\;\pm\;$0.002 &  0.012$\;\pm\;$0.002 &  1.10$\;\pm\;$0.08 & 19\phm{.}\phn\phn$\;\pm\;$1\phm{.}\phn\phn & 22.5\phn$\;\pm\;$0.5\phn & \phn 1.3$\;\pm\;$0.4 & \phn 1.4$\;\pm\;$0.3 \\
\noalign{\vskip 3pt\hrule\vskip -6pt}\\
\multicolumn{11}{c}{CBS 74} \\
\noalign{\vskip 3pt\hrule\vskip -6pt}\\
1998 Jan  \hfill 30 & \phn80$\;\pm\;$20 &  4.0$\;\pm\;$0.1 & --1.81$\;\pm\;$0.05 &  0.33$\;\pm\;$0.06 &  0.54$\;\pm\;$0.09 &  0.61$\;\pm\;$0.02 &  9.43$\;\pm\;$0.09 & 12.4$\;\pm\;$0.2 & \phn 1.1$\;\pm\;$0.2 & \phn 1.3$\;\pm\;$0.4 \\
1998 Apr \hfill 09 & \phn90$\;\pm\;$20 &  3.6$\;\pm\;$0.4 & --1.68$\;\pm\;$0.06 &  0.34$\;\pm\;$0.05 &  0.51$\;\pm\;$0.08 &  0.67$\;\pm\;$0.02 &  9.9\phn$\;\pm\;$0.2\phn & 13.3$\;\pm\;$0.2 & \phn 1.2$\;\pm\;$0.2 & \phn 1.2$\;\pm\;$0.4 \\
1998 Oct  \hfill 14 & \phn90$\;\pm\;$10 &  3.7$\;\pm\;$0.2 & --1.6\phn$\;\pm\;$0.2\phn &  0.35$\;\pm\;$0.01 &  0.54$\;\pm\;$0.02 &  0.66$\;\pm\;$0.03 &  9.7\phn$\;\pm\;$0.2\phn & 13.3$\;\pm\;$0.2 & \phn 1.2$\;\pm\;$0.2 & \phn 0.9$\;\pm\;$0.4 \\
1999 Feb  \hfill 11 & \phn72$\;\pm\;$ 9 &  4.3$\;\pm\;$0.4 & --1.21$\;\pm\;$0.07 &  0.30$\;\pm\;$0.01 &  0.48$\;\pm\;$0.02 &  0.63$\;\pm\;$0.02 &  9.82$\;\pm\;$0.07 & 13.3$\;\pm\;$0.2 & \phn 1.1$\;\pm\;$0.2 & \phn 0.9$\;\pm\;$0.5 \\
1999 Dec \hfill 03 & \phn90$\;\pm\;$20 &  3.8$\;\pm\;$0.1 & --1.25$\;\pm\;$0.09 &  0.5\phn$\;\pm\;$0.1\phn &  0.7\phn$\;\pm\;$0.2\phn &  0.71$\;\pm\;$0.03 &  9.1\phn$\;\pm\;$0.2\phn & 13.1$\;\pm\;$0.3 & \phn 1.0$\;\pm\;$0.1 & \phn 0.3$\;\pm\;$0.4 \\
2000 Mar  \hfill 16 & \phn80$\;\pm\;$10 &  3.7$\;\pm\;$0.1 & --1.68$\;\pm\;$0.07 &  0.37$\;\pm\;$0.02 &  0.57$\;\pm\;$0.04 &  0.65$\;\pm\;$0.02 &  8.7\phn$\;\pm\;$0.1\phn & 12.8$\;\pm\;$0.2 & \phn 0.9$\;\pm\;$0.2 & \phn 0.2$\;\pm\;$0.4 \\
2000 Sep  \hfill 24 & \phn80$\;\pm\;$10 &  3.4$\;\pm\;$0.1 & --1.38$\;\pm\;$0.10 &  0.35$\;\pm\;$0.03 &  0.54$\;\pm\;$0.04 &  0.65$\;\pm\;$0.02 &  9.0\phn$\;\pm\;$0.2\phn & 14.0$\;\pm\;$0.2 & \phn 0.6$\;\pm\;$0.2 & --0.5$\;\pm\;$0.4 \\
2001 Jan  \hfill 24 & \phn80$\;\pm\;$10 &  3.4$\;\pm\;$0.2 & --1.35$\;\pm\;$0.08 &  0.36$\;\pm\;$0.05 &  0.53$\;\pm\;$0.08 &  0.68$\;\pm\;$0.02 &  9.2\phn$\;\pm\;$0.1\phn & 13.2$\;\pm\;$0.2 & \phn 0.5$\;\pm\;$0.2 & \phn 0.2$\;\pm\;$0.4 \\
2002 Oct  \hfill 11 & \phn90$\;\pm\;$10 &  2.9$\;\pm\;$0.1 & --1.41$\;\pm\;$0.09 &  0.43$\;\pm\;$0.03 &  0.61$\;\pm\;$0.04 &  0.69$\;\pm\;$0.02 &  8.7\phn$\;\pm\;$0.1\phn & 15.2$\;\pm\;$0.2 & \phn 0.3$\;\pm\;$0.2 & --1.0$\;\pm\;$0.4 \\
2003 Mar  \hfill 25 & 100$\;\pm\;$10 &  2.5$\;\pm\;$0.2 & --1.7\phn$\;\pm\;$0.4\phn &  0.43$\;\pm\;$0.04 &  0.60$\;\pm\;$0.07 &  0.71$\;\pm\;$0.05 &  9.2\phn$\;\pm\;$0.3\phn & 14.4$\;\pm\;$0.5 & \phn 0.1$\;\pm\;$0.2 & --0.1$\;\pm\;$0.5 \\
2003 Apr \hfill 05 & 110$\;\pm\;$20 &  2.3$\;\pm\;$0.1 & --1.5\phn$\;\pm\;$0.3\phn &  0.47$\;\pm\;$0.06 &  0.66$\;\pm\;$0.08 &  0.72$\;\pm\;$0.03 &  9.2\phn$\;\pm\;$0.3\phn & 14.8$\;\pm\;$0.2 & \phn 0.3$\;\pm\;$0.2 & --0.3$\;\pm\;$0.5 \\
2003 Oct  \hfill 24 & 120$\;\pm\;$20 &  2.7$\;\pm\;$0.2 & --1.38$\;\pm\;$0.09 &  0.52$\;\pm\;$0.04 &  0.76$\;\pm\;$0.06 &  0.68$\;\pm\;$0.02 &  8.7\phn$\;\pm\;$0.2\phn & 13.7$\;\pm\;$0.2 & \phn 0.5$\;\pm\;$0.3 & \phn 0.1$\;\pm\;$0.5 \\
2003 Dec  \hfill 24 & \phn90$\;\pm\;$20 &  2.8$\;\pm\;$0.2 & --1.4\phn$\;\pm\;$0.1\phn &  0.40$\;\pm\;$0.07 &  0.6\phn$\;\pm\;$0.1\phn &  0.73$\;\pm\;$0.02 &  9.2\phn$\;\pm\;$0.1\phn & 14.3$\;\pm\;$0.2 & \phn 0.4$\;\pm\;$0.2 & \phn 0.1$\;\pm\;$0.4 \\
2004 Jan  \hfill 13 & \phn69$\;\pm\;$ 9 &  2.0$\;\pm\;$0.1 & --1.6\phn$\;\pm\;$0.1\phn &  0.32$\;\pm\;$0.02 &  0.44$\;\pm\;$0.02 &  0.73$\;\pm\;$0.02 &  9.2\phn$\;\pm\;$0.1\phn & 14.4$\;\pm\;$0.2 & \phn 0.2$\;\pm\;$0.2 & \phn 0.0$\;\pm\;$0.5 \\
\noalign{\vskip 3pt\hrule\vskip -6pt}\\
\multicolumn{11}{c}{PKS 0921--213} \\
\noalign{\vskip 3pt\hrule\vskip -6pt}\\
1995 Mar  \hfill 24 & 93$\;\pm\;$ 6 &  2.52$\;\pm\;$0.09 & --2.69$\;\pm\;$0.05 &  0.69$\;\pm\;$0.03 &  0.68$\;\pm\;$0.03 &  1.03$\;\pm\;$0.02 &  8.26$\;\pm\;$0.05 & 10.95$\;\pm\;$0.06 & --0.1$\;\pm\;$0.1 & --0.6$\;\pm\;$0.1 \\
1996 Feb  \hfill 15 & 76$\;\pm\;$ 8 &  2.45$\;\pm\;$0.08 & --2.56$\;\pm\;$0.05 &  0.52$\;\pm\;$0.05 &  0.45$\;\pm\;$0.04 &  1.17$\;\pm\;$0.01 &  9.64$\;\pm\;$0.05 & 12.18$\;\pm\;$0.08 & --0.1$\;\pm\;$0.1 & --0.4$\;\pm\;$0.1 \\
1997 Jan \hfill 02 & 65$\;\pm\;$ 7 &  2.2\phn$\;\pm\;$0.1\phn & --2.7\phn$\;\pm\;$0.2\phn &  0.46$\;\pm\;$0.04 &  0.37$\;\pm\;$0.03 &  1.26$\;\pm\;$0.04 &  9.15$\;\pm\;$0.08 & 11.50$\;\pm\;$0.07 & --0.2$\;\pm\;$0.1 & --0.4$\;\pm\;$0.1 \\
1997 Mar  \hfill 24 & 56$\;\pm\;$ 5 &  2.0\phn$\;\pm\;$0.2\phn & --2.55$\;\pm\;$0.08 &  0.46$\;\pm\;$0.03 &  0.35$\;\pm\;$0.02 &  1.31$\;\pm\;$0.01 &  8.21$\;\pm\;$0.07 & 10.29$\;\pm\;$0.06 & --0.1$\;\pm\;$0.1 & --0.2$\;\pm\;$0.1 \\
1998 Jan \hfill 02 & 61$\;\pm\;$ 5 &  1.87$\;\pm\;$0.09 & --2.27$\;\pm\;$0.04 &  0.43$\;\pm\;$0.03 &  0.41$\;\pm\;$0.03 &  1.06$\;\pm\;$0.01 &  9.23$\;\pm\;$0.04 & 11.49$\;\pm\;$0.04 & \phn 0.0$\;\pm\;$0.1 & \phn 0.2$\;\pm\;$0.1 \\
1998 Jan  \hfill 28 & 76$\;\pm\;$ 6 &  2.12$\;\pm\;$0.08 & --2.12$\;\pm\;$0.07 &  0.50$\;\pm\;$0.03 &  0.46$\;\pm\;$0.03 &  1.09$\;\pm\;$0.02 &  9.18$\;\pm\;$0.05 & 11.67$\;\pm\;$0.04 & \phn 0.1$\;\pm\;$0.1 & \phn 0.2$\;\pm\;$0.1 \\
1998 Apr \hfill 07 & 65$\;\pm\;$ 6 &  2.4\phn$\;\pm\;$0.1\phn & --2.58$\;\pm\;$0.10 &  0.41$\;\pm\;$0.03 &  0.37$\;\pm\;$0.03 &  1.11$\;\pm\;$0.02 &  9.77$\;\pm\;$0.06 & 12.20$\;\pm\;$0.08 & \phn 0.0$\;\pm\;$0.1 & \phn 0.0$\;\pm\;$0.1 \\
1998 Dec  \hfill 20 & 56$\;\pm\;$ 4 &  2.67$\;\pm\;$0.10 & --2.39$\;\pm\;$0.05 &  0.35$\;\pm\;$0.02 &  0.35$\;\pm\;$0.02 &  1.00$\;\pm\;$0.02 &  9.63$\;\pm\;$0.04 & 11.90$\;\pm\;$0.04 & --0.1$\;\pm\;$0.1 & --0.4$\;\pm\;$0.1 \\
1999 Dec \hfill 02 & 47$\;\pm\;$ 3 &  2.8\phn$\;\pm\;$0.2\phn & --2.64$\;\pm\;$0.05 &  0.31$\;\pm\;$0.01 &  0.31$\;\pm\;$0.01 &  1.00$\;\pm\;$0.01 &  9.53$\;\pm\;$0.07 & 11.49$\;\pm\;$0.07 & \phn 0.2$\;\pm\;$0.1 & --0.1$\;\pm\;$0.1 \\
2000 Mar  \hfill 16 & 53$\;\pm\;$ 4 &  2.6\phn$\;\pm\;$0.2\phn & --2.1\phn$\;\pm\;$0.1\phn &  0.32$\;\pm\;$0.02 &  0.29$\;\pm\;$0.01 &  1.10$\;\pm\;$0.03 &  9.84$\;\pm\;$0.05 & 12.45$\;\pm\;$0.05 & \phn 0.0$\;\pm\;$0.1 & --0.4$\;\pm\;$0.1 \\
2001 Jan  \hfill 21 & 44$\;\pm\;$ 3 &  2.9\phn$\;\pm\;$0.2\phn & --2.65$\;\pm\;$0.07 &  0.30$\;\pm\;$0.02 &  0.27$\;\pm\;$0.01 &  1.08$\;\pm\;$0.02 &  9.64$\;\pm\;$0.05 & 11.81$\;\pm\;$0.04 & \phn 0.0$\;\pm\;$0.1 & --0.3$\;\pm\;$0.1 \\
2001 Oct  \hfill 24 & 51$\;\pm\;$ 6 &  2.88$\;\pm\;$0.08 & --2.5\phn$\;\pm\;$0.1\phn &  0.39$\;\pm\;$0.04 &  0.35$\;\pm\;$0.04 &  1.11$\;\pm\;$0.02 &  9.01$\;\pm\;$0.04 & 10.90$\;\pm\;$0.04 & --0.1$\;\pm\;$0.1 & --0.3$\;\pm\;$0.1 \\
2003 Jan \hfill 02 & 47$\;\pm\;$ 4 &  2.78$\;\pm\;$0.08 & --2.60$\;\pm\;$0.08 &  0.36$\;\pm\;$0.02 &  0.28$\;\pm\;$0.02 &  1.30$\;\pm\;$0.02 &  8.86$\;\pm\;$0.08 & 10.52$\;\pm\;$0.04 & \phn 0.2$\;\pm\;$0.1 & \phn 0.1$\;\pm\;$0.1 \\
\noalign{\vskip 3pt\hrule\vskip -6pt}\\
\multicolumn{11}{c}{PKS 1020--103} \\
\noalign{\vskip 3pt\hrule\vskip -6pt}\\
1991 Feb \hfill 05 & 230$\;\pm\;$30 &  2.0\phn$\;\pm\;$0.1\phn & --2.2\phn$\;\pm\;$0.1\phn &  1.8$\;\pm\;$0.3 &  1.6$\;\pm\;$0.2 &  1.13$\;\pm\;$0.01 &  8.6\phn$\;\pm\;$0.1\phn & 10.8$\;\pm\;$0.2 & \phn 0.3$\;\pm\;$0.1 & \phn 0.6$\;\pm\;$0.6 \\
1992 Jan  \hfill 16 & 320$\;\pm\;$50 &  1.57$\;\pm\;$0.07 & --2.94$\;\pm\;$0.06 &  2.3$\;\pm\;$0.3 &  2.1$\;\pm\;$0.3 &  1.12$\;\pm\;$0.01 &  8.52$\;\pm\;$0.04 & 11.0$\;\pm\;$0.1 & \phn 0.3$\;\pm\;$0.1 & \phn 0.7$\;\pm\;$0.6 \\
1996 Feb  \hfill 16 & 260$\;\pm\;$30 &  1.83$\;\pm\;$0.08 & --2.71$\;\pm\;$0.05 &  1.8$\;\pm\;$0.2 &  1.7$\;\pm\;$0.2 &  1.09$\;\pm\;$0.01 &  8.81$\;\pm\;$0.04 & 11.3$\;\pm\;$0.1 & \phn 0.4$\;\pm\;$0.1 & \phn 0.8$\;\pm\;$0.6 \\
1997 Jan \hfill 04 & 200$\;\pm\;$30 &  1.85$\;\pm\;$0.07 & --2.74$\;\pm\;$0.05 &  1.4$\;\pm\;$0.2 &  1.3$\;\pm\;$0.2 &  1.02$\;\pm\;$0.02 &  9.31$\;\pm\;$0.04 & 11.0$\;\pm\;$0.1 & \phn 0.6$\;\pm\;$0.1 & \phn 0.6$\;\pm\;$0.6 \\
1998 Jan \hfill 05 & 190$\;\pm\;$30 &  1.90$\;\pm\;$0.07 & --2.2\phn$\;\pm\;$0.1\phn &  1.3$\;\pm\;$0.2 &  1.2$\;\pm\;$0.1 &  1.05$\;\pm\;$0.02 &  9.1\phn$\;\pm\;$0.2\phn & 11.7$\;\pm\;$0.1 & \phn 0.3$\;\pm\;$0.2 & \phn 0.5$\;\pm\;$0.6 \\
1998 Dec  \hfill 20 & 250$\;\pm\;$30 &  2.0\phn$\;\pm\;$0.1\phn & --2.84$\;\pm\;$0.06 &  1.7$\;\pm\;$0.2 &  1.6$\;\pm\;$0.2 &  1.05$\;\pm\;$0.01 &  9.22$\;\pm\;$0.04 & 11.3$\;\pm\;$0.1 & \phn 0.5$\;\pm\;$0.1 & \phn 0.7$\;\pm\;$0.6 \\
1999 Dec \hfill 04 & 190$\;\pm\;$40 &  2.0\phn$\;\pm\;$0.1\phn & --2.50$\;\pm\;$0.05 &  1.3$\;\pm\;$0.3 &  1.4$\;\pm\;$0.3 &  0.96$\;\pm\;$0.02 &  8.51$\;\pm\;$0.04 & 11.2$\;\pm\;$0.1 & \phn 0.4$\;\pm\;$0.1 & \phn 0.9$\;\pm\;$0.6 \\
2001 Jan  \hfill 22 & 210$\;\pm\;$30 &  2.17$\;\pm\;$0.10 & --2.62$\;\pm\;$0.07 &  1.4$\;\pm\;$0.2 &  1.4$\;\pm\;$0.2 &  0.96$\;\pm\;$0.02 &  9.03$\;\pm\;$0.04 & 11.4$\;\pm\;$0.1 & \phn 0.5$\;\pm\;$0.1 & \phn 0.8$\;\pm\;$0.6 \\
2003 Jan \hfill 03 & 180$\;\pm\;$20 &  2.2\phn$\;\pm\;$0.3\phn & --2.64$\;\pm\;$0.07 &  1.1$\;\pm\;$0.1 &  1.2$\;\pm\;$0.1 &  0.95$\;\pm\;$0.03 &  9.37$\;\pm\;$0.05 & 11.8$\;\pm\;$0.1 & \phn 0.2$\;\pm\;$0.1 & \phn 0.7$\;\pm\;$0.6 \\
\noalign{\vskip 3pt\hrule\vskip -6pt}\\
\multicolumn{11}{c}{PKS 1739+18C} \\
\noalign{\vskip 3pt\hrule\vskip -6pt}\\
1992 Jul \hfill 09 & 240$\;\pm\;$\phn20 &  3.45$\;\pm\;$0.08 & --2.17$\;\pm\;$0.09 &  0.91$\;\pm\;$0.05 &  1.14$\;\pm\;$0.07 &  0.80$\;\pm\;$0.01 & 12.81$\;\pm\;$0.06 & 15.84$\;\pm\;$0.07 & --0.1$\;\pm\;$0.1 & --0.4$\;\pm\;$0.1 \\
1996 Jun  \hfill 15 & 450$\;\pm\;$\phn40 &  3.22$\;\pm\;$0.07 & --3.6\phn$\;\pm\;$0.1\phn &  2.0\phn$\;\pm\;$0.1\phn &  2.0\phn$\;\pm\;$0.1\phn &  1.02$\;\pm\;$0.01 & 13.26$\;\pm\;$0.07 & 15.95$\;\pm\;$0.05 & \phn 0.1$\;\pm\;$0.1 & --0.1$\;\pm\;$0.1 \\
1997 Jun \hfill 09 & 590$\;\pm\;$\phn50 &  3.50$\;\pm\;$0.06 & --2.7\phn$\;\pm\;$0.2\phn &  2.7\phn$\;\pm\;$0.2\phn &  2.6\phn$\;\pm\;$0.2\phn &  1.03$\;\pm\;$0.02 & 13.5\phn$\;\pm\;$0.1\phn & 15.82$\;\pm\;$0.07 & \phn 0.2$\;\pm\;$0.1 & --0.2$\;\pm\;$0.1 \\
1997 Sep  \hfill 28 & 580$\;\pm\;$\phn50 &  3.74$\;\pm\;$0.08 & --2.4\phn$\;\pm\;$0.7\phn &  2.7\phn$\;\pm\;$0.2\phn &  2.5\phn$\;\pm\;$0.2\phn &  1.07$\;\pm\;$0.04 & 12.93$\;\pm\;$0.06 & 15.63$\;\pm\;$0.08 & \phn 0.2$\;\pm\;$0.1 & --0.3$\;\pm\;$0.1 \\
1998 Apr \hfill 09 & 450$\;\pm\;$\phn50 &  3.66$\;\pm\;$0.06 & --2.1\phn$\;\pm\;$0.7\phn &  2.2\phn$\;\pm\;$0.2\phn &  2.1\phn$\;\pm\;$0.2\phn &  1.06$\;\pm\;$0.06 & 13.18$\;\pm\;$0.08 & 15.68$\;\pm\;$0.07 & \phn 0.3$\;\pm\;$0.1 & --0.2$\;\pm\;$0.1 \\
1998 Jun  \hfill 27 & 440$\;\pm\;$\phn70 &  3.65$\;\pm\;$0.07 & --2.4\phn$\;\pm\;$0.6\phn &  1.9\phn$\;\pm\;$0.3\phn &  2.0\phn$\;\pm\;$0.3\phn &  0.99$\;\pm\;$0.05 & 13.3\phn$\;\pm\;$0.2\phn & 16.3\phn$\;\pm\;$0.1\phn & \phn 0.1$\;\pm\;$0.1 & --0.3$\;\pm\;$0.1 \\
1998 Oct  \hfill 13 & 410$\;\pm\;$\phn80 &  3.51$\;\pm\;$0.07 & --2.4\phn$\;\pm\;$0.5\phn &  1.9\phn$\;\pm\;$0.3\phn &  1.9\phn$\;\pm\;$0.4\phn &  1.00$\;\pm\;$0.04 & 12.9\phn$\;\pm\;$0.3\phn & 15.7\phn$\;\pm\;$0.1\phn & \phn 0.3$\;\pm\;$0.2 & --0.4$\;\pm\;$0.1 \\
1999 Jun  \hfill 15 & 360$\;\pm\;$\phn90 &  3.69$\;\pm\;$0.09 & --1.4\phn$\;\pm\;$0.2\phn &  1.6\phn$\;\pm\;$0.4\phn &  1.7\phn$\;\pm\;$0.4\phn &  0.96$\;\pm\;$0.01 & 13.09$\;\pm\;$0.07 & 16.34$\;\pm\;$0.05 & \phn 0.3$\;\pm\;$0.2 & \phn 0.0$\;\pm\;$0.1 \\
2000 Jun \hfill 04 & 420$\;\pm\;$\phn90 &  3.5\phn$\;\pm\;$0.1\phn & --3\phm{.}\phn\phn$\;\pm\;$1\phm{.}\phn\phn &  2.1\phn$\;\pm\;$0.4\phn &  1.8\phn$\;\pm\;$0.4\phn &  1.13$\;\pm\;$0.08 & 12.0\phn$\;\pm\;$0.2\phn & 15.40$\;\pm\;$0.09 & \phn 0.2$\;\pm\;$0.1 & --0.2$\;\pm\;$0.2 \\
2000 Sep  \hfill 24 & 600$\;\pm\;$100 &  3.22$\;\pm\;$0.07 & --1.7\phn$\;\pm\;$0.1\phn &  2.9\phn$\;\pm\;$0.5\phn &  2.8\phn$\;\pm\;$0.5\phn &  1.04$\;\pm\;$0.02 & 11.9\phn$\;\pm\;$0.2\phn & 15.7\phn$\;\pm\;$0.2\phn & \phn 0.4$\;\pm\;$0.1 & \phn 0.0$\;\pm\;$0.1 \\
2001 Jul  \hfill 21 & 520$\;\pm\;$100 &  2.91$\;\pm\;$0.06 & --2.5\phn$\;\pm\;$0.6\phn &  2.7\phn$\;\pm\;$0.7\phn &  2.3\phn$\;\pm\;$0.6\phn &  1.19$\;\pm\;$0.06 & 11.6\phn$\;\pm\;$0.1\phn & 14.6\phn$\;\pm\;$0.3\phn & \phn 0.2$\;\pm\;$0.1 & \phn 0.0$\;\pm\;$0.2 \\
2002 Jun  \hfill 14 & 430$\;\pm\;$\phn50 &  2.92$\;\pm\;$0.08 & --3\phm{.}\phn\phn$\;\pm\;$1\phm{.}\phn\phn &  2.4\phn$\;\pm\;$0.2\phn &  1.9\phn$\;\pm\;$0.2\phn &  1.3\phn$\;\pm\;$0.1\phn & 11.5\phn$\;\pm\;$0.1\phn & 14.38$\;\pm\;$0.08 & \phn 0.3$\;\pm\;$0.2 & \phn 0.1$\;\pm\;$0.2 \\
2002 Oct  \hfill 11 & 380$\;\pm\;$100 &  2.63$\;\pm\;$0.06 & --2.6\phn$\;\pm\;$0.7\phn &  2.2\phn$\;\pm\;$0.6\phn &  1.8\phn$\;\pm\;$0.5\phn &  1.23$\;\pm\;$0.07 & 11.5\phn$\;\pm\;$0.1\phn & 14.38$\;\pm\;$0.06 & \phn 0.4$\;\pm\;$0.1 & \phn 0.1$\;\pm\;$0.1 \\
2003 Apr \hfill 02 & 370$\;\pm\;$\phn40 &  2.7\phn$\;\pm\;$0.4\phn & --3\phm{.}\phn\phn$\;\pm\;$1\phm{.}\phn\phn &  1.9\phn$\;\pm\;$0.2\phn &  1.7\phn$\;\pm\;$0.2\phn &  1.12$\;\pm\;$0.07 & 12.19$\;\pm\;$0.07 & 15.1\phn$\;\pm\;$0.1\phn & \phn 0.2$\;\pm\;$0.2 & --0.1$\;\pm\;$0.2 \\
2004 Jun  \hfill 22 & 250$\;\pm\;$\phn50 &  3.3\phn$\;\pm\;$0.4\phn & --2.8\phn$\;\pm\;$0.7\phn &  1.1\phn$\;\pm\;$0.2\phn &  1.2\phn$\;\pm\;$0.2\phn &  0.96$\;\pm\;$0.02 & 13.0\phn$\;\pm\;$0.3\phn & 15.7\phn$\;\pm\;$0.2\phn & \phn 0.2$\;\pm\;$0.1 & --0.3$\;\pm\;$0.1 \\2004 Aug \hfill 05 & 320$\;\pm\;$\phn50 &  2.8\phn$\;\pm\;$0.4\phn & --2.5\phn$\;\pm\;$0.3\phn &  1.5\phn$\;\pm\;$0.2\phn &  1.4\phn$\;\pm\;$0.2\phn &  1.01$\;\pm\;$0.03 & 12.8\phn$\;\pm\;$0.1\phn & 15.9\phn$\;\pm\;$0.2\phn & \phn 0.2$\;\pm\;$0.1 & \phn 0.0$\;\pm\;$0.1 \\
\enddata
\tablecomments{Col. (1): UT Date of Observation. When multiple spectra from one observing run are averaged, the average date
is listed. Col. (2): Total broad line flux divided by the [O{\,\sc i}]$\lambda6300$ line flux. Col. (3): Velocity of the red 
peak in units $10^{3}\;{\rm km\;s^{-1}}$ with respect to the narrow H$\alpha$ line. Col. (4): Velocity of the blue peak in 
units of $10^{3}\;{\rm km\;s^{-1}}$ with respect to the narrow H$\alpha$ line. Col. (5): Flux density ($f_{\nu}$ in mJy) at 
the red peak, relative to the [O{\,\sc i}] flux.  Col. (6): Flux density ($f_{\nu}$ in mJy) at the blue peak, 
relative to the [O{\,\sc i}] flux. Col. (7) Ratio of the flux density of the red and blue peaks. Note that the 
fractional error on the ratio is much smaller than the fractional errors on the peak fluxes, because the error in 
the [O{\,\sc i}] flux does not contribute. Col. (8): Full Width and Half-Maximum (defined by higher of the peaks) in 
units of $10^{3}\;{\rm km\;s^{-1}}$. Col. (9): Full Width and Quarter-Maximum (defined by higher of the peaks) in units 
of $10^{3}\;{\rm km\;s^{-1}}$. Col. (10): Velocity shift of the profile at the Half-Maximum in units 
$10^{3}\;{\rm km\;s^{-1}}$ with respect to the narrow H$\alpha$ line. Col. (11): Velocity shift of the profile at 
the Quarter-Maximum in units $10^{3}\;{\rm km\;s^{-1}}$ with respect to the narrow H$\alpha$ line.  Please see 
\S\ref{variability_parameter_plots} for a decsription of the measurement method and errors.}
\end{deluxetable}
\clearpage
\end{landscape}

\clearpage

\clearpage

\clearpage

\clearpage

\clearpage

\clearpage

\clearpage

\clearpage

\clearpage

\clearpage

\clearpage

\clearpage

\clearpage

\clearpage

\clearpage

\clearpage

\clearpage

\clearpage

\clearpage

\clearpage
\clearpage

\clearpage

\clearpage

\clearpage

\clearpage

\clearpage

\clearpage

\clearpage

\clearpage

\clearpage

\clearpage

\clearpage

\clearpage

\clearpage

\clearpage

\clearpage

\end{document}